\pdfoutput=1

\documentclass[longauth,twocolumn]{aa} 

\usepackage{graphicx}
\usepackage{xcolor}
\usepackage{array}
\newcolumntype{P}[1]{>{\centering\arraybackslash}p{#1}}

\usepackage[bookmarks=false,colorlinks=true,linkcolor=black,citecolor=cyan,filecolor=black,urlcolor=cyan]{hyperref}

\usepackage{amsmath}
\usepackage{multirow,tabularx}

\usepackage{makecell}
\usepackage{newtxtext,newtxmath}

\usepackage[T1]{fontenc}
\usepackage{ae,aecompl}

\usepackage{xspace}
\usepackage{siunitx}
\usepackage{bm}
\usepackage{syntonly}
\usepackage{booktabs}

\usepackage{threeparttable}
\usepackage{pdflscape}
\usepackage{longtable}
\usepackage{makecell}

\usepackage{ltablex}
\usepackage{boxhandler}
\usepackage{graphbox}

\graphicspath{{./figures/}}
\begin{document} 

   \title{Robust AGN and host-galaxy decomposition in optical spectral fitting}
   
    \titlerunning{AGN-host-galaxy decomposition} 
    
   \author{C. Aydar\inst{1, 2}\thanks{\href{mailto:caydar@mpe.mpg.de}{caydar@mpe.mpg.de}} \and
           A. Merloni\inst{1} \and
           G. Zeltyn\inst{3} \and
           C. Andonie\inst{1} \and
           B. Trakhtenbrot\inst{3} \and
           S. Bernal\inst{4, 5} \and
           Q. Wu\inst{6} \and
           J. Buchner\inst{1} \and
           M. Salvato\inst{1} \and
           T. Dwelly\inst{1} \and
           S. F. Anderson\inst{7} \and
           R. J. Assef\inst{8} \and
           F. E. Bauer\inst{9} \and
           W. N. Brandt\inst{10, 11, 12} \and
           S. LaMassa\inst{13} \and
           M. L. Mart\'inez-Aldama\inst{14} \and
           A. L. Rankine\inst{15} \and
           D. P. Schneider\inst{10, 16} \and
           Y. Shen\inst{6, 17} \and
           J. R. Brownstein \inst{18} \and
           H. Javier Ibarra-Medel\inst{19} \and
           A. M. Koekemoer\inst{13} \and
           M. Krumpe\inst{20} \and
           S. Morrison\inst{6} \and
           K. Nandra\inst{1} \and
           C. A. Negrete Pe\~naloza\inst{19} \and
           S. F. Sanchez\inst{21}
            }

   \institute{Max Planck Institute for Extraterrestrial Physics, Gie{\ss}enbachstra{\ss}e 1, 85748 Garching, Germany
        \and
            Excellence Cluster ORIGINS, Boltzmannstrasse 2, D-85748 Garching, Germany   
        \and
            School of Physics and Astronomy, Tel Aviv University, Tel Aviv 69978, Israel 
        \and 
            Departamento de Astronom\'ia, Universidad de Chile, Camino el Observatorio 1515, Santiago, Chile
        \and 
            Millennium Nucleus on Transversal Research and Technology to Explore Supermassive Black Holes (TITANS)
        \and 
            Department of Astronomy, University of Illinois Urbana-Champaign, Urbana, IL 61801, USA
        \and
            Department of Astronomy, University of Washington, Box 351580, Seattle, WA 98195, USA
        \and
            Universidad Diego Portales, Instituto de Estudios Astrof\'sicos, Facultad de Ingenier\'ia y Ciencias, Av. Ej\'ercito Libertador 441, Santiago, Chile 
        \and 
            Instituto de Alta Investigaci\'on, Universidad de Tarapac\'a, Casilla 7D, Arica, Chile
        \and 
            Institute for Gravitation and the Cosmos, The Pennsylvania State University, University Park, PA 16802, USA     
        \and
            Department of Astronomy \& Astrophysics, 525 Davey Lab, The Pennsylvania State University, University Park, PA 16802, USA
        \and
            Department of Physics, 104 Davey Laboratory, The Pennsylvania State University, University Park, PA 16802, USA
        \and
            Space Telescope Science Institute, 3700 San Martin Drive, Baltimore, MD 21218, USA
        \and
            Departamento de Astronom\'ia, Universidad de Concepci\'on, Casilla 160-C, Concepci\'on, Chile
        \and
            Institute for Astronomy, University of Edinburgh, Royal Observatory, Edinburgh EH9 3HJ, UK
        \and
            Department of Astronomy \& Astrophysics, The Pennsylvania State University, University Park, PA 16802, USA
        \and
            National Center for Supercomputing Applications, University of Illinois Urbana-Champaign, Urbana, IL 61801, USA
        \and
            Department of Physics and Astronomy, University of Utah, 270 S. 1400 E. E2108, Salt Lake City, UT 84112, USA
        \and
            Instituto de Astronom\'ia, Universidad Nacional Aut\'onoma de M\'exico, A.P. 70-264, 04510, Mexico, D.F., M\'exico
        \and
            Leibniz-Institut f\"ur Astrophysik Potsdam (AIP), An der Sternwarte 16, 14482 Potsdam, Germany
        \and
            Instituto de Astrof\'isica de Canarias, 38205 La Laguna, Tenerife, Spain
        }
           
   \date{}

  \abstract
  {
  Unraveling the growth of supermassive black holes and their connection to host galaxies requires disentangling the Active Galactic Nuclei (AGN) emission from that of the stellar populations in which they are embedded. 
  When an AGN spectrum is observed at different activity phases, if the spectral decomposition properly recognizes the nuclear and stellar components, key physical properties - such as black-hole mass, stellar mass, and stellar velocity dispersion — should remain consistent.
  These quantities are essential for scaling relations used to study the co-evolution of black holes and galaxies.
  We present a novel optical spectral-fitting approach that combines \texttt{pPXF} and \texttt{PyQSOFit} to robustly decompose spectra into stellar and AGN components.
  We apply this technique to three Sloan Digital Sky Survey samples with repeated optical spectra of the same objects at z$\leq0.55$: 32 changing-look AGN in bright and dim states, and 15 quasars and 15 galaxies with three single-epoch and one stacked spectrum each, spanning a range of signal-to-noise ratios.
  For comparison with the literature, we use SDSS spectra and photometric data from AGN in the eFEDS field, as well as Gemini and VLT observations of some of our selected changing-look AGN.
  This analysis enables us to evaluate the reliability of stellar mass, velocity dispersion, and black-hole mass measurements, especially in relation to the AGN-to-total continuum contribution ($f_{\rm{AGN}}$).
  For host-derived properties, especially when $f_{\rm{AGN}}<0.8$, our method yields consistent results.
  For single-epoch black-hole mass estimates from H$\alpha$ and H$\beta$, 3$\sigma$ confidence in the broad-line flux and full width at half maximum provides effective criteria for selecting reliable measurements.
  After applying these quality cuts, measurements across different epochs agree within uncertainties, and their reliability is confirmed by the alignment with previously established scaling relations.
  Many changing-look AGN in our sample exhibit “breathing” broad-line regions, as determined from H$\alpha$ analysis, while some deviate significantly, suggesting non-virialized systems across the spectral transition.
  }

\maketitle

\section{Introduction}  

Active Galactic Nuclei (AGN) are phases of the evolution of galaxies during which their central supermassive black hole (SMBH) is accreting matter and growing its mass.
These are very energetic processes, and their impact on the host galaxy has been studied both in simulations and in observations alike (see \citealt{Fabian2012, Harrison_RamosAlmeida2024} for a review).
Scaling relations that correlate properties of the SMBH with properties of the host-galaxy and its bulge indicate a possible co-evolution of these systems \citep[e.g.,][]{Magorrian1998, Gebhardt2000, Graham2001, Tremaine2002, Kormendy2013}, driven by feedback from the AGN driving outflows, jets and winds that can interact with the galaxy up to scales that are much larger than the SMBH sphere of influence \citep[e.g.,][]{Fabian2012, ForsterSchreiber2014, Morganti2017}.
Studies with high spatial resolution have been crucial for mapping the influence of outflows and jets in the interstellar medium to either enhance or suppress star formation in the galaxy as a consequence of the SMBH activity \citep[e.g.,][]{Balmaverde2019, Venturi2021, Riffel2023b, Riffel2023a, Speranza2024}.

However, when the AGN is not observed with sufficient spatial resolution, e.g., in single-epoch fiber-fed spectroscopy, it is crucial to disentangle the host galaxy and AGN emission in the combined spectra to determine their relative contributions.
Due to the advent of large spectroscopic surveys such as the Sloan Digital Sky Survey \citep[SDSS,][]{York2000, Kollmeier2026}, and especially considering the new generation of big data from, e.g., the Dark Energy Spectroscopic Instrument \citep[DESI,][]{DESI2026}, the 4-metre Multi-Object Spectroscopic Telescope \citep[4MOST,][]{4MOST2019}, and the Prime Focus Spectrograph \citep[PFS,][]{Tamura2016}, it is necessary to have a reliable method to separate the emission from the stellar populations in the galaxy from the emission from the growth of the SMBH in the large samples of fiber-fed spectra that are and will be available to the scientific community.
Part of the challenges in developing a method applicable in an automated way to thousands or even millions of spectra is the variety of AGN spectra, considering the diversity in obscuration, accretion rate, AGN luminosity, and the influence of the stellar populations \citep[e.g.,][]{Aydar2025}.
With the long-term objective to be able to analyze large samples of AGN, selected in different ways as part of the SDSS-V Black Hole Mapper (BHM) program \citep{Kollmeier2026}, the specific goal of this paper is to devise and validate a flexible and robust optical fitting method, and test its performance on a set of SDSS spectra. 

To test the reliability of the fitting method's results, one could use mock samples, in which the fit measurements should match the input parameters that generated the spectra \citep[e.g.,][]{Santini2012, Mobasher2015, Buchner2024}.
This approach, however, can be highly biased due to its dependence on the templates used to construct the mock spectra, and it becomes unrealistic when comparing the test to observed data, given the difficulty of realistically applying noise \citep[e.g.,][]{Bongiorno2012}.
However, observational data can be used to assess the quality of a fitting model, as observing the same galaxy at different phases of its activity, across various signal-to-noise ratios (S/N), or with different telescopes should yield the same output measurements.

Thus, we selected samples of AGN observed multiple times with SDSS: one sample of `changing-look' AGN \citep[see][for a recent review]{Ricci_Trakhtenbrot2023} in their bright and dim states, and two other samples that allow the comparison of single-epoch and stacked spectra of the same source for (X-ray selected) AGN.
For each object in these samples, some parameters should remain constant, such as galaxy stellar mass, stellar velocity dispersion, and black-hole mass, as we do not expect them to vary on such short timescales.
These quantities are also often used in the scaling relations to understand the possible co-evolution of AGN and their hosts \citep[e.g.,][]{Kormendy2013, Reines2015, Suh2020, Pucha2025}.

This manuscript primarily focuses on validating the proposed methodology.
However, the method can also yield many relevant physical parameters for studying either the AGN or their host galaxies, as it provides continuum, emission, and absorption line measurements for both components separately.

The structure of the paper is as follows.
Section \ref{sec:methodology} describes the methodology for fitting spectra with the host galaxy and AGN decomposition.
The samples used for validating this method are presented in Section \ref{sec:data}.
Section \ref{sec:results:general} presents the method validation by comparing both the outputs of different spectra of the same objects (internal validation) and also the results of our method with analysis from photometric fitting (external validation), with a focus on host galaxy properties (stellar mass and stellar velocity dispersion) and on AGN properties (SMBH mass).
Our results are discussed in the context of scaling relations and `breathing' broad line regions (BLRs) in Section \ref{sec:discussion}, and the paper concludes with recommendations for the use of the fitting method in Section \ref{sec:conclusions}.
Throughout this manuscript, we assume a flat $\Lambda$CDM cosmology with $\Omega_{\rm{M}} = 0.3$ and H$_0 = 70$ km s$^{-1}$ Mpc$^{-1}$.

\section{Methodology}
\label{sec:methodology}

The goal of this work is to provide and validate a reliable method for fitting the contribution of AGN and the host galaxies in optical spectra where both sources of emission are entangled.
Our strategy is to accurately remove the host-galaxy contribution from the observed spectra, and then fit the residual AGN-dominated spectrum with a procedure optimized to account for the AGN continuum and emission lines.
Importantly, removing the host galaxy with high precision enables us to study the properties of even weak emission lines in greater detail, which would otherwise be blended with stellar emission.

To obtain accurate measurements of both the host galaxy emission and the AGN, we use two software packages, developed with a focus on either inactive galaxies or quasars.
We use \texttt{pPXF} \citep{Cappellari2023} for modeling the stellar populations \citep[while still adding the AGN components to avoid overestimates and to account for known degeneracies, e.g.,][]{Tadhunter1996, stasinska2006} and \texttt{PyQSOFit} \citep{PyQSOFit} for fitting the AGN emission.
\texttt{pPXF} was primarily built to focus on low-redshift inactive galaxies, and the adaptations made to the code to account for the AGN emission are not as complex as the fit from \texttt{PyQSOFit}.
On the other hand, \texttt{PyQSOFit} was built with a focus on quasars, and its host galaxy subtraction option often fails to provide accurate fits when accounting for both AGN and stellar populations.
For these reasons, we decided to combine the best aspects of both codes to achieve a reliable decomposition of AGN and host galaxies.
We perform three iterations of the fits with different model components for each spectrum to obtain a trustworthy AGN and host-galaxy decomposition, as described in the following sections.
Regarding computational efficiency, a complete run with both \texttt{pPXF} and \texttt{PyQSOFit} on a single CPU core takes $\sim3$ minutes for each SDSS spectrum.

\subsection{pPXF}
\label{sec:ppxf}

\begin{figure*}[t]
     \includegraphics[width=\textwidth]{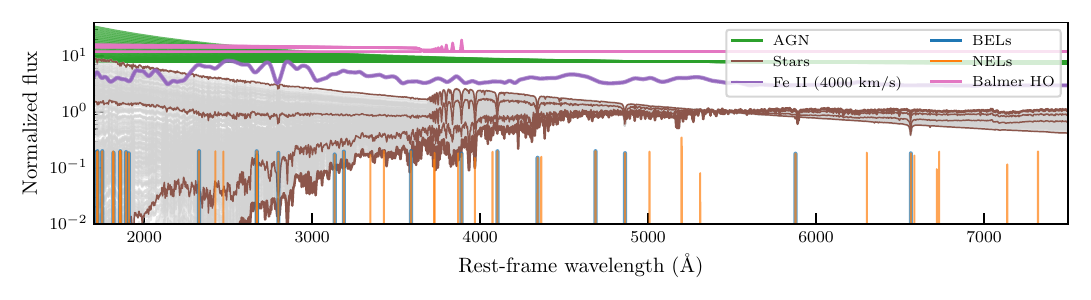}
     \caption{Templates used in \texttt{pPXF} fits to prepare for the galaxy subtraction.
     The AGN power-law continuum is displayed in green, with different power-law indices.
     The E-MILES stellar populations are shown in gray, with some of the SSPs highlighted in brown.
     The Fe II component (broadened with a 4000 km s$^{-1}$ FWHM, as an example) is shown in purple.
     The Balmer continuum and higher-order emission are shown in pink.
     The wavelengths of the broad emission lines are shown in blue, those of the narrow emission lines in orange; they often overlap for lines that have both such components.
     Except for the SSPs, each template is displayed in a different normalization for visualization purposes.}
\label{fig:method:templates_ppxf}
\end{figure*}

The code \texttt{pPXF} \citep[][and references therein]{Cappellari2023} is a \texttt{Python} implementation of the Penalized Pixel-Fitting method for spectral fitting.
It fits the observed spectrum with a set of stellar population templates and associated kinematic components to extract the stellar and gas kinematics, as well as the properties of the predominant stellar populations. 
\texttt{pPXF} was originally built to fit the spectra of inactive galaxies; hence, we use it only to obtain information regarding the stellar populations of the host galaxy (e.g., stellar velocity dispersion, stellar mass, and stellar population age and metallicity), and not to obtain the AGN measurements.
Following \citet{Bernal2025}, we adapted the algorithm to also model AGN in their bright states, where the host galaxy's emission is partly outshone by the quasar.
Figure \ref{fig:method:templates_ppxf} shows the templates used in our fitting procedure with \texttt{pPXF}: power-law featureless continuum (AGN), stellar populations (SSPs), Fe II pseudo-continuum (in the Figure we display only the example with full width at half maximum $\rm{FWHM}=4000$ km s$^{-1}$), Balmer continuum and high-order emission (Balmer HO), broad emission lines (BELs), and narrow emission lines (NELs).

For our implementation of \texttt{pPXF}, we used Simple Stellar Populations (SSPs) from E-MILES \citep{Vazdekis2016} with Padova isochrones \citep{Girardi2000} and a \citet{Salpeter1955} Initial Mass Function.
We adopted a model grid with $\rm log(t/yr) = 7.8-10.2$ in 0.1 dex steps, with the metallicities of -1.71, -1.31, -0.71, -0.4, 0.0, 0.22 in [M/H] (in which 0.0 corresponds to solar metallicity).
The stellar component has four kinematic moments, with the velocity ranging from -300 to 500 km s$^{-1}$, the velocity dispersion from 10 to 500 km s$^{-1}$, and the skewness and kurtosis vary from -0.3 to 0.3.
The description and references for the fitting of the continuum are also described in Table \ref{tab:ppxf_continuum}.

To account for the emission lines in the AGN spectrum, we added templates with Gaussians for the narrow and broad lines.
For the narrow lines, we included the Balmer lines in the optical domain (H$\alpha$, H$\beta$, H$\gamma$, and H$\delta$), [S II], [N II], [O I], [O II], [O III], [Ne III], and [Ne V] optical doublets, He I, and Mg II.
The velocity offset range of the narrow lines goes from -800 to 500 km s$^{-1}$, with velocity dispersion from 1 to 300 km s$^{-1}$.
We also considered Gaussians with broader components for the Balmer lines, He I, and Mg II, which have velocities ranging from -500 to 300 km s$^{-1}$ and velocity dispersions from 400 to 10000 km s$^{-1}$.
We consider a simplified approach of tidying all narrow lines fitted with a single Gaussian and a similar approach for the broad lines, since more complex fitting will be provided by \texttt{PyQSOFit}.
The emission line fitting before the host-galaxy decomposition is not accurate, so the \texttt{pPXF} fit of the lines is only used to guarantee the self-consistency of the fit in each iteration.

We also included components to account for the AGN contribution to the continuum.
We generated featureless power-law continua with
$$f_{\lambda}(\lambda)=f_{\lambda}(\lambda_0)\left(\frac{\lambda}{\lambda_0}\right)^\alpha\,,$$
with $\alpha$ varying from -3.0 to 0.0 in steps of 0.1.
The Balmer continuum emission templates have an optical depth range $\tau$ from 0.1 to 2.0 in steps of 0.1, while the Balmer High Order templates have FWHM from 1000 to 11000 km s$^{-1}$.
The velocity offset of the Balmer emission is tied to that of one of the broad emission lines.

The Fe II pseudo-continuum templates used are a combination of those by \citet{BorosonGreen1992} for optical wavelengths and by \citet{VestergaardWilkes2001} for ultraviolet wavelengths.
We broadened such templates with different values of the FWHM: 1000, 1200, 1400, 1600, 1800, 2000, 2400, 2800, 3400, 4000, 4800, 5800, 7000, 8400, 10000, and 11800 km s$^{-1}$.

The approach to add the appropriate Balmer- and Fe II pseudo-continuum templates required running \texttt{pPXF} (at least) twice.
In the first run, we considered all other components of the fit (i.e., SSPs, BELs, NELs, and the power-law continuum) without including the Balmer and Fe II templates; in particular, we obtained a first estimate of the BEL kinematic width from this run.
Then, we selected the Balmer High-Order and Fe II templates that were broadened to the closest value of the broad-line width, assuming the two are dynamically linked in the BLR of the AGN \citep[e.g.][]{Barth2013, Ji2025}.
In subsequent runs, with the broadened Balmer- and Fe II pseudo-continua and the other components included in the fit, we confirmed that the model converged. 
For all the spectra considered here, only one iteration was enough to reach convergence between the width of the BEL component and the Balmer High Order and Fe II templates.

For the fit of the observed spectra, we first perform the de-reddening of the observed flux due to the Milky Way extinction using the dust map from \citet{Schlegel1998} and the adapted values based on SDSS from \citet{Schlafly2011}, and the \citet{Cardelli1989} extinction curve with $R_V = 3.1$.
Then, we set the spectra to the rest frame using the SDSS pipeline redshift.
To estimate the uncertainties associated with our measurements, we performed Monte Carlo iterations by adding random noise within each pixel's noise range.
The provided estimates and statistical uncertainties correspond to the mean and standard deviations of 25 iterations.

\begin{figure*}[t]
     \includegraphics[width=\textwidth]{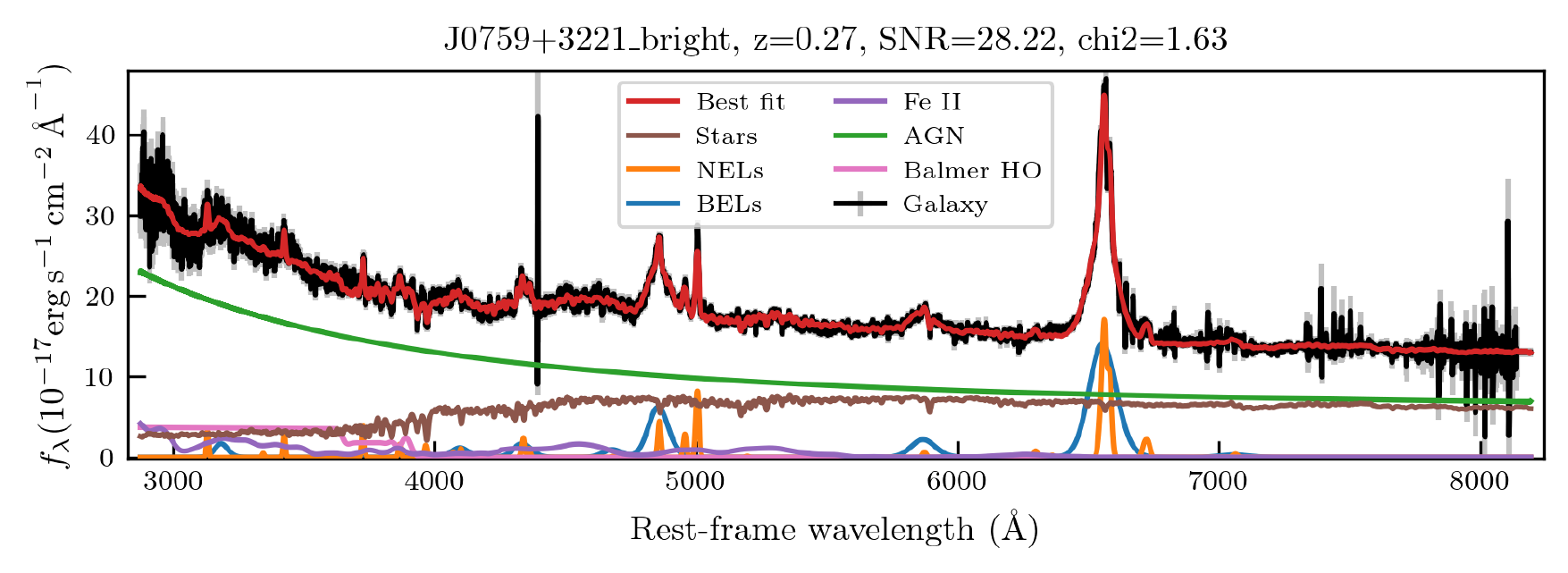}
     \caption{Example \texttt{pPXF} fit.
     The templates are color-coded as in Fig. \ref{fig:method:templates_ppxf}, with the observed spectra in black and the overall fit in red.
     This fit is an example of a successful decomposition of the host galaxy (brown) and the continuum components of the AGN (green, pink, and purple).}
\label{fig:method:ppxf_fit}
\end{figure*}

Figure \ref{fig:method:ppxf_fit} shows an example of a \texttt{pPXF} fit.
Our primary focus is on the output associated with the stellar populations, which is displayed as the brown component.
\texttt{pPXF} provides the weight of the linear combination of the templates used for the fit.
From the weight of the AGN emission (the sum of the weights of the power-law, the Fe II, and the Balmer continuum and high-order emission) in comparison to the contribution of the SSPs to the continuum, we can estimate the fraction of the AGN emission for each spectrum, which will be referred to as AGN continuum contribution ($f_{\rm{AGN}}$).
Since this measurement encompasses the entire wavelength range, it is also affected by redshift.

From the SSP kinematic components, we obtain the stellar velocity dispersion ($\sigma_*$) from the best-fitting line-of-sight velocity distribution returned by pPXF, which is mainly constrained by prominent absorption features  (e.g., Ca H+K, Mg I b).
The output also provides the ages and metallicities of the stellar populations, weighted by both mass and light.
We also estimate the aperture stellar masses ($\overline{M}_*$), i.e., the stellar mass obtained by the spectral fitting without aperture corrections, which corresponds to a lower limit of the stellar mass of the galaxy.
Having derived the properties of the host's stellar populations (e.g., stellar mass, stellar velocity dispersion, stellar population age and metallicity, weights for each SSP template), we subtract the SSP templates from the de-reddened observed spectrum to obtain the isolated AGN emission.

\subsection{PyQSOFit}
\label{sec:pyqsofit}

\begin{figure*}[t]
     \includegraphics[width=\textwidth]{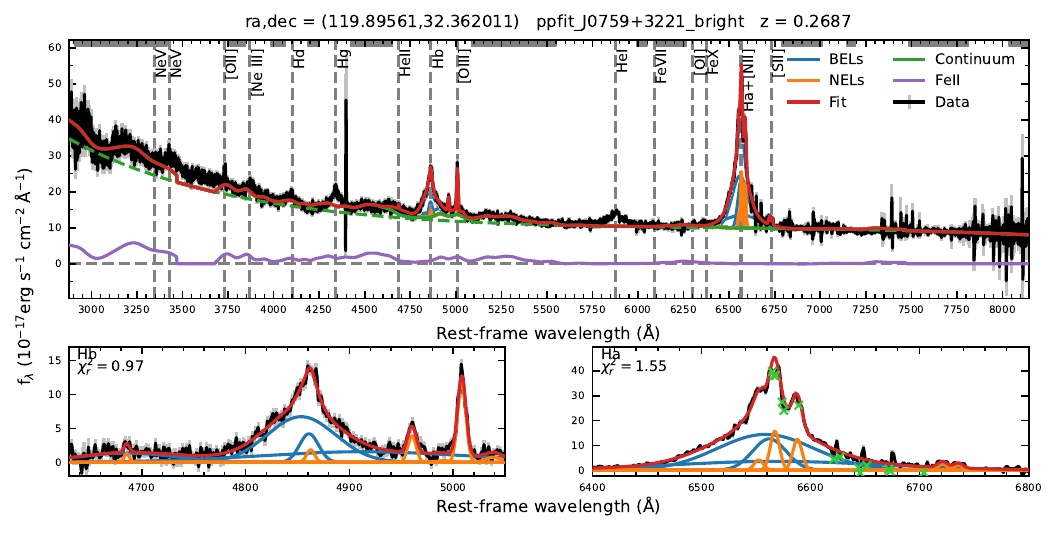}
     \includegraphics[width=\textwidth]{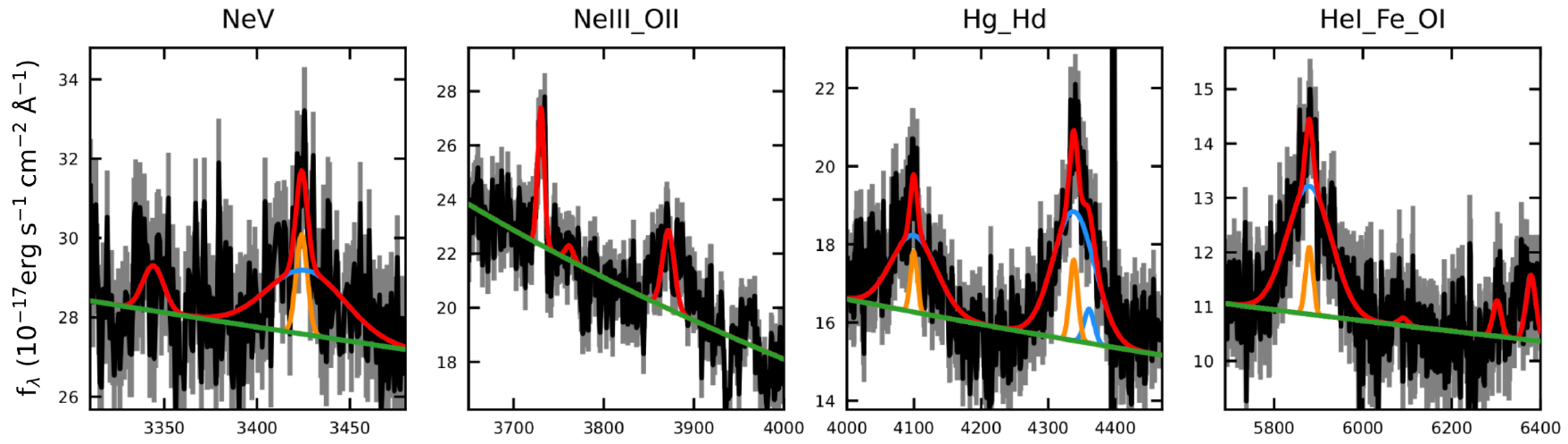}
     \caption{Fit from \texttt{PyQSOFit} of the same spectrum as in Fig. \ref{fig:method:ppxf_fit} after the subtraction of the host galaxy.
     The top panel shows the overall fit in red over the data in black.
     The middle panels display the fits for the H$\alpha$ and H$\beta$ complexes, and the bottom panels present the fits for other emission lines that were treated locally due to being fainter.
     The colors of the templates are as in Fig. \ref{fig:method:templates_ppxf}.
     The gray bands at the top indicate the wavelength regions considered for the continuum fit.}
\label{fig:method:pyqsofit_fit}
\end{figure*}

After the host-galaxy contribution is properly measured and subtracted from the observed spectra, we can fit the isolated AGN emission using \texttt{PyQSOFit} \citep{PyQSOFit}.
\texttt{PyQSOFit} was designed to fit the optical spectra of SDSS quasars, and it includes Monte Carlo estimates of the uncertainties of the fitting results, with a similar setting of 25 runs that was implemented in \texttt{pPXF}.
\texttt{PyQSOFit} provides a fit for each model component, rather than a linear combination as \texttt{pPXF}, and this complexity yields more accurate results for the AGN component.

We use a similar approach to the one described in \citet{Wu2022}.
For the continuum, we fit a combination of a power-law and a third-order polynomial to account for objects with intrinsic dust reddening associated with the AGN environment rather than the host galaxy.
The flexibility of the polynomial fit also includes the contributions of the Balmer continuum and higher-order emission.
We also include Fe II pseudo-continuum templates, but this time the width of the line complexes is fitted to the spectrum, instead of approximating a FWHM as for the \texttt{pPXF} fit.
We use the Fe II template provided by \texttt{PyQSOFit}, which is a combination of the templates from \citet{BorosonGreen1992}, \citet{VestergaardWilkes2001}, \citet{Tsuzuki2006}, and \citet{Salviander2007}.
There is, however, a gap in the Fe II models in the wavelength range $\sim3490-3730$ Å, which is the result of missing high-order Balmer lines and the Balmer edge at 3646 \AA\ \citep[see e.g.,][]{Park2022}.
The parameters and references for the fitting of the AGN continuum by \texttt{PyQSOFit} are also listed in Table \ref{tab:pyqsofit_continuum}.

The emission lines are fitted with 1 to 3 Gaussians per line, with the parameters listed in Table \ref{tab:emission_lines_pyqsofit} and further discussed in Appendix \ref{sec:app:emission_line}.
Using several Gaussians with more constrained parameters is very important to avoid misleading estimates, since we are able to disentangle lines that could be blended, such as [N II] and [S II] from H$\alpha$ (see the middle right panel of Fig. \ref{fig:method:pyqsofit_fit}).
Regarding the subtraction of absorption lines from the SSP contribution, no significant dip was noticed in the visual inspection of the Balmer line profiles in our fits, indicating that the absorption lines are accounted for consistently with the host galaxy templates and do not significantly affect the emission lines.

From these fits, we obtain the main properties of the AGN.
Figure \ref{fig:method:pyqsofit_fit} displays an example of the \texttt{PyQSOFit} fit to the same spectrum as in Fig. \ref{fig:method:ppxf_fit} after its host-galaxy subtraction.
The lower panel of Fig. \ref{fig:method:pyqsofit_fit} shows the fits for faint lines, and although they were measured, some quality cuts as recommended by \citet{Wu2022} should be applied before considering them a reliable detection.
We emphasize the importance of measuring AGN emission lines after decomposing the host-galaxy contribution to obtain unbiased physical parameters of SMBH activity.

\section{Samples}
\label{sec:data}

In this section, we describe the samples used to test and validate our method; all are based on SDSS optical spectra.
In Section \ref{sec:internal_sample}, we present three samples of spectra selected for comparing different observations of the same objects for an internal validation of our methodology.
Section \ref{sec:external_sample_stellarmass} instead  presents four samples from the eFEDS field for the comparison of our method with results from other AGN/host decomposition methods that rely on photometry and either spectral energy distribution (SED) fitting or image decomposition.
At last, Section \ref{sec:external_sample_velocity_dispersion} presents the VLT and Gemini spectra used to compare the stellar velocity dispersion measurements.

The Sloan Digital Sky Survey has been providing optical spectra of all types of astronomical objects since its first generation \citep{York2000} to the most recent release \citep{sdss_dr19}.
The SDSS-I and II performed their observations using a spectrograph \citep{Smee2013} with 3" (diameter) fibers \citep{York2000, Eisenstein2001, Strauss2002, Gunn2006, Abazajian2009}.
SDSS-III introduced the Baryon Oscillation Spectroscopic Survey \citep[BOSS,][]{Eisenstein2011, Dawson2013_boss} spectrographs using 2" (diameter) fibers.
The survey migrated from the plate design to robotic fibers during SDSS-V, while using the same spectrographs \citep{Blanton2017, Abdurrouf_2022_sdss_dr17}, and the ongoing generation SDSS-V \citep{Kollmeier2026} has already announced two data releases \citep{Almeida2023_sdss_dr18, sdss_dr19}, with the next data release also including targets in the Southern Hemisphere using the LCO 2.5-m telescope \citep{Bowen_Vaughan1973}, with a 1.3" (diameter) fiber.

\subsection{Internal comparison}
\label{sec:internal_sample}

For internal comparison, i.e., comparing different observations of the same sources, we constructed three custom-made small samples of SDSS AGN data.
The legacy of decades of SDSS observations is crucial for this analysis, as it allows comparison of data from the same source across years.
This enables, in particular, the analysis of objects that might have changed their emission profile.
For most spectra, we used the \texttt{idlspec2d} pipeline version of SDSS (\citealt{Bolton2012, Dawson2013_boss}; Morrison et al. in prep.).

The sample `CL' consists of 32 changing-look AGN: 20 selected from \citet{Zeltyn24} and 12 newly identified CL-AGN \citep{Zeltyn26} from SDSS-V \citep{Kollmeier2026}.
Two spectra are associated with each AGN, corresponding to their bright (usually quasar-dominated phase, with broad H$\alpha$ and a power-law continuum) and dim (typically host-galaxy-dominated, as seen from absorption lines and a continuum shape consistent with stellar populations, with narrower emission lines) states. 
By selection, there is often a significant dimming of the broad H$\beta$ emission between the bright and dim states. 
The redshift range of the sources is $0.05 \leq z \leq 0.5$. 

We also define two more samples, GAL and QSO, each with 15 AGN detected by eROSITA \citep{Predehl2021, Merloni2024} that the SDSS pipeline classified as \texttt{GALAXY} and \texttt{QSO}, respectively \citep[see the SDSS DR19 Value Added Catalog in][]{sdss_dr19}.
While the \texttt{QSO} class clearly indicates that the AGN is unobscured, the classification as \texttt{GALAXY} indicates the presence of features typical of galaxies, though their point-like X-ray detection suggests the presence of an AGN.
Each source has three spectra from co-added observations (regular observations typically taken within one day) and one stacked spectrum that combines all SDSS observations of this object (from 2020 to 2025), referred to as \texttt{daily} and \texttt{allepoch}, respectively.
The \texttt{daily} spectra of each source comprise a median S/N range through all spectra (\texttt{SN\_MEDIAN\_ALL}) of $3.16<\rm{S/N}<5$, $5\leq\rm{S/N}<10$, and $\rm{S/N}\geq10$.
Unless the S/N bin is explicitly specified, we present results for all S/N ranges.
The sources have $0.02 \leq z \leq 0.55$ and no SDSS pipeline flag indicating a problematic redshift (\texttt{ZWARNING}$=0$).
For both samples, we aimed to select cases in which both host-galaxy and AGN contributions would be present: for sample GAL, we gave preference to spectra with \texttt{SUBCLASS} containing keywords as \texttt{AGN} or \texttt{BROADLINE}, while for sample QSO, we prioritized keywords as \texttt{STARFORMING} or \texttt{STARBURST}.

\begin{figure}[t]
     \includegraphics[width=\columnwidth]{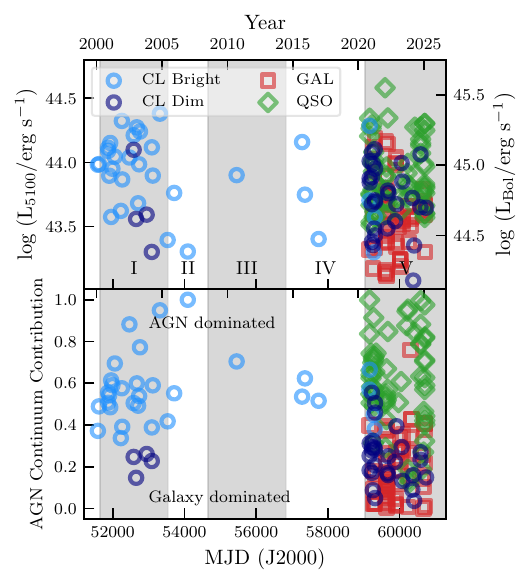}
     \caption{{\it Top panel:} observed luminosity at 5100 \AA\ (left axis) and bolometric luminosity (right axis) vs. time of observation for all the spectra analyzed in this work. {\it Bottom panel:}  fraction of the AGN emission ($f_{\rm AGN}$) as derived from the \texttt{pPXF} decomposition vs. Modified Julian Date.
     Sample CL is indicated in light blue and dark blue circles for spectra taken during the bright and dim stages of the AGN, respectively.
     Sample GAL is shown in red squares and sample QSO in green diamonds.
     For reference, the various SDSS generations are represented by gray or white shaded areas, with numbers I-V as indicators.}
\label{fig:sample:mjd}
\end{figure}

With this sample selection, we have a total of 184 spectra (64 CL, 60 GAL, and 60 QSO) from 62 sources (32 CL, 15 GAL, and 15 QSO), spanning different host-galaxy versus AGN contributions and different S/N.
Figure \ref{fig:sample:mjd} shows the different generations of SDSS spectra that were considered in our three samples with regard to their luminosities (our measurement of $\rm{L}_{5100}$ and bolometric luminosity according to \citealt{Runnoe2012_erratum}) and $f_{\rm{AGN}}$.
$\rm{L}_{5100}$ is estimated from the average flux at 5080-5120 \AA\ and the SDSS pipeline redshift.
Sample CL spans different generations of SDSS, with a similar range of luminosities ($\langle\rm{log (L}_{5100}/\rm{erg\ s}^{-1})\rangle=43.9\pm0.3$ for the bright sample, $43.7\pm0.3$ for the dim sample). 
For most sources in the CL sample, the first observation is brighter than the second.
Sample GAL shows slightly lower luminosities ($\langle\rm{log (L}_{5100}/\rm{erg\ s}^{-1})\rangle=43.6\pm0.3$), as expected for a sample of narrow-line AGN compared to broad-line AGN \citep[e.g.,][]{Koss2017, BarquinGonzalez2024}.
Though the QSO sample has the spectra with the highest luminosities, the mean luminosity is the same as that of sample CL ($\langle\rm{log (L}_{5100}/\rm{erg\ s}^{-1})\rangle=43.9\pm0.3$).
These similar average luminosities are due to their (X-ray) selection, which suggests that an AGN component is present, and the construction of SDSS subclasses that indicate star formation for the QSO sample and AGN activity in the GAL sample. 
Interestingly, most bright CL AGN and QSO have AGN continuum contribution $f_{\rm AGN}$>0.4, while the GAL and dim CL states are mostly $f_{\rm AGN}$<0.4, as expected.

\subsection{Testing against external samples: Stellar mass}
\label{sec:external_sample_stellarmass}

To compare the results of our methodology with other works in the literature, we decided to use the eROSITA Final Equatorial Depth Survey field \citep[eFEDS;][]{Brunner2022}, which provides extensive multi-wavelength coverage.
Out of the 28k X-ray point-like sources detected in eFEDS with eROSITA, 13k were followed up with SDSS as part of the SPectral IDentification of ERosita Sources program \citep[SPIDERS;][]{Aydar2025}.
After removing $\sim3\%$ of stars and doing some quality cuts for guaranteeing reliable counterparts and redshift, we applied our spectral fitting method to sources at z$<1$ which were either red ($g-r>0.4$) or extended according to the Legacy Survey \citep{Dey2019}, which corresponded to 3702 AGN candidates of having a significant host galaxy (compared to the total eFEDS-SPIDERS sample of 11\,864 AGN; Aydar et al., in prep.).

To compare our measurements of the stellar mass with the literature, it is crucial to estimate the total galaxy stellar mass ($M_*$) from the aperture stellar mass ($\overline{M}_*$) by applying aperture corrections.
These corrections were estimated based on the ratio between the total photometric flux of the galaxy in the \textit{r}-band according to Legacy Survey DR10 \citep{Dey2019, Zenteno2025} compared to flux integrated over the same band from each SDSS spectrum (i.e., within the SDSS aperture), and the AGN contribution to the continuum by considering:
\begin{equation}
    M_* = \overline{M}_* \times \frac{(\rm{Total\ flux}/\rm{Aperture\ flux})-f_{\rm{AGN}}}{1-f_{\rm{AGN}}}
\end{equation}
The factor multiplying the aperture stellar mass to obtain the aperture corrected stellar mass will be referred to as \texttt{aperture correction factor}.
By definition, the \texttt{aperture correction factor} must be greater than 1, unless the AGN varied significantly between the two observations (photometric and spectroscopic).
This is a simplistic aperture correction that assumes the light collected by the fiber is proportional to the galaxy's total light, that there are no significant color gradients within the galaxy, and that the AGN luminosity does not change between observations.

To consider external photometry-based validations, we use three well-documented comparison sets:
i) \citet{Li2024_HSC} analyzed AGN in eFEDS using the Subaru Hyper Suprime-Cam (HSC), obtaining images in the \textit{grizy} bands, which enabled them to perform a 2D AGN-host image decomposition, in which they remove the nuclear component from the images and use the residual host to compute the physical parameters;
ii) \citet{Buchner2024} used available X-ray-to-infrared data in the eFEDS field for the SED-fitting code GRAHSP, which aims to provide unbiased and accurate host galaxy parameters of AGN, specifically addressing the overestimation issues found in previous methods, but consequently yielding larger (and more realistic) error bars;
iii) \citet{Yu2023} also used X-ray-to-infrared photometric data in the eFEDS field for SED-fitting, but using the widely used code CIGALE, updated to treat AGN with more detail than the original version that mainly focused on inactive galaxies.
These three methods provide sophisticated approaches for disentangling the host galaxy from the AGN emission using photometry and can serve as a comparison for our spectral method, especially regarding the (aperture-corrected) stellar mass measurements.

All of these samples have their stellar masses calculated using the Chabrier Initial Mass Function, whereas we used the Salpeter Initial Mass Function; therefore, we apply the $+0.25$ dex correction described by \citet{Bernardi2010_IMF}. 
After matching our sources by eFEDS ID, we cleaned the matched samples to avoid problematic cases or changes due to excessive AGN variability by excluding:
\begin{itemize}
    \item $f_{\rm{AGN}}>$0.8;
    \item Redshifts disagreeing with $>0.01$ error;
    \item Unreasonable aperture correction:
        \begin{itemize}
            \item $\texttt{aperture correction factor}<0.5$, i.e., fiber flux larger than the total flux by a factor $>2$;
            \item $\texttt{aperture correction factor}>1.5 \times \left(\dfrac{20\ \rm{kpc}}{2 d_A}\right)^2$, i.e., the aperture correction being larger than required for a galaxy with a 20 kpc radius at the respective redshift, where $d_A$ is the angular distance\footnote{The 1.5 scale for this criterion comes from the attempt to conservatively exclude most of the outliers when considering the aperture correction factor with regards to redshift.};
        \end{itemize}
    \item Difference between high and low estimates of the Bayesian distribution of the logarithm of the stellar mass from GRAHSP $>1$ (more than one order of magnitude, as in \citealt{Buchner2024});
    \item Flag \texttt{above\_Mcut} from HSC set to False, indicating magnitude saturation;
\end{itemize}
It is not straightforward to compare the AGN contribution for the different methods because we estimate this parameter within the SDSS aperture, which should yield higher values than considering the whole galaxy with photometry.
After applying the cuts mentioned above, the matching sample with HSC contains 566 objects, the one with GRAHSP contains 734 sources, and the one with CIGALE contains 634 objects.

For the benchmark samples in Section \ref{sec:internal_sample}, the redshift range is z $\leq0.55$ to guarantee that H$\alpha$ is included in the SDSS wavelength coverage.
For the eFEDS sample, however, we have sources at higher redshifts, with z $\leq1$.
We applied the cut at z $=1$ based on visual inspections \citep[][]{Aydar2025}, since in the SDSS SPIDERS sample, no clear stellar population features (e.g., absorption lines and continuum shape) were visible for z $>1$ sources. 

\subsection{Testing against external samples: Stellar velocity dispersion}
\label{sec:external_sample_velocity_dispersion}

For the comparison with the stellar velocity dispersion, we consider VLT/X-Shooter and Gemini/GMOS observations of the matching sources from the CL sample (Section \ref{sec:internal_sample}), presented in \citet{Zeltyn26}.
Of the total 11 matching sources, 9 were observed with X-Shooter 1.6" slit, while the other 2 were observed with GMOS 0.5" slit.
These observations should provide a larger spectral resolution than the SDSS data.

\section{Method validation}
\label{sec:results:general}

In this Section, we validate our spectral-fitting method by comparing measurements from different observations of the same active galaxies and by comparing spectral-fitting results with SED-fitting and image-decomposition results for the same sources.
The analysis of the properties associated with the host galaxy is in Section \ref{sec:results:host_galaxy}, and the properties of the AGN in Section \ref{sec:results:agn}.
Based on our analysis, we recommend thresholds for measurement reliability.
We also address the S/N dependency in our systematic scatters, which show that the suggested quality cuts are sample-dependent and should be reassessed for studies relying on spectra obtained from other instruments and/or with different S/N and redshift distributions.

\subsection{Host-galaxy properties}
\label{sec:results:host_galaxy}

The main optical spectral measurements associated with the stellar populations of the host galaxy used for scaling relations are the stellar mass ($M_*$) and the stellar velocity dispersion ($\sigma_*$).
Given the known degeneracies between AGN power-law emission and young stellar populations' emission in the blue part of the optical continuum \citep[e.g.,][]{Tadhunter1996, Holt2007, Cardoso2017}, the measurement of $M_*$, which is based on a redder part of the spectra, should be more reliable than the measurement of $\sigma_*$, which is based on the absorption lines in a bluer part of the spectra.

\subsubsection{Stellar Masses}
\label{sec:results:stellar_mass}

The measurement of the aperture stellar masses ($\overline{M}_*$) is based on the \texttt{pPXF} output of the mass-to-light ratio (M/L) from the SSP fitting, to which we associate the flux from the spectra integrated in the DECam-\textit{r} photometric band coverage (see Appendix \ref{sec:app:stellar_mass}).
However, our aperture stellar mass estimates do not necessarily reflect the galaxy's total stellar mass, as at low redshifts, the SDSS fiber may be observing only an inner part of the galaxy.
For samples GAL and QSO, since the observations were all from SDSS-V, the comparison should not be affected by aperture effects; instead, differences in $\overline{M}_*$ could indicate effects of the S/N between the different observations and the stacks of the same object, and also provide systematic uncertainties of the fitting results.
For the CL sample, however, there is a difference between the observations from SDSS-I and -II 3" fiber and the 2" or 1.3" observations from SDSS-III onwards (for APO or LCO, respectively).
We emphasize the dependence of $\overline{M}_*$ with redshift and galaxy morphology, so no conclusions regarding the overall stellar mass of the host galaxy should be drawn without providing aperture corrections.
However, our internal comparison relies on $\overline{M}_*$ to avoid other sources of uncertainty or systematic biases associated with the aperture correction.

\begin{figure}[t]
     \includegraphics[width=\columnwidth]{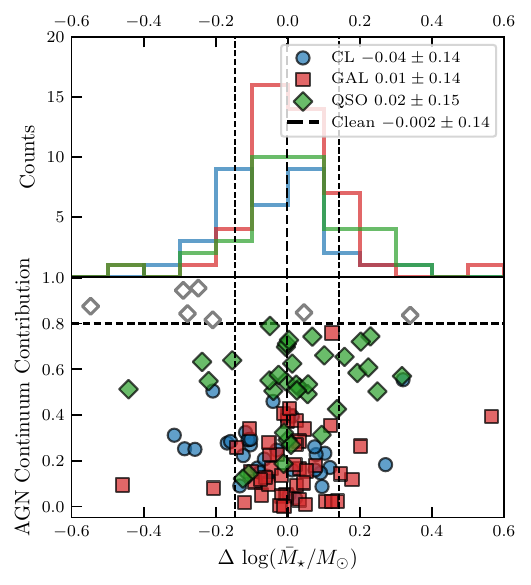}
     \caption{The difference of the aperture stellar masses as a histogram for the different samples on the top and against the fraction of the AGN contribution to the continuum ($f_{\rm{AGN}}$) at the bottom.
     For sample CL (blue), the symbols indicate $\log(\overline{M}_*/M_\odot)_{\rm dim} - \log(\overline{M}_*/M_\odot)_{\rm bright}$, while for samples GAL and QSO (red and green, respectively) we display $ \log(\overline{M}_*/M_\odot)_{\rm allepoch} - \log(\overline{M}_*/M_\odot)_{\rm daily}$.
     The $f_{\rm{AGN}}$ values shown correspond to the bright-state spectra for the CL sample and to the \texttt{daily} spectra for the GAL and QSO samples.
     The inverse-variance-weighted mean and standard deviation for each sample are shown in the legend.
     The lower panel shows a horizontal dashed line at $f_{\rm{AGN}}=0.8$, which can be used as a threshold to exclude spectra whose $\overline{M}_*$ difference is more significant, indicating less reliable measurements.
     The `Clean' sample considers all spectra from samples CL, GAL, and QSO with $f_{\rm{AGN}}<0.8$; we indicate its mean and standard deviation as the vertical dashed and dotted lines in both panels.
     }
\label{fig:results:hist_stellar_mass}
\end{figure}

\begin{figure*}[t]
\sidecaption
  \includegraphics[width=12cm]{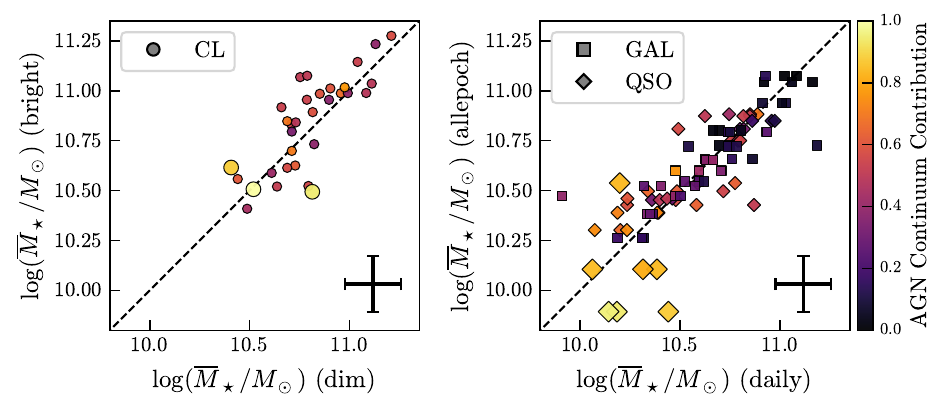}
     \caption{Comparison of aperture stellar masses between different observations of the same sources.
     The diagonal lines indicate the 1:1 relation.
     The left panel compares the bright and dim states of the sample CL (circles), color-coded by the AGN contribution to the continuum ($f_{\rm{AGN}}$) in the bright spectra.
     The right panel presents the comparison fits of individual spectra (\texttt{daily}) to stacked spectra (\texttt{allepoch}) for samples GAL (squares) and QSO (diamonds); color-coded according to the $f_{\rm{AGN}}$ of each \texttt{daily} observation.
     The larger markers indicate $f_{\rm{AGN}}>0.8$; their measurements of host-galaxy properties are less reliable.
     The black crosses indicate the systematic uncertainty derived from Fig. \ref{fig:results:hist_stellar_mass}.
     }
     \label{fig:results:stellar_mass}
\end{figure*}

We compare the results for bright and dim states for sample CL, and for the \texttt{daily} observations with their respective \texttt{allepoch} spectrum for samples GAL and QSO.
The difference in the measurements for $\overline{M}_*$ is shown in Fig. \ref{fig:results:hist_stellar_mass}, with the comparison to $f_{\rm{AGN}}$.
The statistical uncertainties obtained from Monte Carlo iterations are very small, ranging from an average of 3\% of the stellar mass for sample GAL to 12\% for sample QSO, indicating that this is a robust measurement across runs with random noise. 
The differences among the samples are expected, since measuring stellar properties is more challenging in quasar-dominated spectra.
However, when we compare the pairs of observations and calculate the inverse-variance weighted mean and standard deviation (shown in the legend of Fig. \ref{fig:results:hist_stellar_mass}), we see that the samples have a rather stable $\sim$0.14 dex intrinsic scatter in the stellar mass comparisons, which we take as an estimate of the systematic uncertainty of our stellar mass measurements.

Since high QSO contributions to the continuum can substantially outshine the stellar emission, based on Fig. \ref{fig:results:hist_stellar_mass} (and Fig. \ref{fig:app:hist_ML_SSPlum}), we consider only spectra with $f_{\rm{AGN}}<0.8$ to have reliable estimates of the properties of the host galaxy, deriving 0.14 dex for the systematic uncertainty of the combination of the samples with $f_{\rm{AGN}}<0.8$ (Clean).
Accounting only for the S/N$>10$ observations reduces the scatter to 0.10 dex, showing that the optimal cuts to obtain reliable results depend on a sample's S/N, $f_{\rm{AGN}}$, and (potentially) redshift distributions.
The slight negative inverse-variance weighted mean of the CL sample reflects mostly the difference in the fiber apertures between the SDSS generations, since it indicates that the bright-state measurements yield higher masses than the dim state, and for most of the sources, the bright state was observed with a larger fiber than the dim states (see Fig. \ref{fig:sample:mjd}).
Figure~\ref{fig:results:stellar_mass}, which shows a comparison of aperture stellar masses obtained from different observations of the same sources, confirms that sources with $f_{\rm{AGN}}>0.8$ — shown with large markers — are among those that deviate mostly from the 1:1 relation.

\begin{figure*}[t]
     \includegraphics[width=\textwidth]{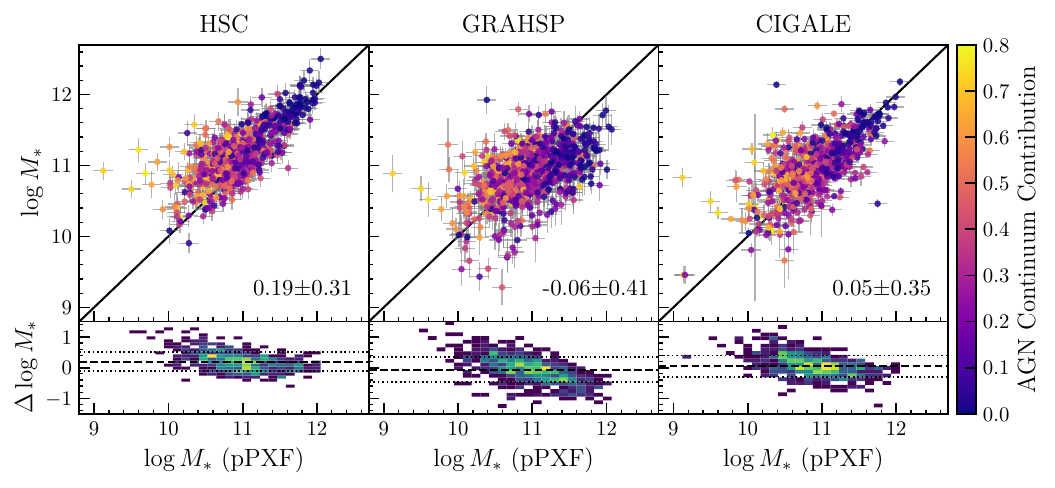}
     \caption{Comparison of (aperture-corrected) stellar masses obtained from our spectral-fitting method (\texttt{pPXF}) with an image decomposition method (HSC on the left, \citealt{Li2024_HSC}) and two SED-fitting methods (GRAHSP in the center, \citealt{Buchner2024}; and CIGALE on the right, \citealt{Yu2023}). 
     In the top panels, the data are color-coded by $f_{\rm{AGN}}$, and the 1:1 relation is shown as a black line.
     The bottom panels show the density distribution of the ratio of stellar mass measurements, with the dashed line indicating the offset and the dotted lines indicating the scatter.
     The mean offset and the scatter (in dex) are displayed in the bottom-right corner of each upper plot.
     }
\label{fig:results:stellarmass_hsc_grahsp}
\end{figure*}

Regarding the testing against external samples, Fig. \ref{fig:results:stellarmass_hsc_grahsp} shows the comparison of the stellar masses from the eFEDS field from our method (\texttt{pPXF}) with the results from \citet[][HSC]{Li2024_HSC}, \citet[][GRAHSP]{Buchner2024}, and \citet[][CIGALE]{Yu2023}. 
We found systematic offsets of 0.19, -0.06, and 0.05 dex for HSC, GRAHSP, and CIGALE, respectively, which are significantly smaller than the dispersion (0.31, 0.41, and 0.35 dex for HSC, GRAHSP, and CIGALE, respectively).
Our results are therefore somewhat intermediate between these methods.
In fact, the comparison between HSC and GRAHSP, regardless of our method, yields a scatter of 0.29 dex, while the comparison of CIGALE with both HSC and GRAHSP separately yields 0.31 dex of scatter; these scatters are all comparable to that between our method and HSC.
Hence, we conclude that a systematic scatter of 0.3-0.4 dex between our results and the literature is within expectations and validates our methodology as competitive with some of the most refined photometric techniques available for dealing with AGN with significant host-galaxy emission.
These results also demonstrate that our `internal' systematic uncertainties of 0.14 dex from Fig. \ref{fig:results:hist_stellar_mass} are smaller than those from comparisons with other observations of the same objects.

\subsubsection{Velocity dispersion}

\begin{figure}[t]
     \includegraphics[width=\columnwidth]{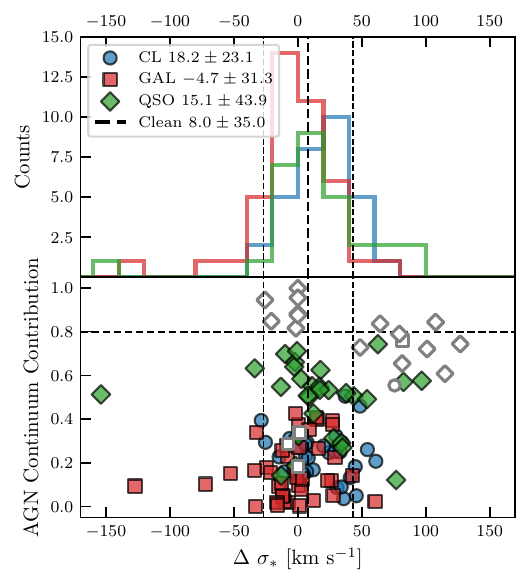}
     \caption{The difference of the stellar velocity dispersion, displayed as in Fig. \ref{fig:results:hist_stellar_mass}.
     Given the instrumental resolution of SDSS, only measurements with $\sigma_*>70$ km s$^{-1}$ are included in the histogram and statistics.
     The pairs of spectra with $f_{\rm AGN}>0.8$ or for which at least one of them has $\sigma_*<70$ km s$^{-1}$ are indicated in gray in the bottom panel.}
\label{fig:results:hist_sigma}
\end{figure}

\begin{figure*}[t]
\sidecaption
  \includegraphics[width=12cm]{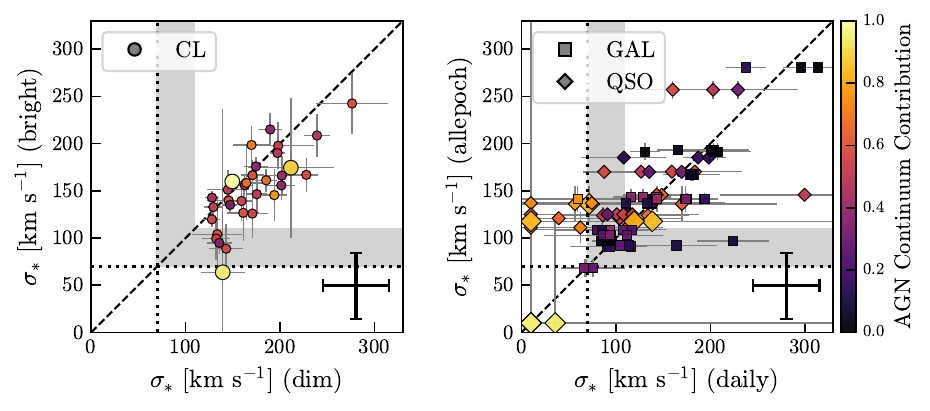}
     \caption{Comparison of stellar velocity dispersions, similar to Fig. \ref{fig:results:stellar_mass}.
     The dotted lines indicate the limits of the instrumental resolution of SDSS (70 km s$^{-1}$).
     The shaded regions indicate the 35 km s$^{-1}$ scatter from Fig. \ref{fig:results:hist_sigma}.
     Results with $\sigma_*<70$ km s$^{-1}$ should not be trusted, and the shaded area indicates cases that should be treated with caution.}
     \label{fig:results:sigma}
\end{figure*}

Figure \ref{fig:results:hist_sigma} displays the comparison of the estimates for the stellar velocity dispersion ($\sigma_*$).
Following the recommendations for velocity dispersion measurements obtained from BHM optical spectra due to the finite resolution of the instrument\footnote{\url{https://www.sdss.org/dr18/bhm/caveats/}}, we excluded the measurements of $\sigma_*<70$ km s$^{-1}$ from the histograms in the top panel of Fig. \ref{fig:results:hist_sigma}.
They are shown in gray in the bottom panel, and excluding such cases in which the measurement is below the instrumental resolution of SDSS data eliminates some, but not all of the most discrepant cases, between two measurements of the same source.
Both GAL and QSO samples have outliers with $\sigma_*>70$ km s$^{-1}$ and an absolute difference between spectra of $>100$ km s$^{-1}$, which makes the scatter larger than the sample CL.
For the CL sample, the positive offset indicates that most dim measurements have a higher $\sigma_*$ than in the bright state, as expected, since in the bright state the host galaxy absorption lines, which are key to constrain $\sigma_*$, are diluted by the stronger AGN contribution.
We also show in gray the cases with $f_{\rm{AGN}}>0.8$, which are not considered for the histograms.
The overall standard deviation for the difference in velocity dispersion, considering all samples and excluding $\sigma_*<70$ km s$^{-1}$, is $35$ km s$^{-1}$, and it improves to 27.5 km s$^{-1}$ when we consider only observations with S/N$>10$.
In Fig. \ref{fig:results:sigma}, we compare the measured velocity dispersions across the various observations for the CL, GAL, and QSO samples. 
Dotted lines are the limits of the instrumental resolution $\sigma_*=70$ km s$^{-1}$, and we highlight in gray shaded areas the dispersion of $35$ km s$^{-1}$ above such threshold, for cases that should be considered with caution.
The spectra with $f_{\rm{AGN}}>0.8$, shown with larger markers, lie mainly within caution or non-reliability regions and exhibit large error bars.

Regarding the comparison of the 11 CL objects observed with SDSS and either X-Shooter or GMOS \citep{Zeltyn26}, we obtain a scatter of 34 km s$^{-1}$, which confirms our systematic uncertainty of 35 km s$^{-1}$, despite the different slit sizes (1.3", 2", or 3" for SDSS, against 1.6" for X-Shooter and 0.5" for GMOS). 
It is worth noting that GMOS and X-Shooter should provide higher spectral resolution, yielding more precise velocity dispersion measurements, but since all measurements matching our sources exceed the SDSS instrumental limit of 70 km s$^{-1}$, this should not significantly affect the comparison. 
We also note that, in our comparisons, the absolute difference between the SDSS and GMOS or X-Shooter observations is always below 70 km s$^{-1}$, demonstrating the efficiency of our method, as it falls within the SDSS spectral resolution limits.

Concluding this Section, based on the reported set of internal and external validation tests, we recommend trusting the measurements of properties from the stellar populations when the AGN contribution to the continuum is $f_{\rm{AGN}}<80\%$.
For the stellar masses, we recommend using a systematic uncertainty of 0.14 dex.
In the case of the stellar velocity dispersion, SDSS BOSS observations with measurements of $\sigma_*<70$ km s$^{-1}$ are unreliable, and those with $70\leq\sigma_*<105$ km s$^{-1}$ should be considered with caution.
Therefore, we recommend using the Monte Carlo statistical uncertainties for each $\sigma_*$ measurement, keeping in mind the SDSS limit and the systematic uncertainty obtained in this work.
If one is interested in measuring $\sigma_*<70$ km s$^{-1}$, data with higher spectral resolution than SDSS should be preferred.

\subsection{AGN properties: black hole mass}
\label{sec:results:agn}

\begin{figure}[t]
     \includegraphics[width=\columnwidth]{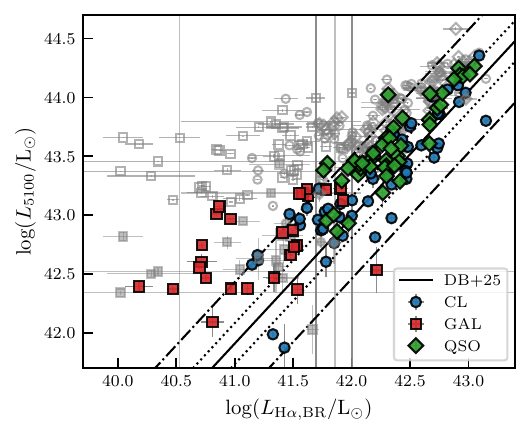}
     \caption{Relation between the AGN continuum luminosity and the broad H$\alpha$ luminosity required to apply the single-epoch method as in \citet{DallaBonta2025} or Eq. \ref{eq:SMBHmass_Ha} (diagonal line, with one sigma in dotted lines and three sigma in dotted-dashed lines).
     The markers in filled gray indicate the sources that do not pass the criteria of Eq. \ref{eq:ew_fwhm_quality}.
     The empty markers indicate the total luminosity at 5100 \AA\ before the decomposition into AGN and host-galaxy contributions.}
\label{fig:results:lum5100_lumHa}
\end{figure}

The main property to describe SMBHs is their mass.
There are different methods to estimate the mass of an SMBH, and in this section, we focus on the single-epoch virial estimates \citep[e.g.,][]{VestergaardPeterson2006, Shen2011, Shen2024}.
A discussion on the validity of such an approach for a sample of changing-look AGN (sample CL) is presented in Section \ref{sec:breathing}.
The single-epoch method relies on reliably measuring the width (and in some cases the luminosity) of broad lines.
Therefore, to guarantee a reliable detection of a broad component in the emission-line fitting, we use the following criteria:

\begin{equation}
    \begin{split}
\rm{FWHM}_{\rm{broad}}/\rm{FWHM}_{\rm{broad}}\_\rm{error}>3, \\
\rm{Flux}_{\rm{broad}}/\rm{Flux}_{\rm{broad}}\_\rm{error}>3.
    \end{split}
\label{eq:ew_fwhm_quality}
\end{equation}

Given the redshift range of our sample, we focus on the SMBH mass estimates from H$\alpha$ and H$\beta$.
The H$\alpha$ SMBH mass estimate is given by \citet{DallaBonta2025}, assuming the virial factor $\langle f \rangle = 0.68\pm0.15$ \citep{Batiste2017}, as
\begin{align}
\label{eq:SMBHmass_Ha}
    \rm{log}\left(\dfrac{M_{\rm{BH, H}\alpha}}{M_{\odot}}\right) =\ & 0.81\ \rm{log}\left(\dfrac{L_{H\alpha,\rm{BR}}}{10^{42} L_{\odot}}\right) + \\
    \nonumber &+ 1.63\ \rm{log}\left(\dfrac{FWHM}{10^{3.5}km\ s^{-1}}\right) + 7.37,
\end{align}
while for H$\beta$ from \citet{Shen2024} we adopt 
\begin{equation}
\label{eq:SMBHmass_Hb}
    \rm{log}\left(\dfrac{M_{\rm{BH, H}\beta}}{M_{\odot}}\right) = 0.5\ \rm{log}\left(\dfrac{L_{5100}}{10^{44} L_{\odot}}\right) + 2.0\ \rm{log}\left(\dfrac{FWHM}{km\ s^{-1}}\right) + 0.85,
\end{equation}
where $\rm{L}_{\rm{H}\alpha,\rm{BR}}$ is the luminosity measured from the broad-component of H$\alpha$, FWHM is measured from the broad component of the emission line in both cases, and $\rm{L}_{5100}$ is the AGN continuum luminosity at 5100\ \AA.
Figure \ref{fig:results:lum5100_lumHa} provides the relation between the AGN continuum and the broad-line luminosities that are assumed for applying the single-epoch method, according to \citet{DallaBonta2025} and Eq. \ref{eq:SMBHmass_Ha}.
We also display the total luminosity at 5100 \AA, showing that our decomposition method can shift the objects towards the expected relation when we consider only the AGN contribution to the continuum.
Many of the sources for which the conditions of Eq. \ref{eq:ew_fwhm_quality} do not apply are outliers in sample GAL, which is a sample of mostly narrow-line AGN.
However, through visual inspection, we find cases in which the GAL observation that lies as an outlier (i.e.,  more than 3$\sigma$ from the luminosity relation from \citealt{DallaBonta2025}) shows examples of AGN with a significant host-galaxy emission and also a significant broad H$\alpha$ component, despite the narrow line being more prominent than the broad component.
These objects are not usually considered for calibrating single-epoch black hole mass estimates, and though our code allows one to measure the properties of their (faint) broad emission lines, these measurements should be considered with caution.

\begin{figure}[t]
     \includegraphics[width=\columnwidth]{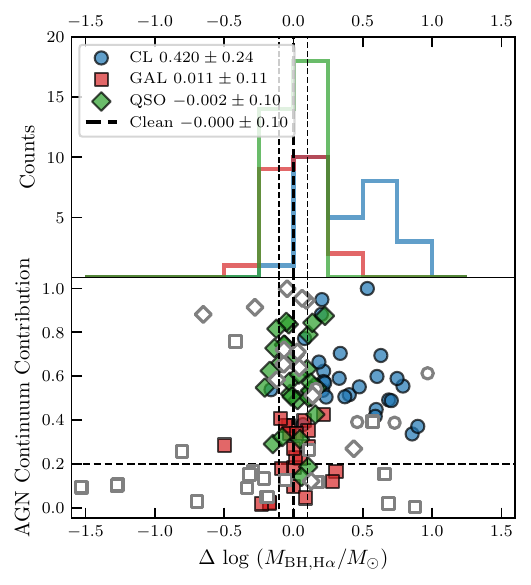}
     \caption{The difference of single-epoch estimates of the SMBH mass from H$\alpha$.
     This is similar to Fig. \ref{fig:results:hist_stellar_mass}, but compares $\rm \log(M_{BH, H\alpha}/M_\odot)_{bright} - \log(M_{BH, H\alpha}/M_\odot)_{dim}$.
     $f_{\rm{AGN}}$ is taken from the dim state for sample CL.
     The gray markers in the bottom panel indicate the cases in which the measurement of a broad component for the fit of H$\alpha$ was not reliable according to Eq. \ref{eq:ew_fwhm_quality}.
     The `Clean' sample statistics do not include these sources nor the CL sample, since changing-look AGN might not be virialized systems (see Section \ref{sec:breathing}).}
\label{fig:results:hist_Mbh_Ha}
\end{figure}

\begin{figure*}[t]
\sidecaption
  \includegraphics[width=12cm]{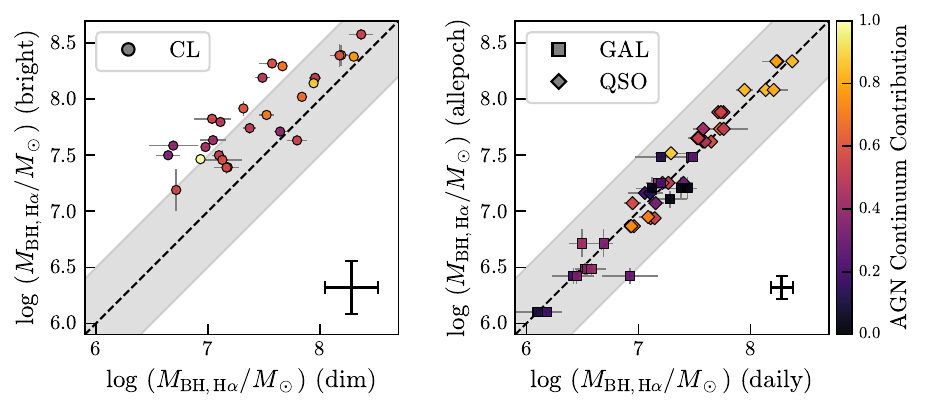}
     \caption{Comparison of SMBH mass according to single-epoch estimates using H$\alpha$, similar to Figure \ref{fig:results:stellar_mass}.
     The gray band represents the systematic 0.5 dex uncertainty inherent in single-epoch black hole mass estimates \citep{Shen2013}.
     The systematic uncertainty shown in the left panel is the 0.24 scatter for the CL sample, while the one in the right panel is the 0.10 scatter for the `Clean' sample, according to Fig. \ref{fig:results:hist_Mbh_Ha}.}
     \label{fig:results:Mbh_Ha}
\end{figure*}

Figure \ref{fig:results:hist_Mbh_Ha} shows the comparison of the results for the SMBH mass from Eq. \ref{eq:SMBHmass_Ha}.
Different from Section \ref{sec:results:host_galaxy}, we show the difference for bright-dim results for sample CL, since the bright spectra should be the most reliable for AGN measurements.
As seen from the gray markers in the bottom panel of Fig. \ref{fig:results:hist_Mbh_Ha} and expected from the sample selection, 51\% of the sample GAL showed the absence of a significant broad component in the H$\alpha$ emission line, although all samples have such cases of non-reliable broad-line measurement (16\% for sample CL and 29\% for sample QSO).
The largest standard deviation is from the CL sample, 0.24 dex, as expected since the BLR of changing-look AGN are not necessarily virialized (a more detailed analysis of the effects on the FWHM and on the line luminosity for the `breathing' of the BLR is given in Section \ref{sec:breathing}).
Therefore, we consider only the QSO and GAL samples for the statistics of the `Clean' sample, since these should be virialized systems for which the single-epoch mass estimate can be reliably applied, yielding a 0.1 dex scatter.
 
The scatter obtained in our sample is in addition to the systematic scatter associated with using the single-epoch mass estimate for large samples of black holes, estimated to be $0.3-0.5$ dex \citep[e.g.,][]{Shen2013}.
We expected our scatter to be smaller than the systematic uncertainty of the virial black-hole mass estimate since the geometry of the systems in samples GAL and QSO should not change significantly between the observations, as seen from their scatter of 0.10 dex in Fig. \ref{fig:results:hist_Mbh_Ha}.
Indeed, as seen in Fig. \ref{fig:results:Mbh_Ha}, by considering the quality cuts of Eq. \ref{eq:ew_fwhm_quality}, samples GAL and QSO are within the expected systematic scatter of the method, shown as gray areas.
For sample CL, there is a strong trend in estimating higher black-hole masses for the bright sources, which could indicate that the continuum and the line width are not behaving as expected for a `breathing' BLR (see more in Section \ref{sec:breathing}).
We believe this trend in sample CL is more closely related to sample selection than to the fitting method, as indicated by the agreement for the GAL and QSO samples.

\begin{figure}[t]
     \includegraphics[width=\columnwidth]{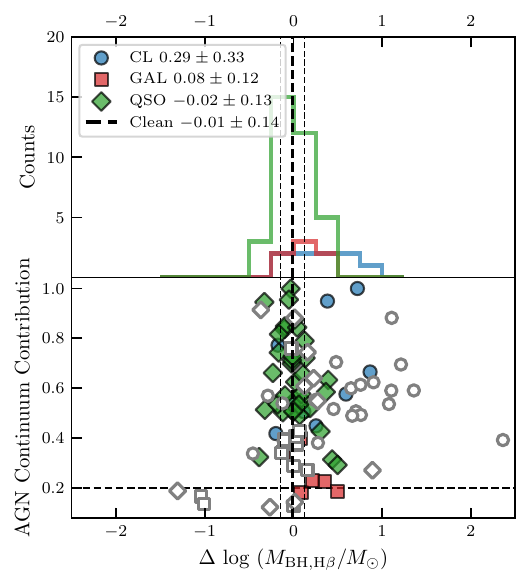}
     \caption{The difference of the SMBH mass according to single-epoch estimates using H$\beta$, as in Fig. \ref{fig:results:hist_Mbh_Ha}.}     
\label{fig:results:hist_Mbh_Hb}
\end{figure}

\begin{figure*}[t]
\sidecaption
  \includegraphics[width=12cm]{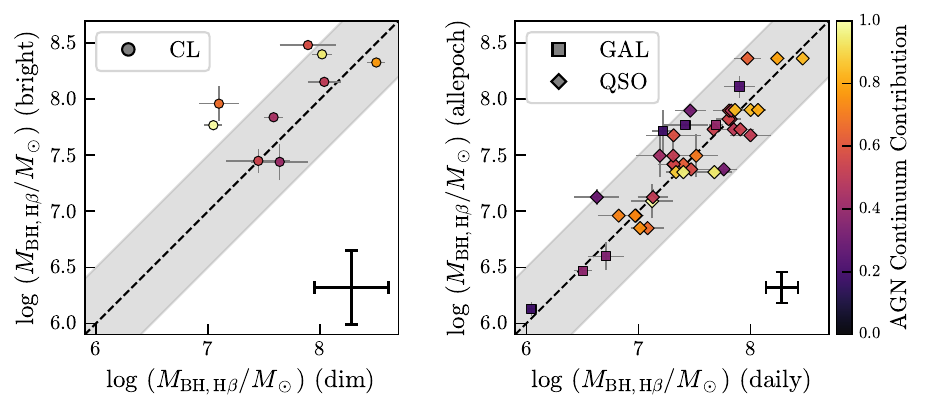}
     \caption{Comparison of SMBH mass according to single-epoch estimates using H$\beta$, as in Figure \ref{fig:results:Mbh_Ha}.
     The systematic uncertainties are shown in the bottom right of the plots, with 0.33 for the CL sample and 0.14 for the GAL and QSO samples, as in Fig. \ref{fig:results:hist_Mbh_Hb}.}
     \label{fig:results:Mbh_Hb}
\end{figure*}

Regarding the SMBH mass estimates from H$\beta$, Fig. \ref{fig:results:hist_Mbh_Hb} shows the comparison of the measurements in a similar fashion to Fig. \ref{fig:results:hist_Mbh_Ha}.
Sample GAL contains very few cases with a significant broad H$\beta$ component, and since sample CL was constructed based on the spectra having broad H$\alpha$ but often with the dimming/brightening based on the disappearance/appearance of a broad component of H$\beta$, many cases did not show a trustworthy H$\beta$ component according to Eq. \ref{eq:ew_fwhm_quality}.
The objects without a significant broad H$\beta$ line correspond to 72\% of the sample CL, 84\% of the sample GAL, and 22\% of the QSO sample. 
For both H$\alpha$ and H$\beta$, the bright CL spectra tend to have higher SMBH mass estimates in comparison to the dim state.
However, by construction, both samples CL and GAL have too few sources with a pair of reliable broad lines for a trustworthy statistical scatter, whereas sample QSO was built based on broad-line detections from the SDSS pipeline classification.
Hence, the selection effects on the construction of the samples are playing a larger role than a general assumption of the validity of equation \ref{eq:SMBHmass_Ha} in comparison to equation \ref{eq:SMBHmass_Hb}.
Nevertheless, the 0.13 dex scatter of the `Clean' sample of reliable broad H$\beta$ measurements for the GAL and QSO samples is consistent with the results for H$\alpha$, though it is slightly larger and based on fewer data points.
Figure \ref{fig:results:Mbh_Hb} shows that overall, the couples of H$\beta$-based mass measurements scatter further from the 1:1 line, with most within the 0.5 dex expected for single-epoch mass measurements.
The objects beyond the scatter are from the CL sample and have larger systematic uncertainties, as in Fig. \ref{fig:results:Mbh_Ha}.
Since the continuum luminosity is used instead of the line luminosity in Eq. \ref{eq:SMBHmass_Hb}, the black-hole mass estimates with H$\beta$ rely more on a correct AGN continuum estimate than H$\alpha$, and therefore the slightly larger scatter seen in Fig. \ref{fig:results:hist_Mbh_Hb} for the QSO sample could be due to these measurements being more sensitive to the host-galaxy decomposition.

\begin{figure}[t]
     \includegraphics[width=\columnwidth]{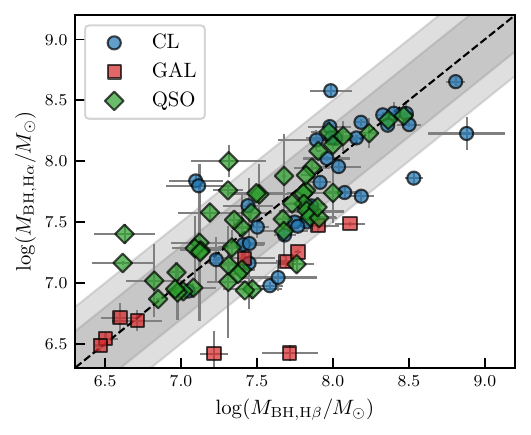}
     \caption{Comparison of SMBH mass single-epoch estimates using H$\alpha$ and H$\beta$ for each source in which both broad lines are in agreement with Eq. \ref{eq:ew_fwhm_quality}.
     The gray-shaded areas around the 1:1 relation indicate systematic scatter of 0.3 and 0.5 dex, as reported by \citet{Shen2013}.}
\label{fig:results:Mbh_Ha_Hb}
\end{figure}

We examined the effect of different S/N ranges on AGN properties, considering only the GAL and QSO samples with reliable broad-line measurements.
For the black hole masses estimated from H$\alpha$, there is no significant change when considering observations with S/N$>5$, and the scatter improves slightly (from 0.10 to 0.09 dex) when accounting for S/N$>10$.
For measurements based on H$\beta$, the influence of the S/N cuts is more evident but still mild, with the scatter of S/N$>10$ observations being 0.08 dex, though we are dealing with fewer sources in general.
Hence, it is important to consider how data quality influences the reliability of the obtained results.

Finally, Fig. \ref{fig:results:Mbh_Ha_Hb} shows the SMBH mass estimate from single-epoch estimates, comparing the measurements from H$\alpha$ and from H$\beta$ for the spectra that have reliable broad components of both lines simultaneously.
Most of the results fall within the expected systematic uncertainty, indicating that the measurements of the FWHM of the broad component and the luminosity are adequate.
However, there are many cases beyond the expected 0.5 dex scatter, which may be due to residual cross-calibration uncertainties from Eqs. \ref{eq:SMBHmass_Ha} and \ref{eq:SMBHmass_Hb}, which use different reference samples.
In this respect, we note here that the sample selection of this manuscript does not allow a systematic evaluation of single-epoch SMBH mass estimates for the H$\alpha$ and H$\beta$ in the sense of attesting which is more reliable, since sample CL is biased for showing broad H$\alpha$ without a broad H$\beta$ component, and sample GAL is biased for not having significant broad emission lines.
Future work will include applying this method to larger, less biased samples, enabling the assessment of the quality of single-epoch estimates.

\section{Further results and discussion}
\label{sec:discussion}

\subsection{Scaling relations}
\label{sec:scaling_relations}

The investigation of whether SMBH growth is coupled to the host-galaxy evolution is supported by empirical scaling relations, in which the mass of the SMBH is related to the stellar velocity dispersion and the stellar mass of the bulge or the host galaxy \citep[e.g.,][]{Gebhardt2000, Tremaine2002, Kormendy2013, Suh2020}.
Once such properties are measured for our sample, we can check if the results match the expected low-redshift scaling relations.
Since we are dealing directly with the spectral data, we do not distinguish between galaxy substructures, such as the bulge, and we also do not differentiate our sources based on the morphological classification of the host galaxies.
Additionally, the aim of this discussion is not to make a general statement about the position of AGN in the scaling-relation planes, since our sample is biased in its selection \citep[e.g.,][]{Lauer2007, Li2025}; in future work, we will use our decomposition in less biased samples to draw conclusions about the co-evolution of AGN and their host galaxies (Aydar et al., in prep.).

\begin{figure*}[t]
     \includegraphics[width=\textwidth]{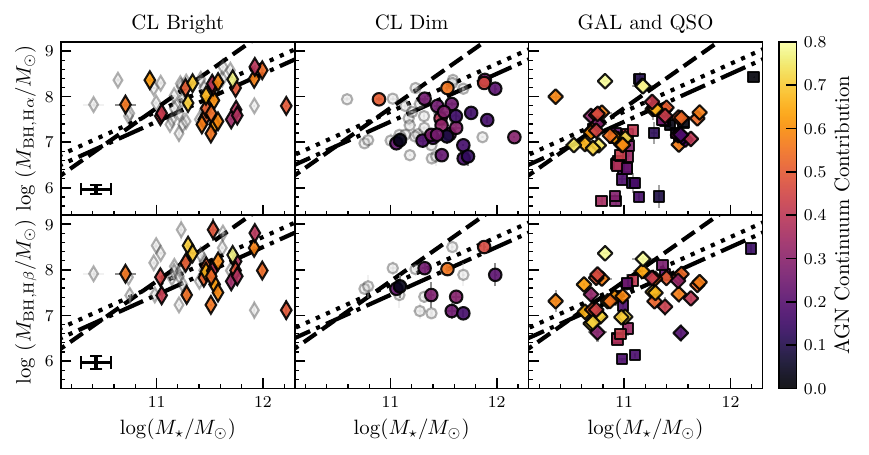}
     \caption{Scaling relation between the mass of the SMBH and the stellar mass, color-coded by the AGN contribution to the continuum in comparison to the host-galaxy emission.
     Upper panels show the estimates of the SMBH from H$\alpha$, while bottom panels display the estimates from H$\beta$.
     Left panels are the results for the CL bright spectra in diamonds, middle panels show the results for the CL dim spectra in circles, and the right panels present samples GAL and QSO in squares and diamonds, respectively.
     We excluded the cases with $f_{\rm{AGN}}>0.8$, since their host-galaxy properties are less reliable.
     We also consider only the cases with a reliable single-epoch SMBH mass estimate according to Eq. \ref{eq:ew_fwhm_quality}.
     For the CL sample in the left and middle panels, for which the aperture corrections are more uncertain, we display the aperture stellar masses (i.e., before aperture correction, $\overline{M}_*$) in gray, equivalent to lower limits of the galaxy stellar mass.
     The lines indicate scaling relations according to the literature, with the dot-dashed lines representing \citet{Reines2015}, the dashed lines representing \citet{Suh2020}, and the dotted lines representing \citet{Pucha2025}.
     The statistical uncertainties from Monte Carlo runs are plotted alongside the color-coded markers, but are often too small to be seen; we also plot the systematic uncertainty for reference on the reliability of the data.}
\label{fig:discussion:MBH_Mstar}
\end{figure*}
 
We show the results for the aperture-corrected stellar mass ($M_*$) in comparison to the SMBH mass in Fig. \ref{fig:discussion:MBH_Mstar}, with the scaling relations from \citet{Reines2015}, \citet{Suh2020}, and \citet{Pucha2025}.
For the CL sample, we show in gray the results for $\overline{M}_*$ (i.e., before aperture correction) because this sample was selected due to the variability of the sources, and therefore the flux from Legacy Survey DR10 cannot be representative of both the dim and the bright states, adding an irreducible uncertainty to the extent of such corrections.
The samples CL dim and GAL show fewer objects in the Figure due to often not exhibiting a significant broad Balmer line, especially for H$\beta$ (see Figs. \ref{fig:results:hist_Mbh_Ha} to \ref{fig:results:Mbh_Hb}).

\begin{figure*}[t]
     \includegraphics[width=\textwidth]{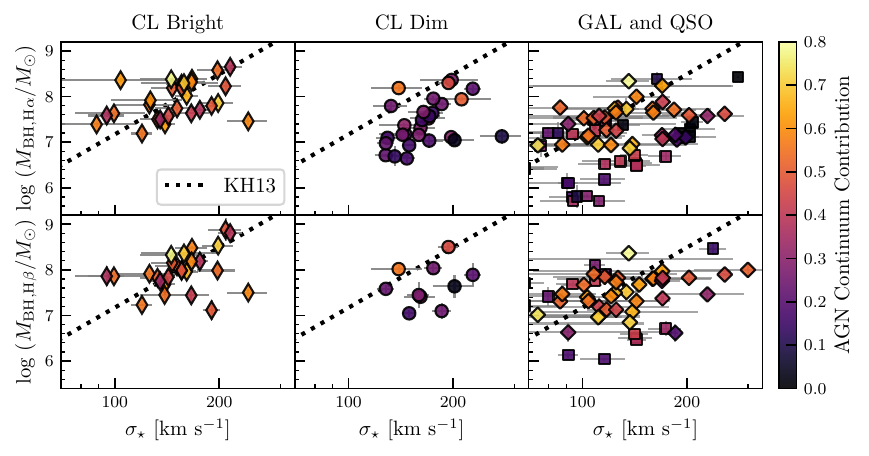}
     \caption{Scaling relation between the mass of the SMBH and the stellar velocity dispersion, similar to Fig. \ref{fig:discussion:MBH_Mstar}.
     The dotted line indicates the scaling relation, as described by \citet{Kormendy2013}.
     }
\label{fig:discussion:M_sigma}
\end{figure*}

For the M$_{\rm BH}$-$\sigma_*$ relation shown in Fig. \ref{fig:discussion:M_sigma}, our data are in broad agreement with the scaling relation proposed by \citet{Kormendy2013}.
On the other hand, sample CL tends to have more overmassive black holes in comparison to sample QSO.
This trend is expected due to the selection of the samples, since sample CL contains mainly bright and massive sources (see Figs. \ref{fig:sample:mjd} and \ref{fig:discussion:MBH_Mstar}).

Overall, the general agreement between our results and the expectations of the scaling relations demonstrates the success of our method in extracting simultaneous information about the AGN and the stellar population from a single spectrum.
This result further indicates that our decomposition method produces reliable information for studying the co-evolution of SMBHs and their host galaxies when applied to larger, less biased samples.

\subsection{Broad Line Region `Breathing'}
\label{sec:breathing}

\begin{figure*}[t]
     \includegraphics[width=0.49\textwidth]{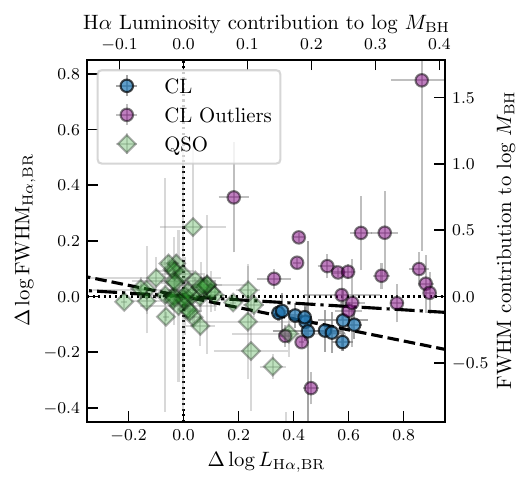}
     \includegraphics[width=0.5\textwidth]{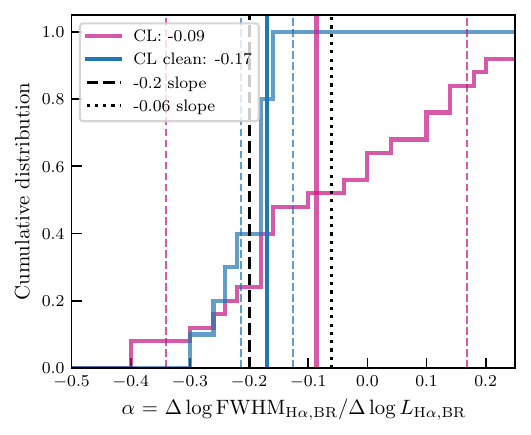}
     \caption{Comparison of the difference in FWHM (y-axis) and luminosity (x-axis) of the broad component of H$\alpha$ for two different observations of the same object.
     The left panel displays the bright-dim for the CL sample (circles) and stack-individual spectra for the QSO sample (green diamonds).
     The objects from the CL sample that are within the standard deviation for the SMBH mass ratio from the clean sample in Fig. \ref{fig:results:hist_Mbh_Ha} are shown in blue, while the outliers that have larger differences in their SMBH mass estimates are displayed in purple.
     The axes show the difference of the measured line property on the left and bottom, and their respective contributions to the black-hole mass according to Eq. \ref{eq:SMBHmass_Ha} on the right and top.
     The left and bottom axes have the same range, so the comparison in dex is straightforward.
     The (vertical and horizontal) dotted lines indicate the  no-difference cases.
     The dashed line indicates the $\alpha=\Delta \log {\rm FWHM}_{\rm H\alpha, BR} / \Delta \log L_{\rm H\alpha, BR}=-0.2$ slope expected for breathing according to Eq. \ref{eq:SMBHmass_Ha}, and the dotted-dashed line is the $\alpha=-0.06$ derived for H$\alpha$ in \citet{Wang2020}.
     The right panel shows the cumulative distribution of the $\alpha$ slope for the overall CL sample in pink, and for the CL sample without the outliers in blue.
     The median distributions of $\alpha$ are vertical lines and in the legend, with the standard deviation shown as dashed lines.
     }
\label{fig:discussion:breathing_ha}
\end{figure*}

Although our methodology is validated by its ability to measure consistent conserved parameters, not all spectral features across different observations are expected to be the same.
For example, while the SMBH mass should be constant, the FWHM of the broad line and the luminosity that are used in the equation for the single-epoch mass derivation are not expected to be the same in different epochs, although they are expected to vary accordingly assuming photoionization equilibrium in a virialized system \citep[e.g.,][]{Park2012, Barth2015}.
This effect of the correlation between the BLR and the AGN luminosity has been analyzed in reverberation mapping studies, which are the basis for the proposal of single-epoch mass estimates, and it is referred to as the BLR `breathing' \citep[e.g.,][]{Runco2016}.
Physically, it means that, for a virialized system in photoionization equilibrium, as the AGN luminosity increases, the average BLR size should increase, and therefore the average broad-line width should decrease \citep{Wang2020}.

The analysis of the breathing properties on broad H$\alpha$ lines in our samples is shown in Figure \ref{fig:discussion:breathing_ha}. 
This figure shows the comparison between the different estimates of FWHM from the broad H$\alpha$ component (a combination of the three Gaussians used to fit the broad component) and the luminosity from the overall H$\alpha$ emission line (a combination of all narrow and broad Gaussians used in the fit), according to Eq. \ref{eq:SMBHmass_Ha}, for each pair of observations\footnote{The results for H$\beta$ are not displayed due to the lower number of sources with a reliable broad component according to Eq. \ref{eq:ew_fwhm_quality}.}.
The lines indicate the expected slope ($\alpha=\Delta\rm{log\ FWHM}/\Delta log\ L$) if the BLR was breathing according to Eq. \ref{eq:SMBHmass_Ha} (dashed) and the relation proposed empirically for the H$\alpha$ breathing by \citet{Wang2020} (dot-dashed).
We do not include sample GAL, since these objects are, by definition, not broad-line galaxies, and when a broad-line component is detected, it is not dominant compared to the narrow component of the emission line.

From the diagram, we see that sample QSO is more concentrated in the locus of very small differences between the observations, indicated by the dotted lines at zero in both axes.
This distribution demonstrates that our methodology is robust, as the sample construction should not result in significant differences across observations.
However, sample CL, also by construction, has larger differences in both width and luminosity.
The objects that showed a significant deviation from the Clean scatter in Fig. \ref{fig:results:hist_Mbh_Ha} are shown in purple and were already candidates for non-virialized systems; this hypothesis is further strengthened by their distribution farther from the expected behavior of a breathing virialized BLR.
The cumulative distributions of the slope $\alpha$ shown in the right panel of Fig. \ref{fig:discussion:breathing_ha} also reinforce that hypothesis.
The mean slope of the overall CL sample (-0.09) lies closer to the empirical expectations for a breathing BLR of -0.06 according to \citealt{Wang2020}; however, after excluding the outliers (purple points), the mean of the slope (-0.17) is in better agreement with the -0.2 model (according to Eq. \ref{eq:SMBHmass_Ha}) expectations, showing the importance of reliable cleaning procedures before stating the validity of one or another approach.
It would be interesting to further test this analysis in larger samples \citep[e.g.,][]{Zeltyn24} to verify if such outliers are indeed non-virialized changing-look AGN, probably due to the appearance or disappearance of a BLR caught in the act by the CL selection \citep{Ricci_Trakhtenbrot2023}.

On the top and left axes of the left-hand panel from Fig. \ref{fig:discussion:breathing_ha} we show the contribution of each parameter to the estimate of the SMBH mass according to Eq. \ref{eq:SMBHmass_Ha}.
The differences in the FWHM are more significant to the overall scatter of the SMBH masses than the luminosity.
Hence, this result shows that most of the uncertainty in the SMBH mass estimates in this work is intrinsically associated with the BLR breathing behavior rather than with limitations of the decomposition method.
A more general discussion of the limitations of the single-epoch SMBH mass estimates is beyond the scope of this manuscript and can be found in, e.g., \citet{Shen2013} and \citet{Shen2024}.

\section{Conclusions and outlook}
\label{sec:conclusions}

In this work, we develop validation tests for high-quality automatic spectral fitting of galaxy-AGN hybrid spectra. 
We present benchmark data samples used to validate stellar mass, stellar velocity dispersion, and SMBH mass estimates from H$\alpha$ and H$\beta$ over a range of signal-to-noise ratios and AGN-to-galaxy contrast ratios.
The `internal' benchmark samples consist of objects observed multiple times by SDSS: 32 changing-look AGN and 30 eROSITA sources, from which 15 are classified as QSO and 15 as GALAXY from the SDSS pipeline (Sect. \ref{sec:data}).
The tests are applied to a two-step spectral fitting method based on \texttt{pPXF} (Sect. \ref{sec:ppxf}) and \texttt{PyQSOFit} (Sect. \ref{sec:pyqsofit}), where galaxy contamination is subtracted before fitting AGN lines.
This methodology is successful and can provide reliable measurements of emission and absorption line properties, as well as the AGN continuum (Sect. \ref{sec:results:general}).

We also provide `external' comparisons with literature samples to validate the stellar mass estimates by comparing photometric data to SDSS spectra from the eFEDS field, and to validate the stellar velocity dispersion by using Gemini and VLT observations of some changing-look AGN from our benchmark sample; all yield consistent results that validate our methodology and derived uncertainties.

Our methodology is more complex than using SDSS pipeline information (which does not differentiate the broad and narrow components of emission lines), than using \texttt{pPXF} alone (which does not have a complex fitting for the AGN contribution for both the line shapes and the continuum), or than using \texttt{PyQSOFit} alone (which does not properly account for the host-galaxy contribution).
As a corollary, this approach enables one to place a galaxy in scaling relations based on stellar and broad-line properties using the same observation, provided it exhibits significant emission from both the stellar populations and the broad-line region.
Despite testing our methodology on the parameters associated with the scaling relations, our method also provides measurements of the emission lines (overall, narrow, and broad components), AGN continuum, SSPs, and absorption lines, which can provide the estimate of many other physical properties and also the selection of subsamples based on, e.g., AGN classification, presence of specific lines such as coronal lines, and presence of outflows. Although the approach was tailored for fitting SDSS data, it can be generalized and applied to any optical spectrum, making this a powerful tool for future large surveys, such as DESI, 4MOST, and PFS.

Based on this work, we advise relying on the results associated with the decomposed host galaxy if $f_{\rm{AGN}}<80\%$, where $f_{\rm{AGN}}$ corresponds to the fraction of the AGN contribution to the continuum (power law featureless continuum, Fe II pseudo-continuum, and Balmer high-order and continuum emission) in comparison to the stellar population (SSPs), calculated from the weight of the templates over the whole wavelength range for each spectrum.
Otherwise, the measurements will be biased by the known degeneracy between the featureless power-law continuum emission from the AGN and the young and blue stellar populations (Sect. \ref{sec:results:host_galaxy}).
For AGN measurements, we advise using the line-quality criteria from Eq. \ref{eq:ew_fwhm_quality} to obtain reliable measurements (Sect. \ref{sec:results:agn}).
However, since we also notice a S/N dependence that often, but not always, yields smaller scatter in the comparisons of different observations of the same objects, we advise each user of our methodology to also check on how the S/N, $f_{\rm{AGN}}$, and redshift affect their data to then tailor the best quality cuts.

The stellar masses are consistent with internal and external comparisons after aperture corrections based on photometric observations from other surveys, although these corrections are only rough approximations for changing-look AGN due to their expected flux changes. 
We find an `internal' systematic uncertainty of 0.14 dex, significantly smaller than the scatter in the comparison with the most robust photometric-based methods for stellar mass estimation (0.3-0.4 dex). 
Stellar velocity dispersion measurements should only be considered above the instrumental resolution (70 km s$^{-1}$ for SDSS) and should be taken with caution if $\sigma_*<105$ km s$^{-1}$ by applying our method to SDSS data, regardless of the S/N of the observation.

The supermassive black-hole mass estimates obtained via the single-epoch method incur an additional systematic uncertainty of $\sim0.1$ dex for H$\alpha$ and H$\beta$. 
This is smaller than the systematic scatter of $0.3-0.5$ dex currently inherent in large samples of virialized systems from single-epoch measurements.
Since we are comparing  multiple observations of the same objects rather than analyzing the scatter of a large, diverse sample, this result confirms the accuracy of our method. 

The results of our technique applied to the testing samples indicate AGN that are broadly consistent with the literature scaling relations, which correlate the SMBH mass with the stellar velocity dispersion and the stellar mass of the host galaxy (Sect. \ref{sec:scaling_relations}).
When applied to the changing-look AGN, the results show some interesting deviations from the expectation of `breathing' virialized BLRs for a subset of sources, which exhibit higher discrepancies in black-hole mass estimates and are candidates for non-virialized systems (Sect. \ref{sec:breathing}).

In conclusion, we have demonstrated that our method of applying \texttt{pPXF} and \texttt{PyQSOFit} for fitting AGN with significant host-galaxy contribution is robust and, given the uncertainties, can provide consistent results for the black-hole and host-galaxy properties that should not change drastically in short periods of time.
This methodology can be applied to other optical spectral samples, thereby improving our understanding of the co-evolution of SMBHs and their host galaxies.
As a future perspective, we intend to apply this method to the overall SPIDERS sample within the Black Hole Mapper collaboration between SDSS-V and eROSITA \citep[see][]{Aydar2025, Kollmeier2026}.
From the sample of 200k X-ray selected AGN with optical spectra, we estimate that $\sim30\%$ of the sample ($\sim60$k spectra) should require a proper decomposition of the host galaxy and the AGN.
The \texttt{python} script of this method will be made available on the SDSS \texttt{GitHub}\footnote{\url{https://github.com/sdss}} together with the 20th Data Release and the Value Added Catalog with the measurements of the emission lines, continuum, and, when detected, absorption lines and stellar population properties of the SPIDERS sample.

\begin{acknowledgements}
Funding for the Sloan Digital Sky Survey V has been provided by the Alfred P. Sloan Foundation, the Heising-Simons Foundation, the National Science Foundation, and the Participating Institutions. SDSS acknowledges support and resources from the Center for High-Performance Computing at the University of Utah. The SDSS website is \url{www.sdss.org}.

SDSS is managed by the Astrophysical Research Consortium for the Participating Institutions of the SDSS Collaboration, including the Carnegie Institution for Science, Chilean National Time Allocation Committee (CNTAC) ratified researchers, Caltech, the Gotham Participation Group, Harvard University, Heidelberg University, The Flatiron Institute, The Johns Hopkins University, L'Ecole polytechnique f\'{e}d\'{e}rale de Lausanne (EPFL), Leibniz-Institut f\"{u}r Astrophysik Potsdam (AIP), Max-Planck-Institut f\"{u}r Astronomie (MPIA Heidelberg), Max-Planck-Institut f\"{u}r Extraterrestrische Physik (MPE), Nanjing University, National Astronomical Observatories of China (NAOC), New Mexico State University, The Ohio State University, Pennsylvania State University, Smithsonian Astrophysical Observatory, Space Telescope Science Institute (STScI), the Stellar Astrophysics Participation Group, Universidad Nacional Aut\'{o}noma de M\'{e}xico, University of Arizona, University of Colorado Boulder, University of Illinois at Urbana-Champaign, University of Toronto, University of Utah, University of Virginia, Yale University, and Yunnan University.

CA acknowledges the support of the Excellence Cluster ORIGINS, which is funded by the Deutsche Forschungsgemeinschaft (DFG, German Research Foundation) under Germany's Excellence Strategy – EXC-2094 – 390783311.

SB acknowledges the National Agency for Research and Development (ANID) grant Gemini-32240014.

RJA was supported by FONDECYT grant number 1231718 and by the ANID BASAL project FB210003.

ALR acknowledges support from a Leverhulme Early Career Fellowship.

YS acknowledges support from NSF grants AST-2009947 and AST-2509424. 

CA acknowledges Mike Eracleous for the Fe II and Balmer Continuum and High Order templates used in \texttt{pPXF}.

CA acknowledges the valuable discussions with Vardha Bennert and Matthew Temple regarding scaling relations, as well as with Rog\'erio Riffel about stellar population fitting.

\end{acknowledgements}

\bibliographystyle{aa}
\bibliography{bib}

@preamble{ " \newcommand{\noop}[1]{} " }

@ARTICLE{Zeltyn24,
       author = {{Zeltyn}, Grisha and {Trakhtenbrot}, Benny and {Eracleous}, Michael and {Yang}, Qian and {Green}, Paul and {Anderson}, Scott F. and {LaMassa}, Stephanie and {Runnoe}, Jessie and {Assef}, Roberto J. and {Bauer}, Franz E. and {Brandt}, W.~N. and {Davis}, Megan C. and {Frederick}, Sara E. and {Fries}, Logan B. and {Graham}, Matthew J. and {Grogin}, Norman A. and {Guolo}, Muryel and {Hern{\'a}ndez-Garc{\'\i}a}, Lorena and {Koekemoer}, Anton M. and {Krumpe}, Mirko and {Liu}, Xin and {Mart{\'\i}nez-Aldama}, Mary Loli and {Ricci}, Claudio and {Schneider}, Donald P. and {Shen}, Yue and {{\'S}niegowska}, Marzena and {Temple}, Matthew J. and {Trump}, Jonathan R. and {Xue}, Yongquan and {Brownstein}, Joel R. and {Dwelly}, Tom and {Morrison}, Sean and {Bizyaev}, Dmitry and {Pan}, Kaike and {Kollmeier}, Juna A.},
        title = "{Exploring Changing-look Active Galactic Nuclei with the Sloan Digital Sky Survey V: First Year Results}",
      journal = {\apj},
         year = 2024,
        month = may,
       volume = {966},
       number = {1},
          eid = {85},
        pages = {85},
          doi = {10.3847/1538-4357/ad2f30},
archivePrefix = {arXiv},
       eprint = {2401.01933},
 primaryClass = {astro-ph.GA},
       adsurl = {https://ui.adsabs.harvard.edu/abs/2024ApJ...966...85Z}
}

@ARTICLE{Zeltyn26,
       author = {{Zeltyn}, Grisha and {Trakhtenbrot}, Benny and {Eracleous}, Michael and {Anderson}, Scott F. and {Ricci}, Claudio and {Merloni}, Andrea and {Runnoe}, Jessie and {Krumpe}, Mirko and {Aird}, James and {Assef}, Roberto J. and {Aydar}, Catarina and {Bauer}, Franz E. and {Brandt}, W.~N. and {Brownstein}, Joel R. and {Buchner}, Johannes and {Chatterjee}, Kaushik and {Duffy}, Laura and {Hern{\'a}ndez-Garc{\'\i}a}, Lorena and  {Hern{\'a}ndez-Toledo}, H{\'\e}ctor and {Koekemoer}, Anton M. and {Morrison}, Sean and {Negrete}, Castalia Alenka and {Salvato}, Mara and {Schneider}, Donald P. and {Shen}, Yue and {{\'S}niegowska}, Marzena},
       title = "{Changing-Look Active Galactic Nuclei in SDSS-V: Host-Galaxy Properties and Black-Hole Scaling Relations}",
      journal = {\apj},
         year = 2026,
        month = apr,
       volume = {1002},
       number = {1},
          eid = {61},
        pages = {61},
          doi = {10.3847/1538-4357/ae5495},
archivePrefix = {arXiv},
       eprint = {2511.07532}
}

@ARTICLE{Bernardi2010_IMF,
       author = {{Bernardi}, M. and {Shankar}, F. and {Hyde}, J.~B. and {Mei}, S. and {Marulli}, F. and {Sheth}, R.~K.},
        title = "{Galaxy luminosities, stellar masses, sizes, velocity dispersions as a function of morphological type}",
      journal = {\mnras},
         year = 2010,
        month = jun,
       volume = {404},
       number = {4},
        pages = {2087-2122},
          doi = {10.1111/j.1365-2966.2010.16425.x},
archivePrefix = {arXiv},
       eprint = {0910.1093},
 primaryClass = {astro-ph.CO},
       adsurl = {https://ui.adsabs.harvard.edu/abs/2010MNRAS.404.2087B}
}

@ARTICLE{Cappellari2023,
    author = {{Cappellari}, M.},
    title = "{Full spectrum fitting with photometry in PPXF: stellar population
        versus dynamical masses, non-parametric star formation history and
        metallicity for 3200 LEGA-C galaxies at redshift $z\approx0.8$}",
    journal = {MNRAS},
    eprint = {2208.14974},
    year = 2023,
    volume = 526,
    pages = {3273-3300},
    doi = {10.1093/mnras/stad2597}
}

@misc{PyQSOFit,
    author = {{Guo}, H. and {Shen}, Y. and {Wang}, S.},
    title = "{PyQSOFit: Python code to fit the spectrum of quasars}",
    howpublished = {Astrophysics Source Code Library},
    year = 2018,
    month = sep,
    archivePrefix = "ascl",
    eprint = {1809.008},
    adsurl = {http://adsabs.harvard.edu/abs/2018ascl.soft09008G}
}

@ARTICLE{Wu2022,
       author = {{Wu}, Qiaoya and {Shen}, Yue},
        title = "{A Catalog of Quasar Properties from Sloan Digital Sky Survey Data Release 16}",
      journal = {\apjs},
         year = 2022,
        month = dec,
       volume = {263},
       number = {2},
          eid = {42},
        pages = {42},
          doi = {10.3847/1538-4365/ac9ead},
archivePrefix = {arXiv},
       eprint = {2209.03987},
 primaryClass = {astro-ph.GA},
       adsurl = {https://ui.adsabs.harvard.edu/abs/2022ApJS..263...42W}
}

@ARTICLE{Vazdekis2016,
       author = {{Vazdekis}, A. and {Koleva}, M. and {Ricciardelli}, E. and {R{\"o}ck}, B. and {Falc{\'o}n-Barroso}, J.},
        title = "{UV-extended E-MILES stellar population models: young components in massive early-type galaxies}",
      journal = {\mnras},
         year = 2016,
        month = dec,
       volume = {463},
       number = {4},
        pages = {3409-3436},
          doi = {10.1093/mnras/stw2231},
archivePrefix = {arXiv},
       eprint = {1612.01187},
 primaryClass = {astro-ph.GA},
       adsurl = {https://ui.adsabs.harvard.edu/abs/2016MNRAS.463.3409V}
}

@ARTICLE{BorosonGreen1992,
       author = {{Boroson}, Todd A. and {Green}, Richard F.},
        title = "{The Emission-Line Properties of Low-Redshift Quasi-stellar Objects}",
      journal = {\apjs},
         year = 1992,
        month = may,
       volume = {80},
        pages = {109},
          doi = {10.1086/191661},
       adsurl = {https://ui.adsabs.harvard.edu/abs/1992ApJS...80..109B}
}

@ARTICLE{VestergaardWilkes2001,
       author = {{Vestergaard}, M. and {Wilkes}, B.~J.},
        title = "{An Empirical Ultraviolet Template for Iron Emission in Quasars as Derived from I Zwicky 1}",
      journal = {\apjs},
         year = 2001,
        month = may,
       volume = {134},
       number = {1},
        pages = {1-33},
          doi = {10.1086/320357},
archivePrefix = {arXiv},
       eprint = {astro-ph/0104320},
 primaryClass = {astro-ph},
       adsurl = {https://ui.adsabs.harvard.edu/abs/2001ApJS..134....1V}
}

@ARTICLE{Shen2011,
       author = {{Shen}, Yue and {Richards}, Gordon T. and {Strauss}, Michael A. and {Hall}, Patrick B. and {Schneider}, Donald P. and {Snedden}, Stephanie and {Bizyaev}, Dmitry and {Brewington}, Howard and {Malanushenko}, Viktor and {Malanushenko}, Elena and {Oravetz}, Dan and {Pan}, Kaike and {Simmons}, Audrey},
        title = "{A Catalog of Quasar Properties from Sloan Digital Sky Survey Data Release 7}",
      journal = {\apjs},
         year = 2011,
        month = jun,
       volume = {194},
       number = {2},
          eid = {45},
        pages = {45},
          doi = {10.1088/0067-0049/194/2/45},
archivePrefix = {arXiv},
       eprint = {1006.5178},
 primaryClass = {astro-ph.CO},
       adsurl = {https://ui.adsabs.harvard.edu/abs/2011ApJS..194...45S}
}

@ARTICLE{Aydar2025,
       author = {{Aydar}, C. and {Merloni}, A. and {Dwelly}, T. and {Comparat}, J. and {Salvato}, M. and {Buchner}, J. and {Brusa}, M. and {Liu}, T. and {Wolf}, J. and {Anderson}, S.~F. and {Andonie}, C.~P. and {Bauer}, F.~E. and {Blanton}, M.~R. and {Brandt}, W.~N. and {D{\'\i}az}, Y. and {Hern{\'a}ndez-Garc{\'\i}a}, L. and {Kim}, D. -W. and {Miyaji}, T. and {Morrison}, S. and {Musiimenta}, B. and {Negrete}, C.~A. and {Ni}, Q. and {Ricci}, C. and {Schneider}, D.~P. and {Schwope}, A. and {Shen}, Y. and {Waddell}, S.~G.~H. and {Arcodia}, R. and {Bizyaev}, D. and {Burchett}, J.~N. and {Chakraborty}, P. and {Covey}, K. and {G{\"a}nsicke}, B.~T. and {Georgakakis}, A. and {Green}, P.~J. and {Ibarra}, H. and {Ider-Chitham}, J. and {Koekemoer}, A.~M. and {Kollmeier}, J.~A. and {Krumpe}, M. and {Lamer}, G. and {Malyali}, A. and {Nandra}, K. and {Pan}, K. and {Pizarro}, C.~R. and {S{\'a}nchez-Gallego}, J. and {Trump}, J.~R. and {Urrutia}, T.},
        title = "{The eROSITA Final Equatorial Depth Survey (eFEDS): SDSS spectroscopic observations of X-ray sources}",
      journal = {\aap},
         year = 2025,
        month = jun,
       volume = {698},
          eid = {A132},
        pages = {A132},
          doi = {10.1051/0004-6361/202554372},
archivePrefix = {arXiv},
       eprint = {2505.03872},
 primaryClass = {astro-ph.HE},
       adsurl = {https://ui.adsabs.harvard.edu/abs/2025A&A...698A.132A}
}

@ARTICLE{Salpeter1955,
       author = {{Salpeter}, Edwin E.},
        title = "{The Luminosity Function and Stellar Evolution.}",
      journal = {\apj},
         year = 1955,
        month = jan,
       volume = {121},
        pages = {161},
          doi = {10.1086/145971},
       adsurl = {https://ui.adsabs.harvard.edu/abs/1955ApJ...121..161S}
}

@ARTICLE{Girardi2000,
       author = {{Girardi}, L. and {Bressan}, A. and {Bertelli}, G. and {Chiosi}, C.},
        title = "{Evolutionary tracks and isochrones for low- and intermediate-mass stars: From 0.15 to 7 M$_{sun}$, and from Z=0.0004 to 0.03}",
      journal = {\aaps},
         year = 2000,
        month = feb,
       volume = {141},
        pages = {371-383},
          doi = {10.1051/aas:2000126},
archivePrefix = {arXiv},
       eprint = {astro-ph/9910164},
 primaryClass = {astro-ph},
       adsurl = {https://ui.adsabs.harvard.edu/abs/2000A&AS..141..371G}
}

@ARTICLE{Cardelli1989,
       author = {{Cardelli}, Jason A. and {Clayton}, Geoffrey C. and {Mathis}, John S.},
        title = "{The Relationship between Infrared, Optical, and Ultraviolet Extinction}",
      journal = {\apj},
         year = 1989,
        month = oct,
       volume = {345},
        pages = {245},
          doi = {10.1086/167900},
       adsurl = {https://ui.adsabs.harvard.edu/abs/1989ApJ...345..245C}
}

@ARTICLE{Schlegel1998,
       author = {{Schlegel}, David J. and {Finkbeiner}, Douglas P. and {Davis}, Marc},
        title = "{Maps of Dust Infrared Emission for Use in Estimation of Reddening and Cosmic Microwave Background Radiation Foregrounds}",
      journal = {\apj},
         year = 1998,
        month = jun,
       volume = {500},
       number = {2},
        pages = {525-553},
          doi = {10.1086/305772},
archivePrefix = {arXiv},
       eprint = {astro-ph/9710327},
 primaryClass = {astro-ph},
       adsurl = {https://ui.adsabs.harvard.edu/abs/1998ApJ...500..525S}
}

@ARTICLE{Schlafly2011,
       author = {{Schlafly}, Edward F. and {Finkbeiner}, Douglas P.},
        title = "{Measuring Reddening with Sloan Digital Sky Survey Stellar Spectra and Recalibrating SFD}",
      journal = {\apj},
         year = 2011,
        month = aug,
       volume = {737},
       number = {2},
          eid = {103},
        pages = {103},
          doi = {10.1088/0004-637X/737/2/103},
archivePrefix = {arXiv},
       eprint = {1012.4804},
 primaryClass = {astro-ph.GA},
       adsurl = {https://ui.adsabs.harvard.edu/abs/2011ApJ...737..103S}
}

@ARTICLE{Tsuzuki2006,
       author = {{Tsuzuki}, Yumihiko and {Kawara}, Kimiaki and {Yoshii}, Yuzuru and {Oyabu}, Shinki and {Tanab{\'e}}, Toshihiko and {Matsuoka}, Yoshiki},
        title = "{Fe II Emission in 14 Low-Redshift Quasars. I. Observations}",
      journal = {\apj},
         year = 2006,
        month = oct,
       volume = {650},
       number = {1},
        pages = {57-79},
          doi = {10.1086/506376},
archivePrefix = {arXiv},
       eprint = {astro-ph/0606040},
 primaryClass = {astro-ph},
       adsurl = {https://ui.adsabs.harvard.edu/abs/2006ApJ...650...57T}
}

@ARTICLE{Salviander2007,
       author = {{Salviander}, S. and {Shields}, G.~A. and {Gebhardt}, K. and {Bonning}, E.~W.},
        title = "{The Black Hole Mass-Galaxy Bulge Relationship for QSOs in the Sloan Digital Sky Survey Data Release 3}",
      journal = {\apj},
         year = 2007,
        month = jun,
       volume = {662},
       number = {1},
        pages = {131-144},
          doi = {10.1086/513086},
archivePrefix = {arXiv},
       eprint = {astro-ph/0612568},
 primaryClass = {astro-ph},
       adsurl = {https://ui.adsabs.harvard.edu/abs/2007ApJ...662..131S}
}

@ARTICLE{Harrison_RamosAlmeida2024,
       author = {{Harrison}, Chris M. and {Ramos Almeida}, Cristina},
        title = "{Observational Tests of Active Galactic Nuclei Feedback: An Overview of Approaches and Interpretation}",
      journal = {Galaxies},
         year = 2024,
        month = apr,
       volume = {12},
       number = {2},
          eid = {17},
        pages = {17},
          doi = {10.3390/galaxies12020017},
archivePrefix = {arXiv},
       eprint = {2404.08050},
 primaryClass = {astro-ph.GA},
       adsurl = {https://ui.adsabs.harvard.edu/abs/2024Galax..12...17H}
}

@ARTICLE{Fabian2012,
       author = {{Fabian}, A.~C.},
        title = "{Observational Evidence of Active Galactic Nuclei Feedback}",
      journal = {\araa},
         year = 2012,
        month = sep,
       volume = {50},
        pages = {455-489},
          doi = {10.1146/annurev-astro-081811-125521},
archivePrefix = {arXiv},
       eprint = {1204.4114},
 primaryClass = {astro-ph.CO},
       adsurl = {https://ui.adsabs.harvard.edu/abs/2012ARA&A..50..455F}
}

@ARTICLE{Morganti2017,
       author = {{Morganti}, Raffaella},
        title = "{The many routes to AGN feedback}",
      journal = {Frontiers in Astronomy and Space Sciences},
         year = 2017,
        month = nov,
       volume = {4},
          eid = {42},
        pages = {42},
          doi = {10.3389/fspas.2017.00042},
archivePrefix = {arXiv},
       eprint = {1712.05301},
 primaryClass = {astro-ph.GA},
       adsurl = {https://ui.adsabs.harvard.edu/abs/2017FrASS...4...42M}
}

@ARTICLE{Riffel2023a,
       author = {{Riffel}, R.~A. and {Storchi-Bergmann}, T. and {Riffel}, R. and {Bianchin}, M. and {Zakamska}, N.~L. and {Ruschel-Dutra}, D. and {Bentz}, M.~C. and {Burtscher}, L. and {Crenshaw}, D.~M. and {Dahmer-Hahn}, L.~G. and {Dametto}, N.~Z. and {Davies}, R.~I. and {Diniz}, M.~R. and {Fischer}, T.~C. and {Harrison}, C.~M. and {Mainieri}, V. and {Revalski}, M. and {Rodriguez-Ardila}, A. and {Rosario}, D.~J. and {Sch{\"o}nell}, A.~J.},
        title = "{The AGNIFS survey: spatially resolved observations of hot molecular and ionized outflows in nearby active galaxies}",
      journal = {\mnras},
         year = 2023,
        month = may,
       volume = {521},
       number = {2},
        pages = {1832-1848},
          doi = {10.1093/mnras/stad599},
archivePrefix = {arXiv},
       eprint = {2302.11324},
 primaryClass = {astro-ph.GA},
       adsurl = {https://ui.adsabs.harvard.edu/abs/2023MNRAS.521.1832R}
}

@ARTICLE{Riffel2023b,
       author = {{Riffel}, Rog{\'e}rio and {Mallmann}, Nicolas D. and {Rembold}, Sandro B. and {Ilha}, Gabriele S. and {Riffel}, Rogemar A. and {Storchi-Bergmann}, Thaisa and {Ruschel-Dutra}, Daniel and {Vazdekis}, Alexandre and {Mart{\'\i}n-Navarro}, Ignacio and {Schimoia}, Jaderson S. and {Ramos Almeida}, Cristina and {da Costa}, Luiz N. and {Vila-Verde}, Glauber C. and {Gatto}, Lara},
        title = "{Mapping the stellar population and gas excitation of MaNGA galaxies with MEGACUBES. Results for AGN versus control sample}",
      journal = {\mnras},
         year = 2023,
        month = oct,
       volume = {524},
       number = {4},
        pages = {5640-5657},
          doi = {10.1093/mnras/stad2234},
archivePrefix = {arXiv},
       eprint = {2307.11474},
 primaryClass = {astro-ph.GA},
       adsurl = {https://ui.adsabs.harvard.edu/abs/2023MNRAS.524.5640R}
}

@ARTICLE{4MOST2019,
       author = {{de Jong}, R.~S. and {Agertz}, O. and {Berbel}, A.~A. and {Aird}, J. and {Alexander}, D.~A. and {Amarsi}, A. and {Anders}, F. and {Andrae}, R. and {Ansarinejad}, B. and {Ansorge}, W. and {Antilogus}, P. and {Anwand-Heerwart}, H. and {Arentsen}, A. and {Arnadottir}, A. and {Asplund}, M. and {Auger}, M. and {Azais}, N. and {Baade}, D. and {Baker}, G. and {Baker}, S. and {Balbinot}, E. and {Baldry}, I.~K. and {Banerji}, M. and {Barden}, S. and {Barklem}, P. and {Barth{\'e}l{\'e}my-Mazot}, E. and {Battistini}, C. and {Bauer}, S. and {Bell}, C.~P.~M. and {Bellido-Tirado}, O. and {Bellstedt}, S. and {Belokurov}, V. and {Bensby}, T. and {Bergemann}, M. and {Bestenlehner}, J.~M. and {Bielby}, R. and {Bilicki}, M. and {Blake}, C. and {Bland-Hawthorn}, J. and {Boeche}, C. and {Boland}, W. and {Boller}, T. and {Bongard}, S. and {Bongiorno}, A. and {Bonifacio}, P. and {Boudon}, D. and {Brooks}, D. and {Brown}, M.~J.~I. and {Brown}, R. and {Br{\"u}ggen}, M. and {Brynnel}, J. and {Brzeski}, J. and {Buchert}, T. and {Buschkamp}, P. and {Caffau}, E. and {Caillier}, P. and {Carrick}, J. and {Casagrande}, L. and {Case}, S. and {Casey}, A. and {Cesarini}, I. and {Cescutti}, G. and {Chapuis}, D. and {Chiappini}, C. and {Childress}, M. and {Christlieb}, N. and {Church}, R. and {Cioni}, M. -R.~L. and {Cluver}, M. and {Colless}, M. and {Collett}, T. and {Comparat}, J. and {Cooper}, A. and {Couch}, W. and {Courbin}, F. and {Croom}, S. and {Croton}, D. and {Daguis{\'e}}, E. and {Dalton}, G. and {Davies}, L.~J.~M. and {Davis}, T. and {de Laverny}, P. and {Deason}, A. and {Dionies}, F. and {Disseau}, K. and {Doel}, P. and {D{\"o}scher}, D. and {Driver}, S.~P. and {Dwelly}, T. and {Eckert}, D. and {Edge}, A. and {Edvardsson}, B. and {Youssoufi}, D.~E. and {Elhaddad}, A. and {Enke}, H. and {Erfanianfar}, G. and {Farrell}, T. and {Fechner}, T. and {Feiz}, C. and {Feltzing}, S. and {Ferreras}, I. and {Feuerstein}, D. and {Feuillet}, D. and {Finoguenov}, A. and {Ford}, D. and {Fotopoulou}, S. and {Fouesneau}, M. and {Frenk}, C. and {Frey}, S. and {Gaessler}, W. and {Geier}, S. and {Gentile Fusillo}, N. and {Gerhard}, O. and {Giannantonio}, T. and {Giannone}, D. and {Gibson}, B. and {Gillingham}, P. and {Gonz{\'a}lez-Fern{\'a}ndez}, C. and {Gonzalez-Solares}, E. and {Gottloeber}, S. and {Gould}, A. and {Grebel}, E.~K. and {Gueguen}, A. and {Guiglion}, G. and {Haehnelt}, M. and {Hahn}, T. and {Hansen}, C.~J. and {Hartman}, H. and {Hauptner}, K. and {Hawkins}, K. and {Haynes}, D. and {Haynes}, R. and {Heiter}, U. and {Helmi}, A. and {Aguayo}, C.~H. and {Hewett}, P. and {Hinton}, S. and {Hobbs}, D. and {Hoenig}, S. and {Hofman}, D. and {Hook}, I. and {Hopgood}, J. and {Hopkins}, A. and {Hourihane}, A. and {Howes}, L. and {Howlett}, C. and {Huet}, T. and {Irwin}, M. and {Iwert}, O. and {Jablonka}, P. and {Jahn}, T. and {Jahnke}, K. and {Jarno}, A. and {Jin}, S. and {Jofre}, P. and {Johl}, D. and {Jones}, D. and {J{\"o}nsson}, H. and {Jordan}, C. and {Karovicova}, I. and {Khalatyan}, A. and {Kelz}, A. and {Kennicutt}, R. and {King}, D. and {Kitaura}, F. and {Klar}, J. and {Klauser}, U. and {Kneib}, J. -P. and {Koch}, A. and {Koposov}, S. and {Kordopatis}, G. and {Korn}, A. and {Kosmalski}, J. and {Kotak}, R. and {Kovalev}, M. and {Kreckel}, K. and {Kripak}, Y. and {Krumpe}, M. and {Kuijken}, K. and {Kunder}, A. and {Kushniruk}, I. and {Lam}, M.~I. and {Lamer}, G. and {Laurent}, F. and {Lawrence}, J. and {Lehmitz}, M. and {Lemasle}, B. and {Lewis}, J. and {Li}, B. and {Lidman}, C. and {Lind}, K. and {Liske}, J. and {Lizon}, J. -L. and {Loveday}, J. and {Ludwig}, H. -G. and {McDermid}, R.~M. and {Maguire}, K. and {Mainieri}, V. and {Mali}, S. and {Mandel}, H.},
        title = "{4MOST: Project overview and information for the First Call for Proposals}",
      journal = {The Messenger},
         year = 2019,
        month = mar,
       volume = {175},
        pages = {3-11},
          doi = {10.18727/0722-6691/5117},
archivePrefix = {arXiv},
       eprint = {1903.02464},
 primaryClass = {astro-ph.IM},
       adsurl = {https://ui.adsabs.harvard.edu/abs/2019Msngr.175....3D}
}

@ARTICLE{DESI2026,
       author = {{DESI Collaboration} and {Abdul Karim}, M. and {Adame}, A.~G. and {Aguado}, D. and {Aguilar}, J. and {Ahlen}, S. and {Alam}, S. and {Aldering}, G. and {Alexander}, D.~M. and {Alfarsy}, R. and {Allen}, L. and {Allende Prieto}, C. and {Alves}, O. and {Anand}, A. and {Andrade}, U. and {Armengaud}, E. and {Avila}, S. and {Aviles}, A. and {Awan}, H. and {Bailey}, S. and {Baleato Lizancos}, A. and {Ballester}, O. and {Bault}, A. and {Bautista}, J. and {Bean}, R. and {Behera}, J. and {BenZvi}, S. and {Beraldo e Silva}, L. and {Bermejo-Climent}, J.~R. and {Beutler}, F. and {Bianchi}, D. and {Blake}, C. and {Blum}, R. and {Bolton}, A.~S. and {Bonici}, M. and {Brieden}, S. and {Brodzeller}, A. and {Brooks}, D. and {Buckley-Geer}, E. and {Burtin}, E. and {Bystr{\"o}m}, A. and {Canning}, R. and {Carnero Rosell}, A. and {Carr}, A. and {Carrilho}, P. and {Casas}, L. and {Castander}, F.~J. and {Cereskaite}, R. and {Cervantes-Cota}, J.~L. and {Chaussidon}, E. and {Chaves-Montero}, J. and {Chen}, S. and {Chen}, X. and {Circosta}, C. and {Claybaugh}, T. and {Cole}, S. and {Cooper}, A.~P. and {Cousinou}, M.-C. and {Cuceu}, A. and {Davis}, T.~M. and {Dawson}, K.~S. and {de Belsunce}, R. and {de la Cruz}, R. and {de la Macorra}, A. and {de Mattia}, A. and {Deiosso}, N. and {Della Costa}, J. and {Demina}, R. and {Demirbozan}, U. and {DeRose}, J. and {Dey}, A. and {Dey}, B. and {Ding}, J. and {Ding}, Z. and {Doel}, P. and {Douglass}, K. and {Dowicz}, M. and {Ebina}, H. and {Edelstein}, J. and {Eisenstein}, D.~J. and {Elbers}, W. and {Emas}, N. and {Escoffier}, S. and {Fagrelius}, P. and {Fan}, X. and {Fanning}, K. and {Favole}, G. and {Fawcett}, V.~A. and {Fern{\'a}ndez-Garc{\'\i}a}, E. and {Ferraro}, S. and {Findlay}, N. and {Font-Ribera}, A. and {Forero-Romero}, J.~E. and {Forero-S{\'a}nchez}, D. and {Frenk}, C.~S. and {G{\"a}nsicke}, B.~T. and {Galbany}, L. and {Garc{\'\i}a-Bellido}, J. and {Garcia-Quintero}, C. and {Garrison}, L.~H. and {Gazta{\~n}aga}, E. and {Gil-Mar{\'\i}n}, H. and {Gloudemans}, A. and {Gnedin}, O.~Y. and {Gontcho A Gontcho}, S. and {Gonzalez}, D. and {Gonzalez-Morales}, A.~X. and {Gonzalez-Perez}, V. and {Gordon}, C. and {Graur}, O. and {Green}, D. and {Gruen}, D. and {Gsponer}, R. and {Guandalin}, C. and {Gutierrez}, G. and {Guy}, J. and {Hahn}, C. and {Han}, J.~J. and {Han}, J. and {He}, S. and {Herrera-Alcantar}, H.~K. and {Heydenreich}, S. and {Honscheid}, K. and {Hou}, J. and {Howlett}, C. and {Huterer}, D. and {Ir{\v{s}}i{\v{c}}}, V. and {Ishak}, M. and {Jacques}, A. and {Jiang}, L. and {Jimenez}, J. and {Jing}, Y.~P. and {Joachimi}, B. and {Joudaki}, S. and {Joyce}, R. and {Jullo}, E. and {Juneau}, S. and {Kara{\c{c}}ayl{\i}}, N.~G. and {Karim}, T. and {Kehoe}, R. and {Kent}, S. and {Khederlarian}, A. and {Kirkby}, D. and {Kisner}, T. and {Kitaura}, F.-S. and {Kizhuprakkat}, N. and {Kong}, H. and {Koposov}, S.~E. and {Kremin}, A. and {Krolewski}, A. and {Lahav}, O. and {Lai}, Y. and {Lamman}, C. and {Lan}, T.-W. and {Landriau}, M. and {Lang}, D. and {Lange}, J.~U. and {Lasker}, J. and {Le Goff}, J.~M. and {Le Guillou}, L. and {Leauthaud}, A. and {Levi}, M.~E. and {Li}, S. and {Li}, T.~S. and {Liu}, W. and {Lodha}, K. and {Lokken}, M. and {Luo}, Y. and {Magneville}, C. and {Manera}, M. and {Manser}, C.~J. and {Margala}, D. and {Martini}, P. and {Maus}, M. and {McCullough}, J. and {McDonald}, P. and {Medina}, G.~E. and {Medina-Varela}, L. and {Meisner}, A. and {Mena-Fern{\'a}ndez}, J. and {Menegas}, A. and {Meneses-Rizo}, J. and {Mezcua}, M. and {Miquel}, R. and {Montero-Camacho}, P. and {Moon}, J. and {Moustakas}, J. and {Mu{\~n}oz-Guti{\'e}rrez}, A. and {Mu noz-Santos}, D. and {Myers}, A.~D. and {Myles}, J. and {Nadathur}, S. and {Najita}, J. and {Napolitano}, L. and {Newman}, J.~A. and {Nikakhtar}, F. and {Nikutta}, R. and {Niz}, G. and {Noriega}, H.~E. and {Nugent}, P.},
        title = "{Data Release 1 of the Dark Energy Spectroscopic Instrument}",
      journal = {\aj},
         year = 2026,
        month = may,
       volume = {171},
       number = {5},
          eid = {285},
        pages = {285},
          doi = {10.3847/1538-3881/ae4c43},
archivePrefix = {arXiv},
       eprint = {2503.14745},
 primaryClass = {astro-ph.CO},
       adsurl = {https://ui.adsabs.harvard.edu/abs/2026AJ....171..285D}
}

@ARTICLE{Venturi2021,
       author = {{Venturi}, G. and {Cresci}, G. and {Marconi}, A. and {Mingozzi}, M. and {Nardini}, E. and {Carniani}, S. and {Mannucci}, F. and {Marasco}, A. and {Maiolino}, R. and {Perna}, M. and {Treister}, E. and {Bland-Hawthorn}, J. and {Gallimore}, J.},
        title = "{MAGNUM survey: Compact jets causing large turmoil in galaxies. Enhanced line widths perpendicular to radio jets as tracers of jet-ISM interaction}",
      journal = {\aap},
         year = 2021,
        month = apr,
       volume = {648},
          eid = {A17},
        pages = {A17},
          doi = {10.1051/0004-6361/202039869},
archivePrefix = {arXiv},
       eprint = {2011.04677},
 primaryClass = {astro-ph.GA},
       adsurl = {https://ui.adsabs.harvard.edu/abs/2021A&A...648A..17V}
}

@ARTICLE{Balmaverde2019,
       author = {{Balmaverde}, B. and {Capetti}, A. and {Marconi}, A. and {Venturi}, G. and {Chiaberge}, M. and {Baldi}, R.~D. and {Baum}, S. and {Gilli}, R. and {Grandi}, P. and {Meyer}, E. and {Miley}, G. and {O'Dea}, C. and {Sparks}, W. and {Torresi}, E. and {Tremblay}, G.},
        title = "{The MURALES survey. II. Presentation of MUSE observations of 20 3C low-z radio galaxies and first results}",
      journal = {\aap},
         year = 2019,
        month = dec,
       volume = {632},
          eid = {A124},
        pages = {A124},
          doi = {10.1051/0004-6361/201935544},
archivePrefix = {arXiv},
       eprint = {1903.10768},
 primaryClass = {astro-ph.GA},
       adsurl = {https://ui.adsabs.harvard.edu/abs/2019A&A...632A.124B}
}

@ARTICLE{Speranza2024,
       author = {{Speranza}, G. and {Ramos Almeida}, C. and {Acosta-Pulido}, J.~A. and {Audibert}, A. and {Holden}, L.~R. and {Tadhunter}, C.~N. and {Lapi}, A. and {Gonz{\'a}lez-Mart{\'\i}n}, O. and {Brusa}, M. and {L{\'o}pez}, I.~E. and {Musiimenta}, B. and {Shankar}, F.},
        title = "{Multiphase characterization of AGN winds in five local type-2 quasars}",
      journal = {\aap},
         year = 2024,
        month = jan,
       volume = {681},
          eid = {A63},
        pages = {A63},
          doi = {10.1051/0004-6361/202347715},
archivePrefix = {arXiv},
       eprint = {2311.10132},
 primaryClass = {astro-ph.GA},
       adsurl = {https://ui.adsabs.harvard.edu/abs/2024A&A...681A..63S}
}

@ARTICLE{Reines2015,
       author = {{Reines}, Amy E. and {Volonteri}, Marta},
        title = "{Relations between Central Black Hole Mass and Total Galaxy Stellar Mass in the Local Universe}",
      journal = {\apj},
         year = 2015,
        month = nov,
       volume = {813},
       number = {2},
          eid = {82},
        pages = {82},
          doi = {10.1088/0004-637X/813/2/82},
archivePrefix = {arXiv},
       eprint = {1508.06274},
 primaryClass = {astro-ph.GA},
       adsurl = {https://ui.adsabs.harvard.edu/abs/2015ApJ...813...82R}
}

@ARTICLE{Suh2020,
       author = {{Suh}, Hyewon and {Civano}, Francesca and {Trakhtenbrot}, Benny and {Shankar}, Francesco and {Hasinger}, G{\"u}nther and {Sanders}, David B. and {Allevato}, Viola},
        title = "{No Significant Evolution of Relations between Black Hole Mass and Galaxy Total Stellar Mass Up to z {\ensuremath{\sim}} 2.5}",
      journal = {\apj},
         year = 2020,
        month = jan,
       volume = {889},
       number = {1},
          eid = {32},
        pages = {32},
          doi = {10.3847/1538-4357/ab5f5f},
archivePrefix = {arXiv},
       eprint = {1912.02824},
 primaryClass = {astro-ph.GA},
       adsurl = {https://ui.adsabs.harvard.edu/abs/2020ApJ...889...32S}
}

@ARTICLE{Pucha2025,
       author = {{Pucha}, Ragadeepika and {Juneau}, S. and {Dey}, Arjun and {Siudek}, M. and {Mezcua}, M. and {Moustakas}, J. and {BenZvi}, S. and {Hainline}, K. and {Hviding}, R. and {Mao}, Yao-Yuan and {Alexander}, D.~M. and {Alfarsy}, R. and {Circosta}, C. and {Guo}, Wei-Jian and {Manwadkar}, V. and {Martini}, P. and {Weaver}, B.~A. and {Aguilar}, J. and {Ahlen}, S. and {Bianchi}, D. and {Brooks}, D. and {Canning}, R. and {Claybaugh}, T. and {Dawson}, K. and {de la Macorra}, A. and {Dey}, Biprateep and {Doel}, P. and {Font-Ribera}, A. and {Forero-Romero}, J.~E. and {Gazta{\~n}aga}, E. and {Gontcho A Gontcho}, S. and {Gutierrez}, G. and {Honscheid}, K. and {Kehoe}, R. and {Koposov}, S.~E. and {Lambert}, A. and {Landriau}, M. and {Le Guillou}, L. and {Meisner}, A. and {Miquel}, R. and {Prada}, F. and {Rossi}, G. and {Sanchez}, E. and {Schlegel}, D. and {Schubnell}, M. and {Seo}, H. and {Sprayberry}, D. and {Tarl{\'e}}, G. and {Zou}, H.},
        title = "{Tripling the Census of Dwarf AGN Candidates Using DESI Early Data}",
      journal = {\apj},
         year = 2025,
        month = mar,
       volume = {982},
       number = {1},
          eid = {10},
        pages = {10},
          doi = {10.3847/1538-4357/adb1dd},
archivePrefix = {arXiv},
       eprint = {2411.00091},
 primaryClass = {astro-ph.GA},
       adsurl = {https://ui.adsabs.harvard.edu/abs/2025ApJ...982...10P}
}

@ARTICLE{Lauer2007,
       author = {{Lauer}, Tod R. and {Tremaine}, Scott and {Richstone}, Douglas and {Faber}, S.~M.},
        title = "{Selection Bias in Observing the Cosmological Evolution of the M$_{{\textbullet}}$-{\ensuremath{\sigma}} and M$_{{\textbullet}}$-L Relationships}",
      journal = {\apj},
         year = 2007,
        month = nov,
       volume = {670},
       number = {1},
        pages = {249-260},
          doi = {10.1086/522083},
archivePrefix = {arXiv},
       eprint = {0705.4103},
 primaryClass = {astro-ph},
       adsurl = {https://ui.adsabs.harvard.edu/abs/2007ApJ...670..249L}
}

@ARTICLE{Barth2013,
       author = {{Barth}, Aaron J. and {Pancoast}, Anna and {Bennert}, Vardha N. and {Brewer}, Brendon J. and {Canalizo}, Gabriela and {Filippenko}, Alexei V. and {Gates}, Elinor L. and {Greene}, Jenny E. and {Li}, Weidong and {Malkan}, Matthew A. and {Sand}, David J. and {Stern}, Daniel and {Treu}, Tommaso and {Woo}, Jong-Hak and {Assef}, Roberto J. and {Bae}, Hyun-Jin and {Buehler}, Tabitha and {Cenko}, S. Bradley and {Clubb}, Kelsey I. and {Cooper}, Michael C. and {Diamond-Stanic}, Aleksandar M. and {H{\"o}nig}, Sebastian F. and {Joner}, Michael D. and {Laney}, C. David and {Lazarova}, Mariana S. and {Nierenberg}, A.~M. and {Silverman}, Jeffrey M. and {Tollerud}, Erik J. and {Walsh}, Jonelle L.},
        title = "{The Lick AGN Monitoring Project 2011: Fe II Reverberation from the Outer Broad-line Region}",
      journal = {\apj},
         year = 2013,
        month = jun,
       volume = {769},
       number = {2},
          eid = {128},
        pages = {128},
          doi = {10.1088/0004-637X/769/2/128},
archivePrefix = {arXiv},
       eprint = {1304.4643},
 primaryClass = {astro-ph.CO},
       adsurl = {https://ui.adsabs.harvard.edu/abs/2013ApJ...769..128B}
}

@ARTICLE{Ji2025,
       author = {{Ji}, Xihan and {Maiolino}, Roberto and {Ferland}, Gary and {D'Eugenio}, Francesco and {Bhatawdekar}, Rachana and {Charlot}, St{\'e}phane and {Chevallard}, Jacopo and {Curti}, Mirko and {Curtis-Lake}, Emma and {Hainline}, Kevin and {Ji}, Zhiyuan and {Robertson}, Brant and {Rodr{\'\i}guez Del Pino}, Bruno and {Scholtz}, Jan and {Tacchella}, Sandro and {Williams}, Christina C. and {Witstok}, Joris},
        title = "{JADES {\textendash} the small blue bump in GN-z11: insights into the nuclear region of a galaxy at z = 10.6}",
      journal = {\mnras},
         year = 2025,
        month = aug,
       volume = {541},
       number = {3},
        pages = {2134-2161},
          doi = {10.1093/mnras/staf1083},
archivePrefix = {arXiv},
       eprint = {2405.05772},
 primaryClass = {astro-ph.GA},
       adsurl = {https://ui.adsabs.harvard.edu/abs/2025MNRAS.541.2134J}
}

@ARTICLE{Li2024_HSC,
       author = {{Li}, Junyao and {Silverman}, John D. and {Merloni}, Andrea and {Salvato}, Mara and {Buchner}, Johannes and {Goulding}, Andy and {Liu}, Teng and {Arcodia}, Riccardo and {Comparat}, Johan and {Ding}, Xuheng and {Ichikawa}, Kohei and {Imanishi}, Masatoshi and {Kawaguchi}, Toshihiro and {Kawinwanichakij}, Lalitwadee and {Toba}, Yoshiki},
        title = "{The eROSITA final equatorial-depth survey (eFEDS): host-galaxy demographics of X-ray AGNs with Subaru Hyper Suprime-Cam}",
      journal = {\mnras},
         year = 2024,
        month = jan,
       volume = {527},
       number = {3},
        pages = {4690-4704},
          doi = {10.1093/mnras/stad3438},
archivePrefix = {arXiv},
       eprint = {2302.12438},
 primaryClass = {astro-ph.GA},
       adsurl = {https://ui.adsabs.harvard.edu/abs/2024MNRAS.527.4690L}
}

@ARTICLE{Li2025,
       author = {{Li}, Junyao and {Silverman}, John D. and {Shen}, Yue and {Volonteri}, Marta and {Jahnke}, Knud and {Zhuang}, Ming-Yang and {Scoggins}, Matthew T. and {Ding}, Xuheng and {Harikane}, Yuichi and {Onoue}, Masafusa and {Tanaka}, Takumi S.},
        title = "{Tip of the Iceberg: Overmassive Black Holes at 4 < z < 7 Found by JWST Are Not Inconsistent with the Local  Relation}",
      journal = {\apj},
         year = 2025,
        month = mar,
       volume = {981},
       number = {1},
          eid = {19},
        pages = {19},
          doi = {10.3847/1538-4357/ada603},
archivePrefix = {arXiv},
       eprint = {2403.00074},
 primaryClass = {astro-ph.GA},
       adsurl = {https://ui.adsabs.harvard.edu/abs/2025ApJ...981...19L}
}

@ARTICLE{DallaBonta2025,
       author = {{Dalla Bont{\`a}}, E. and {Peterson}, B.~M. and {Grier}, C.~J. and {Berton}, M. and {Brandt}, W.~N. and {Ciroi}, S. and {Corsini}, E.~M. and {Dalla Barba}, B. and {Davies}, R. and {Dehghanian}, M. and {Edelson}, R. and {Foschini}, L. and {Gasparri}, D. and {Ho}, L.~C. and {Horne}, K. and {Iodice}, E. and {Morelli}, L. and {Pizzella}, A. and {Portaluri}, E. and {Shen}, Y. and {Schneider}, D.~P. and {Vestergaard}, M.},
        title = "{Estimating masses of supermassive black holes in active galactic nuclei from the H{\ensuremath{\alpha}} emission line}",
      journal = {\aap},
         year = 2025,
        month = apr,
       volume = {696},
          eid = {A48},
        pages = {A48},
          doi = {10.1051/0004-6361/202452746},
archivePrefix = {arXiv},
       eprint = {2410.21387},
 primaryClass = {astro-ph.GA},
       adsurl = {https://ui.adsabs.harvard.edu/abs/2025A&A...696A..48D}
}

@ARTICLE{Batiste2017,
       author = {{Batiste}, Merida and {Bentz}, Misty C. and {Raimundo}, Sandra I. and {Vestergaard}, Marianne and {Onken}, Christopher A.},
        title = "{Recalibration of the M $_{BH}$-{\ensuremath{\sigma}} $_{{\ensuremath{\star}}}$ Relation for AGN}",
      journal = {\apjl},
         year = 2017,
        month = mar,
       volume = {838},
       number = {1},
          eid = {L10},
        pages = {L10},
          doi = {10.3847/2041-8213/aa6571},
archivePrefix = {arXiv},
       eprint = {1612.02815},
 primaryClass = {astro-ph.GA},
       adsurl = {https://ui.adsabs.harvard.edu/abs/2017ApJ...838L..10B}
}

@ARTICLE{Kollmeier2026,
       author = {{Kollmeier}, Juna A. and {Rix}, Hans-Walter and {Aerts}, Conny and {Aird}, James and {Vera Alfaro}, Pablo and {Almeida}, Andr{\'e}s and {Anderson}, Scott F. and {Arseneau}, Stefan M. and {Assef}, Roberto J. and {Aviram}, Shir and {Aydar}, Catarina and {Badenes}, Carles and {Bandyopadhyay}, Avrajit and {Barger}, Kat and {Barkhouser}, Robert H. and {Bauer}, Franz E. and {Behmard}, Aida and {Bender}, Chad and {Besser}, Felipe and {Bhattarai}, Binod and {Bilgi}, Pavaman and {Bird}, Jonathan and {Bizyaev}, Dmitry and {Blanc}, Guillermo A. and {Blanton}, Michael R. and {Bochanski}, John and {Bovy}, Jo and {Brandon}, Christopher and {Brandt}, William Nielsen and {Brownstein}, Joel R. and {Buchner}, Johannes and {Burchett}, Joseph N. and {Carlberg}, Joleen and {Casey}, Andrew R. and {Castaneda-Carlos}, Lesly and {Chakraborty}, Priyanka and {Chanam{\'e}}, Julio and {Chandra}, Vedant and {Cherinka}, Brian and {Chilingarian}, Igor and {Comparat}, Johan and {Cosens}, Maren and {Covey}, Kevin and {Crane}, Jeffrey D. and {Crumpler}, Nicole R. and {Cruz-Gonzalez}, Irene and {Cunha}, Katia and {Cunningham}, Tim and {Dai}, Xinyu and {Darling}, Jeremy and {Davidson}, Jr., James W. and {Davis}, Megan C. and {De Lee}, Nathan and {Deacon}, Niall and {M{\'e}ndez Delgado}, Jos{\'e} Eduardo and {Demasi}, Sebastian and {Demianenko}, Mariia and {Derwent}, Mark and {D'Onghia}, Elena and {Di Mille}, Francesco and {Dias}, Bruno and {Donor}, John and {Dow}, Peter N. and {Drory}, Niv and {Dwelly}, Tom and {Egorov}, Oleg and {Egorova}, Evgeniya and {El-Badry}, Kareem and {Engelman}, Mike and {Eracleous}, Mike and {Fan}, Xiaohui and {Farr}, Emily and {Fries}, Logan and {Frinchaboy}, Peter and {Froning}, Cynthia S. and {G{\"a}nsicke}, Boris T. and {Garc{\'\i}a}, Pablo and {Gelfand}, Joseph and {Gentile Fusillo}, Nicola Pietro and {Glover}, Simon and {Grabowski}, Katie and {Grebel}, Eva K. and {Green}, Paul J. and {Grier}, Catherine and {Gupta}, Pramod and {Gray}, Aidan C. and {H{\"a}berle}, Maximilian and {Hall}, Patrick B. and {Hammond}, Randolph P. and {Hawkins}, Keith and {Harding}, Albert C. and {Heged{\H{u}}s}, Viola and {Herbst}, Tom and {Hermes}, J.~J. and {Rodr{\'\i}guez Hidalgo}, Paola and {Hilder}, Thomas and {Hogg}, David W. and {Holtzman}, Jon A. and {Horta}, Danny and {Huang}, Yang and {Hwang}, Hsiang-Chih and {Ibarra-Medel}, Hector Javier and {Imig}, Julie and {Inight}, Keith and {Jana}, Arghajit and {Ji}, Alexander P. and {Jim{\'e}nez-Arranz}, {\'O}scar and {Jofre}, Paula and {Johns}, Matt and {Johnson}, Jennifer and {Johnson}, James W. and {Johnston}, Evelyn J. and {Jones}, Amy M. and {Katkov}, Ivan and {Knapp}, Gillian R. and {Koekemoer}, Anton M. and {Kounkel}, Marina and {Kreckel}, Kathryn and {Krishnarao}, Dhanesh and {Krumpe}, Mirko and {Kumari}, Nimisha and {Kupfer}, Thomas and {Lacerna}, Ivan and {Laporte}, Chervin and {Lepine}, Sebastien and {Li}, Jing and {Liu}, Xin and {Loebman}, Sarah and {Long}, Knox and {Roman-Lopes}, Alexandre and {Lu}, Yuxi and {Majewski}, Steven Raymond and {Maoz}, Dan and {McKinnon}, Kevin A. and {Medan}, Ilija and {Merloni}, Andrea and {Minniti}, Dante and {Morrison}, Sean and {Myers}, Natalie and {M{\'e}sz{\'a}ros}, Szabolcs and {Nandra}, Kirpal and {Nayak}, Prasanta K. and {Ness}, Melissa K. and {Nidever}, David L. and {O'Brien}, Thomas and {Oeur}, Micah and {Oravetz}, Audrey and {Oravetz}, Daniel and {Otto}, Jonah and {Pallathadka}, Gautham Adamane and {Palunas}, Povilas and {Pan}, Kaike and {Pappalardo}, Daniel and {Pandey}, Rakesh and {Negrete Pe{\~n}aloza}, Castalia Alenka and {Pinsonneault}, Marc H. and {Pogge}, Richard W. and {Taghizadeh Popp}, Manuchehr and {Price-Whelan}, Adrian M. and {Pulatova}, Nadiia and {Qiu}, Dan and {Ramirez}, Solange and {Rankine}, Amy and {Ricci}, Claudio and {Runnoe}, Jessie C. and {Sanchez}, Sebastian and {Salvato}, Mara and {Sarbadhicary}, Sumit K. and {Sattler}, Natascha and {Saydjari}, Andrew K. and {Sayres}, Conor and {Schinnerer}, Eva and {Schlaufman}, Kevin C. and {Schneider}, Donald P. and {Schreiber}, Matthias R. and {Schwope}, Axel and {Serna}, Javier and {Shen}, Yue and {Sif{\'o}n}, Crist{\'o}bal and {Singh}, Amrita and {Sinha}, Amaya and {Smee}, Stephen and {Song}, Ying-Yi and {Souto}, Diogo and {Stassun}, Keivan G. and {Steinmetz}, Matthias and {Stone-Martinez}, Alexander and {Stringfellow}, Guy and {Stutz}, Amelia and {S{\'a}nchez-Gallego}, Jos{\'e} and {Tan}, Jonathan C. and {Tayar}, Jamie and {Thai}, Riley and {Thakar}, Ani and {Ting}, Yuan-Sen and {Tkachenko}, Andrew and {Tovmassian}, Gagik and {Trakhtenbrot}, Benny and {Fern{\'a}ndez-Trincado}, Jos{\'e} G. and {Troup}, Nicholas},
        title = "{Sloan Digital Sky Survey. V. Pioneering Panoptic Spectroscopy}",
      journal = {\aj},
         year = 2026,
        month = jan,
       volume = {171},
       number = {1},
          eid = {52},
        pages = {52},
          doi = {10.3847/1538-3881/ae0576},
archivePrefix = {arXiv},
       eprint = {2507.06989},
 primaryClass = {astro-ph.IM},
       adsurl = {https://ui.adsabs.harvard.edu/abs/2026AJ....171...52K}
}

@ARTICLE{Strauss2002,
       author = {{Strauss}, Michael A. and {Weinberg}, David H. and {Lupton}, Robert H. and {Narayanan}, Vijay K. and {Annis}, James and {Bernardi}, Mariangela and {Blanton}, Michael and {Burles}, Scott and {Connolly}, A.~J. and {Dalcanton}, Julianne and {Doi}, Mamoru and {Eisenstein}, Daniel and {Frieman}, Joshua A. and {Fukugita}, Masataka and {Gunn}, James E. and {Ivezi{\'c}}, {\v{Z}}eljko and {Kent}, Stephen and {Kim}, Rita S.~J. and {Knapp}, G.~R. and {Kron}, Richard G. and {Munn}, Jeffrey A. and {Newberg}, Heidi Jo and {Nichol}, R.~C. and {Okamura}, Sadanori and {Quinn}, Thomas R. and {Richmond}, Michael W. and {Schlegel}, David J. and {Shimasaku}, Kazuhiro and {SubbaRao}, Mark and {Szalay}, Alexander S. and {Vanden Berk}, Dan and {Vogeley}, Michael S. and {Yanny}, Brian and {Yasuda}, Naoki and {York}, Donald G. and {Zehavi}, Idit},
        title = "{Spectroscopic Target Selection in the Sloan Digital Sky Survey: The Main Galaxy Sample}",
      journal = {\aj},
         year = 2002,
        month = sep,
       volume = {124},
       number = {3},
        pages = {1810-1824},
          doi = {10.1086/342343},
archivePrefix = {arXiv},
       eprint = {astro-ph/0206225},
 primaryClass = {astro-ph},
       adsurl = {https://ui.adsabs.harvard.edu/abs/2002AJ....124.1810S}
}

@ARTICLE{Eisenstein2001,
       author = {{Eisenstein}, Daniel J. and {Annis}, James and {Gunn}, James E. and {Szalay}, Alexander S. and {Connolly}, Andrew J. and {Nichol}, R.~C. and {Bahcall}, Neta A. and {Bernardi}, Mariangela and {Burles}, Scott and {Castander}, Francisco J. and {Fukugita}, Masataka and {Hogg}, David W. and {Ivezi{\'c}}, {\v{Z}}eljko and {Knapp}, G.~R. and {Lupton}, Robert H. and {Narayanan}, Vijay and {Postman}, Marc and {Reichart}, Daniel E. and {Richmond}, Michael and {Schneider}, Donald P. and {Schlegel}, David J. and {Strauss}, Michael A. and {SubbaRao}, Mark and {Tucker}, Douglas L. and {Vanden Berk}, Daniel and {Vogeley}, Michael S. and {Weinberg}, David H. and {Yanny}, Brian},
        title = "{Spectroscopic Target Selection for the Sloan Digital Sky Survey: The Luminous Red Galaxy Sample}",
      journal = {\aj},
         year = 2001,
        month = nov,
       volume = {122},
       number = {5},
        pages = {2267-2280},
          doi = {10.1086/323717},
archivePrefix = {arXiv},
       eprint = {astro-ph/0108153},
 primaryClass = {astro-ph},
       adsurl = {https://ui.adsabs.harvard.edu/abs/2001AJ....122.2267E}
}

@ARTICLE{Abazajian2009,
       author = {{Abazajian}, Kevork N. and {Adelman-McCarthy}, Jennifer K. and {Ag{\"u}eros}, Marcel A. and {Allam}, Sahar S. and {Allende Prieto}, Carlos and {An}, Deokkeun and {Anderson}, Kurt S.~J. and {Anderson}, Scott F. and {Annis}, James and {Bahcall}, Neta A. and {Bailer-Jones}, C.~A.~L. and {Barentine}, J.~C. and {Bassett}, Bruce A. and {Becker}, Andrew C. and {Beers}, Timothy C. and {Bell}, Eric F. and {Belokurov}, Vasily and {Berlind}, Andreas A. and {Berman}, Eileen F. and {Bernardi}, Mariangela and {Bickerton}, Steven J. and {Bizyaev}, Dmitry and {Blakeslee}, John P. and {Blanton}, Michael R. and {Bochanski}, John J. and {Boroski}, William N. and {Brewington}, Howard J. and {Brinchmann}, Jarle and {Brinkmann}, J. and {Brunner}, Robert J. and {Budav{\'a}ri}, Tam{\'a}s and {Carey}, Larry N. and {Carliles}, Samuel and {Carr}, Michael A. and {Castander}, Francisco J. and {Cinabro}, David and {Connolly}, A.~J. and {Csabai}, Istv{\'a}n and {Cunha}, Carlos E. and {Czarapata}, Paul C. and {Davenport}, James R.~A. and {de Haas}, Ernst and {Dilday}, Ben and {Doi}, Mamoru and {Eisenstein}, Daniel J. and {Evans}, Michael L. and {Evans}, N.~W. and {Fan}, Xiaohui and {Friedman}, Scott D. and {Frieman}, Joshua A. and {Fukugita}, Masataka and {G{\"a}nsicke}, Boris T. and {Gates}, Evalyn and {Gillespie}, Bruce and {Gilmore}, G. and {Gonzalez}, Belinda and {Gonzalez}, Carlos F. and {Grebel}, Eva K. and {Gunn}, James E. and {Gy{\"o}ry}, Zsuzsanna and {Hall}, Patrick B. and {Harding}, Paul and {Harris}, Frederick H. and {Harvanek}, Michael and {Hawley}, Suzanne L. and {Hayes}, Jeffrey J.~E. and {Heckman}, Timothy M. and {Hendry}, John S. and {Hennessy}, Gregory S. and {Hindsley}, Robert B. and {Hoblitt}, J. and {Hogan}, Craig J. and {Hogg}, David W. and {Holtzman}, Jon A. and {Hyde}, Joseph B. and {Ichikawa}, Shin-ichi and {Ichikawa}, Takashi and {Im}, Myungshin and {Ivezi{\'c}}, {\v{Z}}eljko and {Jester}, Sebastian and {Jiang}, Linhua and {Johnson}, Jennifer A. and {Jorgensen}, Anders M. and {Juri{\'c}}, Mario and {Kent}, Stephen M. and {Kessler}, R. and {Kleinman}, S.~J. and {Knapp}, G.~R. and {Konishi}, Kohki and {Kron}, Richard G. and {Krzesinski}, Jurek and {Kuropatkin}, Nikolay and {Lampeitl}, Hubert and {Lebedeva}, Svetlana and {Lee}, Myung Gyoon and {Lee}, Young Sun and {French Leger}, R. and {L{\'e}pine}, S{\'e}bastien and {Li}, Nolan and {Lima}, Marcos and {Lin}, Huan and {Long}, Daniel C. and {Loomis}, Craig P. and {Loveday}, Jon and {Lupton}, Robert H. and {Magnier}, Eugene and {Malanushenko}, Olena and {Malanushenko}, Viktor and {Mandelbaum}, Rachel and {Margon}, Bruce and {Marriner}, John P. and {Mart{\'\i}nez-Delgado}, David and {Matsubara}, Takahiko and {McGehee}, Peregrine M. and {McKay}, Timothy A. and {Meiksin}, Avery and {Morrison}, Heather L. and {Mullally}, Fergal and {Munn}, Jeffrey A. and {Murphy}, Tara and {Nash}, Thomas and {Nebot}, Ada and {Neilsen}, Jr., Eric H. and {Newberg}, Heidi Jo and {Newman}, Peter R. and {Nichol}, Robert C. and {Nicinski}, Tom and {Nieto-Santisteban}, Maria and {Nitta}, Atsuko and {Okamura}, Sadanori and {Oravetz}, Daniel J. and {Ostriker}, Jeremiah P. and {Owen}, Russell and {Padmanabhan}, Nikhil and {Pan}, Kaike and {Park}, Changbom and {Pauls}, George and {Peoples}, Jr., John and {Percival}, Will J. and {Pier}, Jeffrey R. and {Pope}, Adrian C. and {Pourbaix}, Dimitri and {Price}, Paul A. and {Purger}, Norbert and {Quinn}, Thomas and {Raddick}, M. Jordan and {Re Fiorentin}, Paola and {Richards}, Gordon T. and {Richmond}, Michael W. and {Riess}, Adam G. and {Rix}, Hans-Walter and {Rockosi}, Constance M. and {Sako}, Masao and {Schlegel}, David J. and {Schneider}, Donald P. and {Scholz}, Ralf-Dieter and {Schreiber}, Matthias R. and {Schwope}, Axel D. and {Seljak}, Uro{\v{s}} and {Sesar}, Branimir and {Sheldon}, Erin and {Shimasaku}, Kazu and {Sibley}, Valena C. and {Simmons}, A.~E. and {Sivarani}, Thirupathi and {Allyn Smith}, J. and {Smith}, Martin C. and {Smol{\v{c}}i{\'c}}, Vernesa and {Snedden}, Stephanie A. and {Stebbins}, Albert and {Steinmetz}, Matthias and {Stoughton}, Chris and {Strauss}, Michael A. and {SubbaRao}, Mark and {Suto}, Yasushi and {Szalay}, Alexander S. and {Szapudi}, Istv{\'a}n and {Szkody}, Paula and {Tanaka}, Masayuki and {Tegmark}, Max and {Teodoro}, Luis F.~A. and {Thakar}, Aniruddha R. and {Tremonti}, Christy A. and {Tucker}, Douglas L. and {Uomoto}, Alan and {Vanden Berk}, Daniel E. and {Vandenberg}, Jan and {Vidrih}, S. and {Vogeley}, Michael S. and {Voges}, Wolfgang and {Vogt}, Nicole P. and {Wadadekar}, Yogesh and {Watters}, Shannon and {Weinberg}, David H. and {West}, Andrew A. and {White}, Simon D.~M. and {Wilhite}, Brian C. and {Wonders}, Alainna C. and {Yanny}, Brian and {Yocum}, D.~R.},
        title = "{The Seventh Data Release of the Sloan Digital Sky Survey}",
      journal = {\apjs},
         year = 2009,
        month = jun,
       volume = {182},
       number = {2},
        pages = {543-558},
          doi = {10.1088/0067-0049/182/2/543},
archivePrefix = {arXiv},
       eprint = {0812.0649},
 primaryClass = {astro-ph},
       adsurl = {https://ui.adsabs.harvard.edu/abs/2009ApJS..182..543A}
}

@ARTICLE{Eisenstein2011,
       author = {{Eisenstein}, Daniel J. and {Weinberg}, David H. and {Agol}, Eric and {Aihara}, Hiroaki and {Allende Prieto}, Carlos and {Anderson}, Scott F. and {Arns}, James A. and {Aubourg}, {\'E}ric and {Bailey}, Stephen and {Balbinot}, Eduardo and {Barkhouser}, Robert and {Beers}, Timothy C. and {Berlind}, Andreas A. and {Bickerton}, Steven J. and {Bizyaev}, Dmitry and {Blanton}, Michael R. and {Bochanski}, John J. and {Bolton}, Adam S. and {Bosman}, Casey T. and {Bovy}, Jo and {Brandt}, W.~N. and {Breslauer}, Ben and {Brewington}, Howard J. and {Brinkmann}, J. and {Brown}, Peter J. and {Brownstein}, Joel R. and {Burger}, Dan and {Busca}, Nicolas G. and {Campbell}, Heather and {Cargile}, Phillip A. and {Carithers}, William C. and {Carlberg}, Joleen K. and {Carr}, Michael A. and {Chang}, Liang and {Chen}, Yanmei and {Chiappini}, Cristina and {Comparat}, Johan and {Connolly}, Natalia and {Cortes}, Marina and {Croft}, Rupert A.~C. and {Cunha}, Katia and {da Costa}, Luiz N. and {Davenport}, James R.~A. and {Dawson}, Kyle and {De Lee}, Nathan and {Porto de Mello}, Gustavo F. and {de Simoni}, Fernando and {Dean}, Janice and {Dhital}, Saurav and {Ealet}, Anne and {Ebelke}, Garrett L. and {Edmondson}, Edward M. and {Eiting}, Jacob M. and {Escoffier}, Stephanie and {Esposito}, Massimiliano and {Evans}, Michael L. and {Fan}, Xiaohui and {Femen{\'\i}a Castell{\'a}}, Bruno and {Dutra Ferreira}, Leticia and {Fitzgerald}, Greg and {Fleming}, Scott W. and {Font-Ribera}, Andreu and {Ford}, Eric B. and {Frinchaboy}, Peter M. and {Garc{\'\i}a P{\'e}rez}, Ana Elia and {Gaudi}, B. Scott and {Ge}, Jian and {Ghezzi}, Luan and {Gillespie}, Bruce A. and {Gilmore}, G. and {Girardi}, L{\'e}o and {Gott}, J. Richard and {Gould}, Andrew and {Grebel}, Eva K. and {Gunn}, James E. and {Hamilton}, Jean-Christophe and {Harding}, Paul and {Harris}, David W. and {Hawley}, Suzanne L. and {Hearty}, Frederick R. and {Hennawi}, Joseph F. and {Gonz{\'a}lez Hern{\'a}ndez}, Jonay I. and {Ho}, Shirley and {Hogg}, David W. and {Holtzman}, Jon A. and {Honscheid}, Klaus and {Inada}, Naohisa and {Ivans}, Inese I. and {Jiang}, Linhua and {Jiang}, Peng and {Johnson}, Jennifer A. and {Jordan}, Cathy and {Jordan}, Wendell P. and {Kauffmann}, Guinevere and {Kazin}, Eyal and {Kirkby}, David and {Klaene}, Mark A. and {Knapp}, G.~R. and {Kneib}, Jean-Paul and {Kochanek}, C.~S. and {Koesterke}, Lars and {Kollmeier}, Juna A. and {Kron}, Richard G. and {Lampeitl}, Hubert and {Lang}, Dustin and {Lawler}, James E. and {Le Goff}, Jean-Marc and {Lee}, Brian L. and {Lee}, Young Sun and {Leisenring}, Jarron M. and {Lin}, Yen-Ting and {Liu}, Jian and {Long}, Daniel C. and {Loomis}, Craig P. and {Lucatello}, Sara and {Lundgren}, Britt and {Lupton}, Robert H. and {Ma}, Bo and {Ma}, Zhibo and {MacDonald}, Nicholas and {Mack}, Claude and {Mahadevan}, Suvrath and {Maia}, Marcio A.~G. and {Majewski}, Steven R. and {Makler}, Martin and {Malanushenko}, Elena and {Malanushenko}, Viktor and {Mandelbaum}, Rachel and {Maraston}, Claudia and {Margala}, Daniel and {Maseman}, Paul and {Masters}, Karen L. and {McBride}, Cameron K. and {McDonald}, Patrick and {McGreer}, Ian D. and {McMahon}, Richard G. and {Mena Requejo}, Olga and {M{\'e}nard}, Brice and {Miralda-Escud{\'e}}, Jordi and {Morrison}, Heather L. and {Mullally}, Fergal and {Muna}, Demitri and {Murayama}, Hitoshi and {Myers}, Adam D. and {Naugle}, Tracy and {Neto}, Angelo Fausti and {Nguyen}, Duy Cuong and {Nichol}, Robert C. and {Nidever}, David L. and {O'Connell}, Robert W. and {Ogando}, Ricardo L.~C. and {Olmstead}, Matthew D. and {Oravetz}, Daniel J. and {Padmanabhan}, Nikhil and {Paegert}, Martin and {Palanque-Delabrouille}, Nathalie and {Pan}, Kaike and {Pandey}, Parul and {Parejko}, John K. and {P{\^a}ris}, Isabelle and {Pellegrini}, Paulo and {Pepper}, Joshua and {Percival}, Will J. and {Petitjean}, Patrick and {Pfaffenberger}, Robert and {Pforr}, Janine and {Phleps}, Stefanie and {Pichon}, Christophe and {Pieri}, Matthew M. and {Prada}, Francisco and {Price-Whelan}, Adrian M. and {Raddick}, M. Jordan and {Ramos}, Beatriz H.~F. and {Reid}, I. Neill and {Reyle}, Celine and {Rich}, James and {Richards}, Gordon T. and {Rieke}, George H. and {Rieke}, Marcia J. and {Rix}, Hans-Walter and {Robin}, Annie C. and {Rocha-Pinto}, Helio J. and {Rockosi}, Constance M. and {Roe}, Natalie A. and {Rollinde}, Emmanuel and {Ross}, Ashley J. and {Ross}, Nicholas P. and {Rossetto}, Bruno and {S{\'a}nchez}, Ariel G. and {Santiago}, Basilio and {Sayres}, Conor and {Schiavon}, Ricardo and {Schlegel}, David J. and {Schlesinger}, Katharine J. and {Schmidt}, Sarah J. and {Schneider}, Donald P. and {Sellgren}, Kris and {Shelden}, Alaina and {Sheldon}, Erin and {Shetrone}, Matthew and {Shu}, Yiping and {Silverman}, John D. and {Simmerer}, Jennifer and {Simmons}, Audrey E. and {Sivarani}, Thirupathi and {Skrutskie}, M.~F. and {Slosar}, An{\v{z}}e and {Smee}, Stephen and {Smith}, Verne V. and {Snedden}, Stephanie A. and {Stassun}, Keivan G. and {Steele}, Oliver and {Steinmetz}, Matthias and {Stockett}, Mark H. and {Stollberg}, Todd and {Strauss}, Michael A. and {Szalay}, Alexander S. and {Tanaka}, Masayuki and {Thakar}, Aniruddha R. and {Thomas}, Daniel and {Tinker}, Jeremy L. and {Tofflemire}, Benjamin M. and {Tojeiro}, Rita and {Tremonti}, Christy A. and {Vargas Maga{\~n}a}, Mariana and {Verde}, Licia and {Vogt}, Nicole P. and {Wake}, David A. and {Wan}, Xiaoke and {Wang}, Ji and {Weaver}, Benjamin A. and {White}, Martin and {White}, Simon D.~M. and {Wilson}, John C. and {Wisniewski}, John P. and {Wood-Vasey}, W. Michael and {Yanny}, Brian and {Yasuda}, Naoki and {Y{\`e}che}, Christophe and {York}, Donald G. and {Young}, Erick and {Zasowski}, Gail and {Zehavi}, Idit and {Zhao}, Bo},
        title = "{SDSS-III: Massive Spectroscopic Surveys of the Distant Universe, the Milky Way, and Extra-Solar Planetary Systems}",
      journal = {\aj},
         year = 2011,
        month = sep,
       volume = {142},
       number = {3},
          eid = {72},
        pages = {72},
          doi = {10.1088/0004-6256/142/3/72},
archivePrefix = {arXiv},
       eprint = {1101.1529},
 primaryClass = {astro-ph.IM},
       adsurl = {https://ui.adsabs.harvard.edu/abs/2011AJ....142...72E}
}

@ARTICLE{Dawson2013_boss,
       author = {{Dawson}, Kyle S. and {Schlegel}, David J. and {Ahn}, Christopher P. and {Anderson}, Scott F. and {Aubourg}, {\'E}ric and {Bailey}, Stephen and {Barkhouser}, Robert H. and {Bautista}, Julian E. and {Beifiori}, Alessandra and {Berlind}, Andreas A. and {Bhardwaj}, Vaishali and {Bizyaev}, Dmitry and {Blake}, Cullen H. and {Blanton}, Michael R. and {Blomqvist}, Michael and {Bolton}, Adam S. and {Borde}, Arnaud and {Bovy}, Jo and {Brandt}, W.~N. and {Brewington}, Howard and {Brinkmann}, Jon and {Brown}, Peter J. and {Brownstein}, Joel R. and {Bundy}, Kevin and {Busca}, N.~G. and {Carithers}, William and {Carnero}, Aurelio R. and {Carr}, Michael A. and {Chen}, Yanmei and {Comparat}, Johan and {Connolly}, Natalia and {Cope}, Frances and {Croft}, Rupert A.~C. and {Cuesta}, Antonio J. and {da Costa}, Luiz N. and {Davenport}, James R.~A. and {Delubac}, Timoth{\'e}e and {de Putter}, Roland and {Dhital}, Saurav and {Ealet}, Anne and {Ebelke}, Garrett L. and {Eisenstein}, Daniel J. and {Escoffier}, S. and {Fan}, Xiaohui and {Filiz Ak}, N. and {Finley}, Hayley and {Font-Ribera}, Andreu and {G{\'e}nova-Santos}, R. and {Gunn}, James E. and {Guo}, Hong and {Haggard}, Daryl and {Hall}, Patrick B. and {Hamilton}, Jean-Christophe and {Harris}, Ben and {Harris}, David W. and {Ho}, Shirley and {Hogg}, David W. and {Holder}, Diana and {Honscheid}, Klaus and {Huehnerhoff}, Joe and {Jordan}, Beatrice and {Jordan}, Wendell P. and {Kauffmann}, Guinevere and {Kazin}, Eyal A. and {Kirkby}, David and {Klaene}, Mark A. and {Kneib}, Jean-Paul and {Le Goff}, Jean-Marc and {Lee}, Khee-Gan and {Long}, Daniel C. and {Loomis}, Craig P. and {Lundgren}, Britt and {Lupton}, Robert H. and {Maia}, Marcio A.~G. and {Makler}, Martin and {Malanushenko}, Elena and {Malanushenko}, Viktor and {Mandelbaum}, Rachel and {Manera}, Marc and {Maraston}, Claudia and {Margala}, Daniel and {Masters}, Karen L. and {McBride}, Cameron K. and {McDonald}, Patrick and {McGreer}, Ian D. and {McMahon}, Richard G. and {Mena}, Olga and {Miralda-Escud{\'e}}, Jordi and {Montero-Dorta}, Antonio D. and {Montesano}, Francesco and {Muna}, Demitri and {Myers}, Adam D. and {Naugle}, Tracy and {Nichol}, Robert C. and {Noterdaeme}, Pasquier and {Nuza}, Sebasti{\'a}n E. and {Olmstead}, Matthew D. and {Oravetz}, Audrey and {Oravetz}, Daniel J. and {Owen}, Russell and {Padmanabhan}, Nikhil and {Palanque-Delabrouille}, Nathalie and {Pan}, Kaike and {Parejko}, John K. and {P{\^a}ris}, Isabelle and {Percival}, Will J. and {P{\'e}rez-Fournon}, Ismael and {P{\'e}rez-R{\`a}fols}, Ignasi and {Petitjean}, Patrick and {Pfaffenberger}, Robert and {Pforr}, Janine and {Pieri}, Matthew M. and {Prada}, Francisco and {Price-Whelan}, Adrian M. and {Raddick}, M. Jordan and {Rebolo}, Rafael and {Rich}, James and {Richards}, Gordon T. and {Rockosi}, Constance M. and {Roe}, Natalie A. and {Ross}, Ashley J. and {Ross}, Nicholas P. and {Rossi}, Graziano and {Rubi{\~n}o-Martin}, J.~A. and {Samushia}, Lado and {S{\'a}nchez}, Ariel G. and {Sayres}, Conor and {Schmidt}, Sarah J. and {Schneider}, Donald P. and {Sc{\'o}ccola}, C.~G. and {Seo}, Hee-Jong and {Shelden}, Alaina and {Sheldon}, Erin and {Shen}, Yue and {Shu}, Yiping and {Slosar}, An{\v{z}}e and {Smee}, Stephen A. and {Snedden}, Stephanie A. and {Stauffer}, Fritz and {Steele}, Oliver and {Strauss}, Michael A. and {Streblyanska}, Alina and {Suzuki}, Nao and {Swanson}, Molly E.~C. and {Tal}, Tomer and {Tanaka}, Masayuki and {Thomas}, Daniel and {Tinker}, Jeremy L. and {Tojeiro}, Rita and {Tremonti}, Christy A. and {Vargas Maga{\~n}a}, M. and {Verde}, Licia and {Viel}, Matteo and {Wake}, David A. and {Watson}, Mike and {Weaver}, Benjamin A. and {Weinberg}, David H. and {Weiner}, Benjamin J. and {West}, Andrew A. and {White}, Martin and {Wood-Vasey}, W.~M. and {Yeche}, Christophe and {Zehavi}, Idit and {Zhao}, Gong-Bo and {Zheng}, Zheng},
        title = "{The Baryon Oscillation Spectroscopic Survey of SDSS-III}",
      journal = {\aj},
         year = 2013,
        month = jan,
       volume = {145},
       number = {1},
          eid = {10},
        pages = {10},
          doi = {10.1088/0004-6256/145/1/10},
archivePrefix = {arXiv},
       eprint = {1208.0022},
 primaryClass = {astro-ph.CO},
       adsurl = {https://ui.adsabs.harvard.edu/abs/2013AJ....145...10D}
}

@ARTICLE{Abdurrouf_2022_sdss_dr17,
       author = {{Abdurro'uf} and {Accetta}, Katherine and {Aerts}, Conny and {Silva Aguirre}, V{\'\i}ctor and {Ahumada}, Romina and {Ajgaonkar}, Nikhil and {Filiz Ak}, N. and {Alam}, Shadab and {Allende Prieto}, Carlos and {Almeida}, Andr{\'e}s and {Anders}, Friedrich and {Anderson}, Scott F. and {Andrews}, Brett H. and {Anguiano}, Borja and {Aquino-Ort{\'\i}z}, Erik and {Arag{\'o}n-Salamanca}, Alfonso and {Argudo-Fern{\'a}ndez}, Maria and {Ata}, Metin and {Aubert}, Marie and {Avila-Reese}, Vladimir and {Badenes}, Carles and {Barb{\'a}}, Rodolfo H. and {Barger}, Kat and {Barrera-Ballesteros}, Jorge K. and {Beaton}, Rachael L. and {Beers}, Timothy C. and {Belfiore}, Francesco and {Bender}, Chad F. and {Bernardi}, Mariangela and {Bershady}, Matthew A. and {Beutler}, Florian and {Bidin}, Christian Moni and {Bird}, Jonathan C. and {Bizyaev}, Dmitry and {Blanc}, Guillermo A. and {Blanton}, Michael R. and {Boardman}, Nicholas Fraser and {Bolton}, Adam S. and {Boquien}, M{\'e}d{\'e}ric and {Borissova}, Jura and {Bovy}, Jo and {Brandt}, W.~N. and {Brown}, Jordan and {Brownstein}, Joel R. and {Brusa}, Marcella and {Buchner}, Johannes and {Bundy}, Kevin and {Burchett}, Joseph N. and {Bureau}, Martin and {Burgasser}, Adam and {Cabang}, Tuesday K. and {Campbell}, Stephanie and {Cappellari}, Michele and {Carlberg}, Joleen K. and {Wanderley}, F{\'a}bio Carneiro and {Carrera}, Ricardo and {Cash}, Jennifer and {Chen}, Yan-Ping and {Chen}, Wei-Huai and {Cherinka}, Brian and {Chiappini}, Cristina and {Choi}, Peter Doohyun and {Chojnowski}, S. Drew and {Chung}, Haeun and {Clerc}, Nicolas and {Cohen}, Roger E. and {Comerford}, Julia M. and {Comparat}, Johan and {da Costa}, Luiz and {Covey}, Kevin and {Crane}, Jeffrey D. and {Cruz-Gonzalez}, Irene and {Culhane}, Connor and {Cunha}, Katia and {Dai}, Y. Sophia and {Damke}, Guillermo and {Darling}, Jeremy and {Davidson}, James W., Jr. and {Davies}, Roger and {Dawson}, Kyle and {De Lee}, Nathan and {Diamond-Stanic}, Aleksandar M. and {Cano-D{\'\i}az}, Mariana and {S{\'a}nchez}, Helena Dom{\'\i}nguez and {Donor}, John and {Duckworth}, Chris and {Dwelly}, Tom and {Eisenstein}, Daniel J. and {Elsworth}, Yvonne P. and {Emsellem}, Eric and {Eracleous}, Mike and {Escoffier}, Stephanie and {Fan}, Xiaohui and {Farr}, Emily and {Feng}, Shuai and {Fern{\'a}ndez-Trincado}, Jos{\'e} G. and {Feuillet}, Diane and {Filipp}, Andreas and {Fillingham}, Sean P. and {Frinchaboy}, Peter M. and {Fromenteau}, Sebastien and {Galbany}, Llu{\'\i}s and {Garc{\'\i}a}, Rafael A. and {Garc{\'\i}a-Hern{\'a}ndez}, D.~A. and {Ge}, Junqiang and {Geisler}, Doug and {Gelfand}, Joseph and {G{\'e}ron}, Tobias and {Gibson}, Benjamin J. and {Goddy}, Julian and {Godoy-Rivera}, Diego and {Grabowski}, Kathleen and {Green}, Paul J. and {Greener}, Michael and {Grier}, Catherine J. and {Griffith}, Emily and {Guo}, Hong and {Guy}, Julien and {Hadjara}, Massinissa and {Harding}, Paul and {Hasselquist}, Sten and {Hayes}, Christian R. and {Hearty}, Fred and {Hern{\'a}ndez}, Jes{\'u}s and {Hill}, Lewis and {Hogg}, David W. and {Holtzman}, Jon A. and {Horta}, Danny and {Hsieh}, Bau-Ching and {Hsu}, Chin-Hao and {Hsu}, Yun-Hsin and {Huber}, Daniel and {Huertas-Company}, Marc and {Hutchinson}, Brian and {Hwang}, Ho Seong and {Ibarra-Medel}, H{\'e}ctor J. and {Chitham}, Jacob Ider and {Ilha}, Gabriele S. and {Imig}, Julie and {Jaekle}, Will and {Jayasinghe}, Tharindu and {Ji}, Xihan and {Johnson}, Jennifer A. and {Jones}, Amy and {J{\"o}nsson}, Henrik and {Katkov}, Ivan and {Khalatyan}, Arman, Dr. and {Kinemuchi}, Karen and {Kisku}, Shobhit and {Knapen}, Johan H. and {Kneib}, Jean-Paul and {Kollmeier}, Juna A. and {Kong}, Miranda and {Kounkel}, Marina and {Kreckel}, Kathryn and {Krishnarao}, Dhanesh and {Lacerna}, Ivan and {Lane}, Richard R. and {Langgin}, Rachel and {Lavender}, Ramon and {Law}, David R. and {Lazarz}, Daniel and {Leung}, Henry W. and {Leung}, Ho-Hin and {Lewis}, Hannah M. and {Li}, Cheng and {Li}, Ran and {Lian}, Jianhui and {Liang}, Fu-Heng and {Lin}, Lihwai and {Lin}, Yen-Ting and {Lin}, Sicheng and {Lintott}, Chris and {Long}, Dan and {Longa-Pe{\~n}a}, Pen{\'e}lope and {L{\'o}pez-Cob{\'a}}, Carlos and {Lu}, Shengdong and {Lundgren}, Britt F. and {Luo}, Yuanze and {Mackereth}, J. Ted and {de la Macorra}, Axel and {Mahadevan}, Suvrath and {Majewski}, Steven R. and {Manchado}, Arturo and {Mandeville}, Travis and {Maraston}, Claudia and {Margalef-Bentabol}, Berta and {Masseron}, Thomas and {Masters}, Karen L. and {Mathur}, Savita and {McDermid}, Richard M. and {Mckay}, Myles and {Merloni}, Andrea and {Merrifield}, Michael and {Meszaros}, Szabolcs and {Miglio}, Andrea and {Di Mille}, Francesco and {Minniti}, Dante and {Minsley}, Rebecca and {Monachesi}, Antonela and {Moon}, Jeongin and {Mosser}, Benoit and {Mulchaey}, John and {Muna}, Demitri and {Mu{\~n}oz}, Ricardo R. and {Myers}, Adam D. and {Myers}, Natalie and {Nadathur}, Seshadri and {Nair}, Preethi and {Nandra}, Kirpal and {Neumann}, Justus and {Newman}, Jeffrey A. and {Nidever}, David L. and {Nikakhtar}, Farnik and {Nitschelm}, Christian and {O'Connell}, Julia E. and {Garma-Oehmichen}, Luis and {Luan Souza de Oliveira}, Gabriel and {Olney}, Richard and {Oravetz}, Daniel and {Ortigoza-Urdaneta}, Mario and {Osorio}, Yeisson and {Otter}, Justin and {Pace}, Zachary J. and {Padilla}, Nelson and {Pan}, Kaike and {Pan}, Hsi-An and {Parikh}, Taniya and {Parker}, James and {Peirani}, Sebastien and {Pe{\~n}a Ram{\'\i}rez}, Karla and {Penny}, Samantha and {Percival}, Will J. and {Perez-Fournon}, Ismael and {Pinsonneault}, Marc and {Poidevin}, Fr{\'e}d{\'e}rick and {Poovelil}, Vijith Jacob and {Price-Whelan}, Adrian M. and {B{\'a}rbara de Andrade Queiroz}, Anna and {Raddick}, M. Jordan and {Ray}, Amy and {Rembold}, Sandro Barboza and {Riddle}, Nicole and {Riffel}, Rogemar A. and {Riffel}, Rog{\'e}rio and {Rix}, Hans-Walter and {Robin}, Annie C. and {Rodr{\'\i}guez-Puebla}, Aldo and {Roman-Lopes}, Alexandre and {Rom{\'a}n-Z{\'u}{\~n}iga}, Carlos and {Rose}, Benjamin and {Ross}, Ashley J. and {Rossi}, Graziano and {Rubin}, Kate H.~R. and {Salvato}, Mara and {S{\'a}nchez}, Seb{\'a}stian F. and {S{\'a}nchez-Gallego}, Jos{\'e} R. and {Sanderson}, Robyn and {Santana Rojas}, Felipe Antonio and {Sarceno}, Edgar and {Sarmiento}, Regina and {Sayres}, Conor and {Sazonova}, Elizaveta and {Schaefer}, Adam L. and {Schiavon}, Ricardo and {Schlegel}, David J. and {Schneider}, Donald P. and {Schultheis}, Mathias and {Schwope}, Axel and {Serenelli}, Aldo and {Serna}, Javier and {Shao}, Zhengyi and {Shapiro}, Griffin and {Sharma}, Anubhav and {Shen}, Yue and {Shetrone}, Matthew and {Shu}, Yiping and {Simon}, Joshua D. and {Skrutskie}, M.~F. and {Smethurst}, Rebecca and {Smith}, Verne and {Sobeck}, Jennifer and {Spoo}, Taylor and {Sprague}, Dani and {Stark}, David V. and {Stassun}, Keivan G. and {Steinmetz}, Matthias and {Stello}, Dennis and {Stone-Martinez}, Alexander and {Storchi-Bergmann}, Thaisa and {Stringfellow}, Guy S. and {Stutz}, Amelia and {Su}, Yung-Chau and {Taghizadeh-Popp}, Manuchehr and {Talbot}, Michael S. and {Tayar}, Jamie and {Telles}, Eduardo and {Teske}, Johanna and {Thakar}, Ani and {Theissen}, Christopher and {Tkachenko}, Andrew and {Thomas}, Daniel and {Tojeiro}, Rita and {Hernandez Toledo}, Hector and {Troup}, Nicholas W. and {Trump}, Jonathan R. and {Trussler}, James and {Turner}, Jacqueline and {Tuttle}, Sarah and {Unda-Sanzana}, Eduardo and {V{\'a}zquez-Mata}, Jos{\'e} Antonio and {Valentini}, Marica and {Valenzuela}, Octavio and {Vargas-Gonz{\'a}lez}, Jaime and {Vargas-Maga{\~n}a}, Mariana and {Alfaro}, Pablo Vera and {Villanova}, Sandro and {Vincenzo}, Fiorenzo and {Wake}, David and {Warfield}, Jack T. and {Washington}, Jessica Diane and {Weaver}, Benjamin Alan and {Weijmans}, Anne-Marie and {Weinberg}, David H. and {Weiss}, Achim and {Westfall}, Kyle B. and {Wild}, Vivienne and {Wilde}, Matthew C. and {Wilson}, John C. and {Wilson}, Robert F. and {Wilson}, Mikayla and {Wolf}, Julien and {Wood-Vasey}, W.~M. and {Yan}, Renbin and {Zamora}, Olga and {Zasowski}, Gail and {Zhang}, Kai and {Zhao}, Cheng and {Zheng}, Zheng and {Zheng}, Zheng and {Zhu}, Kai},
        title = "{The Seventeenth Data Release of the Sloan Digital Sky Surveys: Complete Release of MaNGA, MaStar, and APOGEE-2 Data}",
      journal = {\apjs},
         year = 2022,
        month = apr,
       volume = {259},
       number = {2},
          eid = {35},
        pages = {35},
          doi = {10.3847/1538-4365/ac4414},
archivePrefix = {arXiv},
       eprint = {2112.02026},
 primaryClass = {astro-ph.GA},
       adsurl = {https://ui.adsabs.harvard.edu/abs/2022ApJS..259...35A}
}

@INPROCEEDINGS{Tamura2016,
       author = {{Tamura}, Naoyuki and {Takato}, Naruhisa and {Shimono}, Atsushi and {Moritani}, Yuki and {Yabe}, Kiyoto and {Ishizuka}, Yuki and {Ueda}, Akitoshi and {Kamata}, Yukiko and {Aghazarian}, Hrand and {Arnouts}, St{\'e}phane and {Barban}, Gabriel and {Barkhouser}, Robert H. and {Borges}, Renato C. and {Braun}, David F. and {Carr}, Michael A. and {Chabaud}, Pierre-Yves and {Chang}, Yin-Chang and {Chen}, Hsin-Yo and {Chiba}, Masashi and {Chou}, Richard C.~Y. and {Chu}, You-Hua and {Cohen}, Judith and {de Almeida}, Rodrigo P. and {de Oliveira}, Antonio C. and {de Oliveira}, Ligia S. and {Dekany}, Richard G. and {Dohlen}, Kjetil and {dos Santos}, Jesulino B. and {dos Santos}, Leandro H. and {Ellis}, Richard and {Fabricius}, Maximilian and {Ferrand}, Didier and {Ferreira}, D{\'e}cio and {Golebiowski}, Mirek and {Greene}, Jenny E. and {Gross}, Johannes and {Gunn}, James E. and {Hammond}, Randolph and {Harding}, Albert and {Hart}, Murdock and {Heckman}, Timothy M. and {Hirata}, Christopher M. and {Ho}, Paul and {Hope}, Stephen C. and {Hovland}, Larry and {Hsu}, Shu-Fu and {Hu}, Yen-Shan and {Huang}, Ping-Jie and {Jaquet}, Marc and {Jing}, Yipeng and {Karr}, Jennifer and {Kimura}, Masahiko and {King}, Matthew E. and {Komatsu}, Eiichiro and {Le Brun}, Vincent and {Le F{\`e}vre}, Olivier and {Le Fur}, Arnaud and {Le Mignant}, David and {Ling}, Hung-Hsu and {Loomis}, Craig P. and {Lupton}, Robert H. and {Madec}, Fabrice and {Mao}, Peter and {Marrara}, Lucas S. and {Mendes de Oliveira}, Claudia and {Minowa}, Yosuke and {Morantz}, Chaz and {Murayama}, Hitoshi and {Murray}, Graham J. and {Ohyama}, Youichi and {Orndorff}, Joseph and {Pascal}, Sandrine and {Pereira}, Jefferson M. and {Reiley}, Daniel and {Reinecke}, Martin and {Ritter}, Andreas and {Roberts}, Mitsuko and {Schwochert}, Mark A. and {Seiffert}, Michael D. and {Smee}, Stephen A. and {Sodre}, Laerte and {Spergel}, David N. and {Steinkraus}, Aaron J. and {Strauss}, Michael A. and {Surace}, Christian and {Suto}, Yasushi and {Suzuki}, Nao and {Swinbank}, John and {Tait}, Philip J. and {Takada}, Masahiro and {Tamura}, Tomonori and {Tanaka}, Yoko and {Tresse}, Laurence and {Verducci}, Orlando and {Vibert}, Didier and {Vidal}, Clement and {Wang}, Shiang-Yu and {Wen}, Chih-Yi and {Yan}, Chi-Hung and {Yasuda}, Naoki},
        title = "{Prime Focus Spectrograph (PFS) for the Subaru telescope: overview, recent progress, and future perspectives}",
    booktitle = {Ground-based and Airborne Instrumentation for Astronomy VI},
         year = 2016,
       editor = {{Evans}, Christopher J. and {Simard}, Luc and {Takami}, Hideki},
       series = {Society of Photo-Optical Instrumentation Engineers (SPIE) Conference Series},
       volume = {9908},
        month = aug,
          eid = {99081M},
        pages = {99081M},
          doi = {10.1117/12.2232103},
archivePrefix = {arXiv},
       eprint = {1608.01075},
 primaryClass = {astro-ph.IM},
       adsurl = {https://ui.adsabs.harvard.edu/abs/2016SPIE.9908E..1MT}
}

@ARTICLE{sdss_dr19,
       author = {{SDSS Collaboration} and {Adamane Pallathadka}, Gautham and {Aghakhanloo}, Mojgan and {Aird}, James and {Almeida}, Andr{\'e}s and {Amrita}, Singh and {Anders}, Friedrich and {Anderson}, Scott F. and {Arseneau}, Stefan and {Gonz{\'a}lez Avila}, Consuelo and {Aviram}, Shir and {Aydar}, Catarina and {Badenes}, Carles and {Barrera-Ballesteros}, Jorge K. and {Bauer}, Franz E. and {Behmard}, Aida and {Berg}, Michelle and {Besser}, F. and {Moni Bidin}, Christian and {Bizyaev}, Dmitry and {Blanc}, Guillermo and {Blanton}, Michael R. and {Bovy}, Jo and {Brandt}, William Nielsen and {Brownstein}, Joel R. and {Buchner}, Johannes and {Bulbul}, Esra and {Burchett}, Joseph N. and {Carigi}, Leticia and {Carlberg}, Joleen K. and {Casey}, Andrew R. and {Chakraborty}, Priyanka and {Chanam{\'e}}, Julio and {Chandra}, Vedant and {Chiappini}, Cristina and {Chilingarian}, Igor and {Comparat}, Johan and {Covey}, Kevin and {Crumpler}, Nicole and {Cunha}, Katia and {D'Onghia}, Elena and {Dai}, Xinyu and {Darling}, Jeremy and {Davis}, Megan and {De Lee}, Nathan and {Deacon}, Niall and {M{\'e}ndez Delgado}, Jos{\'e} Eduardo and {Demasi}, Sebastian and {Demianenko}, Mariia and {Demke}, Delvin and {Donor}, John and {Drory}, Niv and {Villa Durango}, Monica Alejandra and {Dwelly}, Tom and {Egorov}, Oleg and {Egorova}, Evgeniya and {El-Badry}, Kareem and {Eracleous}, Mike and {Fan}, Xiaohui and {Farr}, Emily and {Finkbeiner}, Douglas P. and {Fries}, Logan and {Frinchaboy}, Peter and {Gentile Fusillo}, Nicola Pietro and {Serrano F{\'e}lix}, Luis Daniel and {Gaensicke}, Boris and {Galligan}, Emma and {Garc{\'\i}a}, Pablo and {Gelfand}, Joseph and {Grabowski}, Katie and {Grebel}, Eva and {Green}, Paul J and {Greve}, Hannah and {Grier}, Catherine and {Griffith}, Emily and {Guetzoyan}, Paloma and {Gupta}, Pramod and {Hackshaw}, Zoe and {Hall}, Patrick B. and {Hawkins}, Keith and {Heged{\H{u}}s}, Viola and {Hekker}, Saskia and {Herbst}, T.~M. and {Hermes}, J.~J. and {Hern{\'a}ndez-Garc{\'\i}a}, Lorena and {Hiremath}, Pranavi and {Hogg}, David W and {Holtzman}, Jon and {Horne}, Keith and {Horta}, Danny and {Huang}, Yang and {Hutchinson}, Brian and {H{\"a}berle}, Maximilian and {Ibarra-Medel}, Hector Javier and {Ji}, Alexander P. and {Jofre}, Paula and {Johnson}, James W. and {Johnson}, Jennifer and {Johnston}, Evelyn J. and {Kaldor}, Mary and {Katkov}, Ivan and {Khalatyan}, Arman and {Khoperskov}, Sergey and {Klessen}, Ralf and {Kluge}, Matthias and {Koekemoer}, Anton M. and {Kollmeier}, Juna A. and {Kounkel}, Marina and {Kreckel}, Kathryn and {Krishnarao}, Dhanesh and {Krumpe}, Mirko and {Lacerna}, Ivan and {Laporte}, Chervin and {Lepine}, Sebastien and {Li}, Jing and {Liang}, Fu-Heng and {Limberg}, Guilherme and {Liu}, Xin and {Loebman}, Sarah and {Long}, Knox and {Lu}, Yuxi and {Lucey}, Madeline and {Lugo-Aranda}, Alejandra Z. and {Mart{\'\i}nez Martinez-Aldama}, Mary Loli and {McKinnon}, Kevin and {Medan}, Ilija and {Merloni}, Andrea and {Morrison}, Sean and {Myers}, Natalie and {M{\'e}sz{\'a}ros}, Szabolcs and {M{\"u}ller-Horn}, Johanna and {Nepal}, Samir and {Ness}, Melissa and {Nidever}, David and {Nitschelm}, Christian and {Oravetz}, Audrey and {Otto}, Jonah and {Pan}, Kaike and {P{\'e}rez Paolino}, Facundo and {Negrete Pe{\~n}aloza}, Castalia Alenka and {Pinsonneault}, Marc and {Taghizadeh Popp}, Manuchehr and {Price-Whelan}, Adrian and {Pulatova}, Nadiia and {Queiroz}, Anna Barbara and {Raddick}, Jordan and {Rankine}, Amy and {Rix}, Hans-Walter and {Rom{\'a}n-Z{\'u}{\~n}iga}, Carlos and {Fern{\'a}ndez Rosso}, Daniela and {Runnoe}, Jessie and {Mahmud Saad}, Serat and {Salvato}, Mara and {Sanchez}, Sebastian F. and {Sattler}, Natascha and {Saydjari}, Andrew and {Sayres}, Conor and {Schlaufman}, Kevin and {Schneider}, Donald P. and {Schwope}, Axel and {Seaton}, Lucas M. and {Seeburger}, Rhys and {Serna}, Javier and {Sharma}, Sanjib and {Shen}, Yue and {Sinha}, Amaya and {Sizemore}, Brian and {Sniegowska}, Marzena and {Song}, Yingyi and {Souto}, Diogo and {Stassun}, Keivan and {Steinmetz}, Matthias and {Stone}, Zachary and {Stone-Martinez}, Alexander and {Stringfellow}, Guy S. and {Mata S{\'a}nchez}, Aurora and {S{\'a}nchez-Gallego}, Jos{\'e} and {Tan}, Jonathan and {Tayar}, Jamie and {Thai}, Riley and {Thakar}, Ani and {Thibodeaux}, Pierre and {Ting}, Yuan-Sen and {Tkachenko}, Andrew and {Trakhtenbrot}, Benny and {Fernandez Trincado}, Jose G. and {Troup}, Nicholas and {Trump}, Jonathan R. and {Ulloa}, Natalie and {Van der Marel}, Roeland P. and {Vera}, Pablo and {Villanova}, Sandro and {Villase{\~n}or}, Jaime and {Wang}, Ji and {Way}, Zachary and {Weijmans}, Anne-Marie and {Wheeler}, Adam and {Wilson}, John C. and {Wofford}, Aida and {Wong}, Tony},
        title = "{The Nineteenth Data Release of the Sloan Digital Sky Survey}",
      journal = {AASJournals},
         year = 2025,
        month = jul,
          eid = {arXiv:2507.07093},
        pages = {submitted},
          doi = {10.48550/arXiv.2507.07093},
archivePrefix = {arXiv},
       eprint = {2507.07093},
 primaryClass = {astro-ph.GA},
       adsurl = {https://ui.adsabs.harvard.edu/abs/2025arXiv250707093S}
}

@ARTICLE{Predehl2021,
       author = {{Predehl}, P. and {Andritschke}, R. and {Arefiev}, V. and {Babyshkin}, V. and {Batanov}, O. and {Becker}, W. and {B{\"o}hringer}, H. and {Bogomolov}, A. and {Boller}, T. and {Borm}, K. and {Bornemann}, W. and {Br{\"a}uninger}, H. and {Br{\"u}ggen}, M. and {Brunner}, H. and {Brusa}, M. and {Bulbul}, E. and {Buntov}, M. and {Burwitz}, V. and {Burkert}, W. and {Clerc}, N. and {Churazov}, E. and {Coutinho}, D. and {Dauser}, T. and {Dennerl}, K. and {Doroshenko}, V. and {Eder}, J. and {Emberger}, V. and {Eraerds}, T. and {Finoguenov}, A. and {Freyberg}, M. and {Friedrich}, P. and {Friedrich}, S. and {F{\"u}rmetz}, M. and {Georgakakis}, A. and {Gilfanov}, M. and {Granato}, S. and {Grossberger}, C. and {Gueguen}, A. and {Gureev}, P. and {Haberl}, F. and {H{\"a}lker}, O. and {Hartner}, G. and {Hasinger}, G. and {Huber}, H. and {Ji}, L. and {Kienlin}, A. v. and {Kink}, W. and {Korotkov}, F. and {Kreykenbohm}, I. and {Lamer}, G. and {Lomakin}, I. and {Lapshov}, I. and {Liu}, T. and {Maitra}, C. and {Meidinger}, N. and {Menz}, B. and {Merloni}, A. and {Mernik}, T. and {Mican}, B. and {Mohr}, J. and {M{\"u}ller}, S. and {Nandra}, K. and {Nazarov}, V. and {Pacaud}, F. and {Pavlinsky}, M. and {Perinati}, E. and {Pfeffermann}, E. and {Pietschner}, D. and {Ramos-Ceja}, M.~E. and {Rau}, A. and {Reiffers}, J. and {Reiprich}, T.~H. and {Robrade}, J. and {Salvato}, M. and {Sanders}, J. and {Santangelo}, A. and {Sasaki}, M. and {Scheuerle}, H. and {Schmid}, C. and {Schmitt}, J. and {Schwope}, A. and {Shirshakov}, A. and {Steinmetz}, M. and {Stewart}, I. and {Str{\"u}der}, L. and {Sunyaev}, R. and {Tenzer}, C. and {Tiedemann}, L. and {Tr{\"u}mper}, J. and {Voron}, V. and {Weber}, P. and {Wilms}, J. and {Yaroshenko}, V.},
        title = "{The eROSITA X-ray telescope on SRG}",
      journal = {\aap},
         year = 2021,
        month = mar,
       volume = {647},
          eid = {A1},
        pages = {A1},
          doi = {10.1051/0004-6361/202039313},
archivePrefix = {arXiv},
       eprint = {2010.03477},
 primaryClass = {astro-ph.HE},
       adsurl = {https://ui.adsabs.harvard.edu/abs/2021A&A...647A...1P}
}

@ARTICLE{Merloni2024,
       author = {{Merloni}, A. and {Lamer}, G. and {Liu}, T. and {Ramos-Ceja}, M.~E. and {Brunner}, H. and {Bulbul}, E. and {Dennerl}, K. and {Doroshenko}, V. and {Freyberg}, M.~J. and {Friedrich}, S. and {Gatuzz}, E. and {Georgakakis}, A. and {Haberl}, F. and {Igo}, Z. and {Kreykenbohm}, I. and {Liu}, A. and {Maitra}, C. and {Malyali}, A. and {Mayer}, M.~G.~F. and {Nandra}, K. and {Predehl}, P. and {Robrade}, J. and {Salvato}, M. and {Sanders}, J.~S. and {Stewart}, I. and {Tub{\'\i}n-Arenas}, D. and {Weber}, P. and {Wilms}, J. and {Arcodia}, R. and {Artis}, E. and {Aschersleben}, J. and {Avakyan}, A. and {Aydar}, C. and {Bahar}, Y.~E. and {Balzer}, F. and {Becker}, W. and {Berger}, K. and {Boller}, T. and {Bornemann}, W. and {Br{\"u}ggen}, M. and {Brusa}, M. and {Buchner}, J. and {Burwitz}, V. and {Camilloni}, F. and {Clerc}, N. and {Comparat}, J. and {Coutinho}, D. and {Czesla}, S. and {Dannhauer}, S.~M. and {Dauner}, L. and {Dauser}, T. and {Dietl}, J. and {Dolag}, K. and {Dwelly}, T. and {Egg}, K. and {Ehl}, E. and {Freund}, S. and {Friedrich}, P. and {Gaida}, R. and {Garrel}, C. and {Ghirardini}, V. and {Gokus}, A. and {Gr{\"u}nwald}, G. and {Grandis}, S. and {Grotova}, I. and {Gruen}, D. and {Gueguen}, A. and {H{\"a}mmerich}, S. and {Hamaus}, N. and {Hasinger}, G. and {Haubner}, K. and {Homan}, D. and {Ider Chitham}, J. and {Joseph}, W.~M. and {Joyce}, A. and {K{\"o}nig}, O. and {Kaltenbrunner}, D.~M. and {Khokhriakova}, A. and {Kink}, W. and {Kirsch}, C. and {Kluge}, M. and {Knies}, J. and {Krippendorf}, S. and {Krumpe}, M. and {Kurpas}, J. and {Li}, P. and {Liu}, Z. and {Locatelli}, N. and {Lorenz}, M. and {M{\"u}ller}, S. and {Magaudda}, E. and {Mannes}, C. and {McCall}, H. and {Meidinger}, N. and {Michailidis}, M. and {Migkas}, K. and {Mu{\~n}oz-Giraldo}, D. and {Musiimenta}, B. and {Nguyen-Dang}, N.~T. and {Ni}, Q. and {Olechowska}, A. and {Ota}, N. and {Pacaud}, F. and {Pasini}, T. and {Perinati}, E. and {Pires}, A.~M. and {Pommranz}, C. and {Ponti}, G. and {Poppenhaeger}, K. and {P{\"u}hlhofer}, G. and {Rau}, A. and {Reh}, M. and {Reiprich}, T.~H. and {Roster}, W. and {Saeedi}, S. and {Santangelo}, A. and {Sasaki}, M. and {Schmitt}, J. and {Schneider}, P.~C. and {Schrabback}, T. and {Schuster}, N. and {Schwope}, A. and {Seppi}, R. and {Serim}, M.~M. and {Shreeram}, S. and {Sokolova-Lapa}, E. and {Starck}, H. and {Stelzer}, B. and {Stierhof}, J. and {Suleimanov}, V. and {Tenzer}, C. and {Traulsen}, I. and {Tr{\"u}mper}, J. and {Tsuge}, K. and {Urrutia}, T. and {Veronica}, A. and {Waddell}, S.~G.~H. and {Willer}, R. and {Wolf}, J. and {Yeung}, M.~C.~H. and {Zainab}, A. and {Zangrandi}, F. and {Zhang}, X. and {Zhang}, Y. and {Zheng}, X.},
        title = "{The SRG/eROSITA all-sky survey. First X-ray catalogues and data release of the western Galactic hemisphere}",
      journal = {\aap},
         year = 2024,
        month = feb,
       volume = {682},
          eid = {A34},
        pages = {A34},
          doi = {10.1051/0004-6361/202347165},
archivePrefix = {arXiv},
       eprint = {2401.17274},
 primaryClass = {astro-ph.HE},
       adsurl = {https://ui.adsabs.harvard.edu/abs/2024A&A...682A..34M}
}

@ARTICLE{Gunn2006,
       author = {{Gunn}, James E. and {Siegmund}, Walter A. and {Mannery}, Edward J. and {Owen}, Russell E. and {Hull}, Charles L. and {Leger}, R. French and {Carey}, Larry N. and {Knapp}, Gillian R. and {York}, Donald G. and {Boroski}, William N. and {Kent}, Stephen M. and {Lupton}, Robert H. and {Rockosi}, Constance M. and {Evans}, Michael L. and {Waddell}, Patrick and {Anderson}, John E. and {Annis}, James and {Barentine}, John C. and {Bartoszek}, Larry M. and {Bastian}, Steven and {Bracker}, Stephen B. and {Brewington}, Howard J. and {Briegel}, Charles I. and {Brinkmann}, Jon and {Brown}, Yorke J. and {Carr}, Michael A. and {Czarapata}, Paul C. and {Drennan}, Craig C. and {Dombeck}, Thomas and {Federwitz}, Glenn R. and {Gillespie}, Bruce A. and {Gonzales}, Carlos and {Hansen}, Sten U. and {Harvanek}, Michael and {Hayes}, Jeffrey and {Jordan}, Wendell and {Kinney}, Ellyne and {Klaene}, Mark and {Kleinman}, S.~J. and {Kron}, Richard G. and {Kresinski}, Jurek and {Lee}, Glenn and {Limmongkol}, Siriluk and {Lindenmeyer}, Carl W. and {Long}, Daniel C. and {Loomis}, Craig L. and {McGehee}, Peregrine M. and {Mantsch}, Paul M. and {Neilsen}, Jr., Eric H. and {Neswold}, Richard M. and {Newman}, Peter R. and {Nitta}, Atsuko and {Peoples}, Jr., John and {Pier}, Jeffrey R. and {Prieto}, Peter S. and {Prosapio}, Angela and {Rivetta}, Claudio and {Schneider}, Donald P. and {Snedden}, Stephanie and {Wang}, Shu-i.},
        title = "{The 2.5 m Telescope of the Sloan Digital Sky Survey}",
      journal = {\aj},
         year = 2006,
        month = apr,
       volume = {131},
       number = {4},
        pages = {2332-2359},
          doi = {10.1086/500975},
archivePrefix = {arXiv},
       eprint = {astro-ph/0602326},
 primaryClass = {astro-ph},
       adsurl = {https://ui.adsabs.harvard.edu/abs/2006AJ....131.2332G}
}

@ARTICLE{Bowen_Vaughan1973,
       author = {{Bowen}, I.~S. and {Vaughan}, Jr., A.~H.},
        title = "{The optical design of the 40-in. telescope and of the Ir{\'e}n{\'e}e DuPont telescope at Las Campanas Observatory, Chile.}",
      journal = {\ao},
         year = 1973,
        month = jan,
       volume = {12},
        pages = {1430-1434},
          doi = {10.1364/AO.12.001430},
       adsurl = {https://ui.adsabs.harvard.edu/abs/1973ApOpt..12.1430B}
}

@ARTICLE{Smee2013,
       author = {{Smee}, Stephen A. and {Gunn}, James E. and {Uomoto}, Alan and {Roe}, Natalie and {Schlegel}, David and {Rockosi}, Constance M. and {Carr}, Michael A. and {Leger}, French and {Dawson}, Kyle S. and {Olmstead}, Matthew D. and {Brinkmann}, Jon and {Owen}, Russell and {Barkhouser}, Robert H. and {Honscheid}, Klaus and {Harding}, Paul and {Long}, Dan and {Lupton}, Robert H. and {Loomis}, Craig and {Anderson}, Lauren and {Annis}, James and {Bernardi}, Mariangela and {Bhardwaj}, Vaishali and {Bizyaev}, Dmitry and {Bolton}, Adam S. and {Brewington}, Howard and {Briggs}, John W. and {Burles}, Scott and {Burns}, James G. and {Castander}, Francisco Javier and {Connolly}, Andrew and {Davenport}, James R.~A. and {Ebelke}, Garrett and {Epps}, Harland and {Feldman}, Paul D. and {Friedman}, Scott D. and {Frieman}, Joshua and {Heckman}, Timothy and {Hull}, Charles L. and {Knapp}, Gillian R. and {Lawrence}, David M. and {Loveday}, Jon and {Mannery}, Edward J. and {Malanushenko}, Elena and {Malanushenko}, Viktor and {Merrelli}, Aronne James and {Muna}, Demitri and {Newman}, Peter R. and {Nichol}, Robert C. and {Oravetz}, Daniel and {Pan}, Kaike and {Pope}, Adrian C. and {Ricketts}, Paul G. and {Shelden}, Alaina and {Sandford}, Dale and {Siegmund}, Walter and {Simmons}, Audrey and {Smith}, D. Shane and {Snedden}, Stephanie and {Schneider}, Donald P. and {SubbaRao}, Mark and {Tremonti}, Christy and {Waddell}, Patrick and {York}, Donald G.},
        title = "{The Multi-object, Fiber-fed Spectrographs for the Sloan Digital Sky Survey and the Baryon Oscillation Spectroscopic Survey}",
      journal = {\aj},
         year = 2013,
        month = aug,
       volume = {146},
       number = {2},
          eid = {32},
        pages = {32},
          doi = {10.1088/0004-6256/146/2/32},
archivePrefix = {arXiv},
       eprint = {1208.2233},
 primaryClass = {astro-ph.IM},
       adsurl = {https://ui.adsabs.harvard.edu/abs/2013AJ....146...32S}
}

@ARTICLE{Zenteno2025,
       author = {{Zenteno}, A. and {Kluge}, M. and {Kharkrang}, R. and {Hernandez-Lang}, D. and {Damke}, G. and {Saro}, A. and {Monteiro-Oliveira}, R. and {Carrasco}, E.~R. and {Salvato}, M. and {Comparat}, J. and {Fabricius}, M. and {Snigula}, J. and {Arevalo}, P. and {Cuevas}, H. and {Nilo Castellon}, J.~L. and {Ramirez}, A. and {V{\'e}liz Astudillo}, S. and {Landriau}, M. and {Myers}, A.~D. and {Schlafly}, E. and {Valdes}, F. and {Weaver}, B.~A. and {Mohr}, J.~J. and {Grandis}, S. and {Klein}, M. and {Liu}, A. and {Bulbul}, E. and {Zhang}, X. and {Sanders}, J.~S. and {Bahar}, Y.~E. and {Ghirardini}, V. and {Ramos}, M.~E. and {Balzer}, F.},
        title = "{The dynamical state of eROSITA clusters and its impact on the brightest cluster galaxy luminosity}",
      journal = {\aap},
         year = 2025,
        month = jun,
       volume = {698},
          eid = {A171},
        pages = {A171},
          doi = {10.1051/0004-6361/202452440},
archivePrefix = {arXiv},
       eprint = {2503.21066},
 primaryClass = {astro-ph.CO},
       adsurl = {https://ui.adsabs.harvard.edu/abs/2025A&A...698A.171Z}
}

@ARTICLE{Bolton2012,
       author = {{Bolton}, Adam S. and {Schlegel}, David J. and {Aubourg}, {\'E}ric and {Bailey}, Stephen and {Bhardwaj}, Vaishali and {Brownstein}, Joel R. and {Burles}, Scott and {Chen}, Yan-Mei and {Dawson}, Kyle and {Eisenstein}, Daniel J. and {Gunn}, James E. and {Knapp}, G.~R. and {Loomis}, Craig P. and {Lupton}, Robert H. and {Maraston}, Claudia and {Muna}, Demitri and {Myers}, Adam D. and {Olmstead}, Matthew D. and {Padmanabhan}, Nikhil and {P{\^a}ris}, Isabelle and {Percival}, Will J. and {Petitjean}, Patrick and {Rockosi}, Constance M. and {Ross}, Nicholas P. and {Schneider}, Donald P. and {Shu}, Yiping and {Strauss}, Michael A. and {Thomas}, Daniel and {Tremonti}, Christy A. and {Wake}, David A. and {Weaver}, Benjamin A. and {Wood-Vasey}, W. Michael},
        title = "{Spectral Classification and Redshift Measurement for the SDSS-III Baryon Oscillation Spectroscopic Survey}",
      journal = {\aj},
         year = 2012,
        month = nov,
       volume = {144},
       number = {5},
          eid = {144},
        pages = {144},
          doi = {10.1088/0004-6256/144/5/144},
archivePrefix = {arXiv},
       eprint = {1207.7326},
 primaryClass = {astro-ph.CO},
       adsurl = {https://ui.adsabs.harvard.edu/abs/2012AJ....144..144B}
}

@ARTICLE{Graham2001,
       author = {{Graham}, Alister W. and {Erwin}, Peter and {Caon}, N. and {Trujillo}, I.},
        title = "{A Correlation between Galaxy Light Concentration and Supermassive Black Hole Mass}",
      journal = {\apjl},
         year = 2001,
        month = dec,
       volume = {563},
       number = {1},
        pages = {L11-L14},
          doi = {10.1086/338500},
archivePrefix = {arXiv},
       eprint = {astro-ph/0111152},
 primaryClass = {astro-ph},
       adsurl = {https://ui.adsabs.harvard.edu/abs/2001ApJ...563L..11G}
}

@ARTICLE{Magorrian1998,
       author = {{Magorrian}, John and {Tremaine}, Scott and {Richstone}, Douglas and {Bender}, Ralf and {Bower}, Gary and {Dressler}, Alan and {Faber}, S.~M. and {Gebhardt}, Karl and {Green}, Richard and {Grillmair}, Carl and {Kormendy}, John and {Lauer}, Tod},
        title = "{The Demography of Massive Dark Objects in Galaxy Centers}",
      journal = {\aj},
         year = 1998,
        month = jun,
       volume = {115},
       number = {6},
        pages = {2285-2305},
          doi = {10.1086/300353},
archivePrefix = {arXiv},
       eprint = {astro-ph/9708072},
 primaryClass = {astro-ph},
       adsurl = {https://ui.adsabs.harvard.edu/abs/1998AJ....115.2285M}
}

@ARTICLE{Bernal2025,
       author = {{Bernal}, S. and {S{\'a}nchez-S{\'a}ez}, P. and {Ar{\'e}valo}, P. and {Bauer}, F.~E. and {Lira}, P. and {Sotomayor}, B.},
        title = "{The success of optical variability in uncovering active galactic nuclei in low stellar mass galaxies}",
      journal = {\aap},
         year = 2025,
        month = feb,
       volume = {694},
          eid = {A127},
        pages = {A127},
          doi = {10.1051/0004-6361/202451870},
archivePrefix = {arXiv},
       eprint = {2412.14298},
 primaryClass = {astro-ph.GA},
       adsurl = {https://ui.adsabs.harvard.edu/abs/2025A&A...694A.127B}
}

@ARTICLE{Runnoe2012_erratum,
       author = {{Runnoe}, Jessie C. and {Brotherton}, Michael S. and {Shang}, Zhaohui},
        title = "{Erratum: Updating quasar bolometric luminosity corrections}",
      journal = {\mnras},
         year = 2012,
        month = dec,
       volume = {427},
       number = {2},
        pages = {1800-1800},
          doi = {10.1111/j.1365-2966.2012.21878.x},
       adsurl = {https://ui.adsabs.harvard.edu/abs/2012MNRAS.427.1800R}
}

@ARTICLE{Buchner2024,
       author = {{Buchner}, Johannes and {Starck}, Hattie and {Salvato}, Mara and {Netzer}, Hagai and {Igo}, Zsofi and {Laloux}, Brivael and {Georgakakis}, Antonis and {Gauger}, Isabelle and {Olechowska}, Anna and {Lopez}, Nicolas and {Shankar}, Suraj D. and {Li}, Junyao and {Nandra}, Kirpal and {Merloni}, Andrea},
        title = "{Genuine Retrieval of the AGN Host Stellar Population (GRAHSP)}",
      journal = {\aap},
         year = 2024,
        month = dec,
       volume = {692},
          eid = {A161},
        pages = {A161},
          doi = {10.1051/0004-6361/202449372},
archivePrefix = {arXiv},
       eprint = {2405.19297},
 primaryClass = {astro-ph.GA},
       adsurl = {https://ui.adsabs.harvard.edu/abs/2024A&A...692A.161B}
}

@ARTICLE{Bongiorno2012,
       author = {{Bongiorno}, A. and {Merloni}, A. and {Brusa}, M. and {Magnelli}, B. and {Salvato}, M. and {Mignoli}, M. and {Zamorani}, G. and {Fiore}, F. and {Rosario}, D. and {Mainieri}, V. and {Hao}, H. and {Comastri}, A. and {Vignali}, C. and {Balestra}, I. and {Bardelli}, S. and {Berta}, S. and {Civano}, F. and {Kampczyk}, P. and {Le Floc'h}, E. and {Lusso}, E. and {Lutz}, D. and {Pozzetti}, L. and {Pozzi}, F. and {Riguccini}, L. and {Shankar}, F. and {Silverman}, J.},
        title = "{Accreting supermassive black holes in the COSMOS field and the connection to their host galaxies}",
      journal = {\mnras},
         year = 2012,
        month = dec,
       volume = {427},
       number = {4},
        pages = {3103-3133},
          doi = {10.1111/j.1365-2966.2012.22089.x},
archivePrefix = {arXiv},
       eprint = {1209.1640},
 primaryClass = {astro-ph.CO},
       adsurl = {https://ui.adsabs.harvard.edu/abs/2012MNRAS.427.3103B}
}

@ARTICLE{Koss2017,
       author = {{Koss}, Michael and {Trakhtenbrot}, Benny and {Ricci}, Claudio and {Lamperti}, Isabella and {Oh}, Kyuseok and {Berney}, Simon and {Schawinski}, Kevin and {Balokovi{\'c}}, Mislav and {Baronchelli}, Linda and {Crenshaw}, D. Michael and {Fischer}, Travis and {Gehrels}, Neil and {Harrison}, Fiona and {Hashimoto}, Yasuhiro and {Hogg}, Drew and {Ichikawa}, Kohei and {Masetti}, Nicola and {Mushotzky}, Richard and {Sartori}, Lia and {Stern}, Daniel and {Treister}, Ezequiel and {Ueda}, Yoshihiro and {Veilleux}, Sylvain and {Winter}, Lisa},
        title = "{BAT AGN Spectroscopic Survey. I. Spectral Measurements, Derived Quantities, and AGN Demographics}",
      journal = {\apj},
         year = 2017,
        month = nov,
       volume = {850},
       number = {1},
          eid = {74},
        pages = {74},
          doi = {10.3847/1538-4357/aa8ec9},
archivePrefix = {arXiv},
       eprint = {1707.08123},
 primaryClass = {astro-ph.HE},
       adsurl = {https://ui.adsabs.harvard.edu/abs/2017ApJ...850...74K}
}

@ARTICLE{BarquinGonzalez2024,
       author = {{Barqu{\'\i}n-Gonz{\'a}lez}, L. and {Mateos}, S. and {Carrera}, F.~J. and {Ordov{\'a}s-Pascual}, I. and {Alonso-Herrero}, A. and {Caccianiga}, A. and {Cardiel}, N. and {Corral}, A. and {Dom{\'\i}nguez}, R.~M. and {Garc{\'\i}a-Bernete}, I. and {Mountrichas}, G. and {Severgnini}, P.},
        title = "{Extinction and AGN over host galaxy contrast effects on the optical spectroscopic classification of AGN}",
      journal = {\aap},
         year = 2024,
        month = jul,
       volume = {687},
          eid = {A159},
        pages = {A159},
          doi = {10.1051/0004-6361/202348948},
archivePrefix = {arXiv},
       eprint = {2404.19544},
 primaryClass = {astro-ph.GA},
       adsurl = {https://ui.adsabs.harvard.edu/abs/2024A&A...687A.159B}
}

@ARTICLE{Santini2012,
       author = {{Santini}, P. and {Rosario}, D.~J. and {Shao}, L. and {Lutz}, D. and {Maiolino}, R. and {Alexander}, D.~M. and {Altieri}, B. and {Andreani}, P. and {Aussel}, H. and {Bauer}, F.~E. and {Berta}, S. and {Bongiovanni}, A. and {Brandt}, W.~N. and {Brusa}, M. and {Cepa}, J. and {Cimatti}, A. and {Daddi}, E. and {Elbaz}, D. and {Fontana}, A. and {F{\"o}rster Schreiber}, N.~M. and {Genzel}, R. and {Grazian}, A. and {Le Floc'h}, E. and {Magnelli}, B. and {Mainieri}, V. and {Nordon}, R. and {P{\'e}rez Garcia}, A.~M. and {Poglitsch}, A. and {Popesso}, P. and {Pozzi}, F. and {Riguccini}, L. and {Rodighiero}, G. and {Salvato}, M. and {Sanchez-Portal}, M. and {Sturm}, E. and {Tacconi}, L.~J. and {Valtchanov}, I. and {Wuyts}, S.},
        title = "{Enhanced star formation rates in AGN hosts with respect to inactive galaxies from PEP-Herschel observations}",
      journal = {\aap},
         year = 2012,
        month = apr,
       volume = {540},
          eid = {A109},
        pages = {A109},
          doi = {10.1051/0004-6361/201118266},
archivePrefix = {arXiv},
       eprint = {1201.4394},
 primaryClass = {astro-ph.CO},
       adsurl = {https://ui.adsabs.harvard.edu/abs/2012A&A...540A.109S}
}

@ARTICLE{Mobasher2015,
       author = {{Mobasher}, Bahram and {Dahlen}, Tomas and {Ferguson}, Henry C. and {Acquaviva}, Viviana and {Barro}, Guillermo and {Finkelstein}, Steven L. and {Fontana}, Adriano and {Gruetzbauch}, Ruth and {Johnson}, Seth and {Lu}, Yu and {Papovich}, Casey J. and {Pforr}, Janine and {Salvato}, Mara and {Somerville}, Rachel S. and {Wiklind}, Tommy and {Wuyts}, Stijn and {Ashby}, Matthew L.~N. and {Bell}, Eric and {Conselice}, Christopher J. and {Dickinson}, Mark E. and {Faber}, Sandra M. and {Fazio}, Giovanni and {Finlator}, Kristian and {Galametz}, Audrey and {Gawiser}, Eric and {Giavalisco}, Mauro and {Grazian}, Andrea and {Grogin}, Norman A. and {Guo}, Yicheng and {Hathi}, Nimish and {Kocevski}, Dale and {Koekemoer}, Anton M. and {Koo}, David C. and {Newman}, Jeffrey A. and {Reddy}, Naveen and {Santini}, Paola and {Wechsler}, Risa H.},
        title = "{A Critical Assessment of Stellar Mass Measurement Methods}",
      journal = {\apj},
         year = 2015,
        month = jul,
       volume = {808},
       number = {1},
          eid = {101},
        pages = {101},
          doi = {10.1088/0004-637X/808/1/101},
archivePrefix = {arXiv},
       eprint = {1505.01501},
 primaryClass = {astro-ph.GA},
       adsurl = {https://ui.adsabs.harvard.edu/abs/2015ApJ...808..101M}
}

@ARTICLE{ForsterSchreiber2014,
       author = {{F{\"o}rster Schreiber}, N.~M. and {Genzel}, R. and {Newman}, S.~F. and {Kurk}, J.~D. and {Lutz}, D. and {Tacconi}, L.~J. and {Wuyts}, S. and {Bandara}, K. and {Burkert}, A. and {Buschkamp}, P. and {Carollo}, C.~M. and {Cresci}, G. and {Daddi}, E. and {Davies}, R. and {Eisenhauer}, F. and {Hicks}, E.~K.~S. and {Lang}, P. and {Lilly}, S.~J. and {Mainieri}, V. and {Mancini}, C. and {Naab}, T. and {Peng}, Y. and {Renzini}, A. and {Rosario}, D. and {Shapiro Griffin}, K. and {Shapley}, A.~E. and {Sternberg}, A. and {Tacchella}, S. and {Vergani}, D. and {Wisnioski}, E. and {Wuyts}, E. and {Zamorani}, G.},
        title = "{The Sins/zC-Sinf Survey of z \raisebox{-0.5ex}\textasciitilde 2 Galaxy Kinematics: Evidence for Powerful Active Galactic Nucleus-Driven Nuclear Outflows in Massive Star-Forming Galaxies}",
      journal = {\apj},
         year = 2014,
        month = may,
       volume = {787},
       number = {1},
          eid = {38},
        pages = {38},
          doi = {10.1088/0004-637X/787/1/38},
archivePrefix = {arXiv},
       eprint = {1311.2596},
 primaryClass = {astro-ph.CO},
       adsurl = {https://ui.adsabs.harvard.edu/abs/2014ApJ...787...38F}
}

@ARTICLE{Dey2019,
       author = {{Dey}, Arjun and {Schlegel}, David J. and {Lang}, Dustin and {Blum}, Robert and {Burleigh}, Kaylan and {Fan}, Xiaohui and {Findlay}, Joseph R. and {Finkbeiner}, Doug and {Herrera}, David and {Juneau}, St{\'e}phanie and {Landriau}, Martin and {Levi}, Michael and {McGreer}, Ian and {Meisner}, Aaron and {Myers}, Adam D. and {Moustakas}, John and {Nugent}, Peter and {Patej}, Anna and {Schlafly}, Edward F. and {Walker}, Alistair R. and {Valdes}, Francisco and {Weaver}, Benjamin A. and {Y{\`e}che}, Christophe and {Zou}, Hu and {Zhou}, Xu and {Abareshi}, Behzad and {Abbott}, T.~M.~C. and {Abolfathi}, Bela and {Aguilera}, C. and {Alam}, Shadab and {Allen}, Lori and {Alvarez}, A. and {Annis}, James and {Ansarinejad}, Behzad and {Aubert}, Marie and {Beechert}, Jacqueline and {Bell}, Eric F. and {BenZvi}, Segev Y. and {Beutler}, Florian and {Bielby}, Richard M. and {Bolton}, Adam S. and {Brice{\~n}o}, C{\'e}sar and {Buckley-Geer}, Elizabeth J. and {Butler}, Karen and {Calamida}, Annalisa and {Carlberg}, Raymond G. and {Carter}, Paul and {Casas}, Ricard and {Castander}, Francisco J. and {Choi}, Yumi and {Comparat}, Johan and {Cukanovaite}, Elena and {Delubac}, Timoth{\'e}e and {DeVries}, Kaitlin and {Dey}, Sharmila and {Dhungana}, Govinda and {Dickinson}, Mark and {Ding}, Zhejie and {Donaldson}, John B. and {Duan}, Yutong and {Duckworth}, Christopher J. and {Eftekharzadeh}, Sarah and {Eisenstein}, Daniel J. and {Etourneau}, Thomas and {Fagrelius}, Parker A. and {Farihi}, Jay and {Fitzpatrick}, Mike and {Font-Ribera}, Andreu and {Fulmer}, Leah and {G{\"a}nsicke}, Boris T. and {Gaztanaga}, Enrique and {George}, Koshy and {Gerdes}, David W. and {Gontcho}, Satya Gontcho A. and {Gorgoni}, Claudio and {Green}, Gregory and {Guy}, Julien and {Harmer}, Diane and {Hernandez}, M. and {Honscheid}, Klaus and {Huang}, Lijuan Wendy and {James}, David J. and {Jannuzi}, Buell T. and {Jiang}, Linhua and {Joyce}, Richard and {Karcher}, Armin and {Karkar}, Sonia and {Kehoe}, Robert and {Kneib}, Jean-Paul and {Kueter-Young}, Andrea and {Lan}, Ting-Wen and {Lauer}, Tod R. and {Le Guillou}, Laurent and {Le Van Suu}, Auguste and {Lee}, Jae Hyeon and {Lesser}, Michael and {Perreault Levasseur}, Laurence and {Li}, Ting S. and {Mann}, Justin L. and {Marshall}, Robert and {Mart{\'\i}nez-V{\'a}zquez}, C.~E. and {Martini}, Paul and {du Mas des Bourboux}, H{\'e}lion and {McManus}, Sean and {Meier}, Tobias Gabriel and {M{\'e}nard}, Brice and {Metcalfe}, Nigel and {Mu{\~n}oz-Guti{\'e}rrez}, Andrea and {Najita}, Joan and {Napier}, Kevin and {Narayan}, Gautham and {Newman}, Jeffrey A. and {Nie}, Jundan and {Nord}, Brian and {Norman}, Dara J. and {Olsen}, Knut A.~G. and {Paat}, Anthony and {Palanque-Delabrouille}, Nathalie and {Peng}, Xiyan and {Poppett}, Claire L. and {Poremba}, Megan R. and {Prakash}, Abhishek and {Rabinowitz}, David and {Raichoor}, Anand and {Rezaie}, Mehdi and {Robertson}, A.~N. and {Roe}, Natalie A. and {Ross}, Ashley J. and {Ross}, Nicholas P. and {Rudnick}, Gregory and {Safonova}, Sasha and {Saha}, Abhijit and {S{\'a}nchez}, F. Javier and {Savary}, Elodie and {Schweiker}, Heidi and {Scott}, Adam and {Seo}, Hee-Jong and {Shan}, Huanyuan and {Silva}, David R. and {Slepian}, Zachary and {Soto}, Christian and {Sprayberry}, David and {Staten}, Ryan and {Stillman}, Coley M. and {Stupak}, Robert J. and {Summers}, David L. and {Sien Tie}, Suk and {Tirado}, H. and {Vargas-Maga{\~n}a}, Mariana and {Vivas}, A. Katherina and {Wechsler}, Risa H. and {Williams}, Doug and {Yang}, Jinyi and {Yang}, Qian and {Yapici}, Tolga and {Zaritsky}, Dennis and {Zenteno}, A. and {Zhang}, Kai and {Zhang}, Tianmeng and {Zhou}, Rongpu and {Zhou}, Zhimin},
        title = "{Overview of the DESI Legacy Imaging Surveys}",
      journal = {\aj},
         year = 2019,
        month = may,
       volume = {157},
       number = {5},
          eid = {168},
        pages = {168},
          doi = {10.3847/1538-3881/ab089d},
archivePrefix = {arXiv},
       eprint = {1804.08657},
 primaryClass = {astro-ph.IM},
       adsurl = {https://ui.adsabs.harvard.edu/abs/2019AJ....157..168D}
}

@ARTICLE{Tadhunter1996,
       author = {{Tadhunter}, C.~N. and {Dickson}, R.~C. and {Shaw}, M.~A.},
        title = "{Young stars and scattered light in the powerful radio galaxy 3C 321}",
      journal = {\mnras},
         year = 1996,
        month = jul,
       volume = {281},
       number = {2},
        pages = {591-603},
          doi = {10.1093/mnras/281.2.591},
       adsurl = {https://ui.adsabs.harvard.edu/abs/1996MNRAS.281..591T}
}

@ARTICLE{stasinska2006,
       author = {{Stasi{\'n}ska}, Gra{\.z}yna and {Cid Fernandes}, Roberto and {Mateus}, Ab{\'\i}lio and {Sodr{\'e}}, Laerte and {Asari}, Natalia V.},
        title = "{Semi-empirical analysis of Sloan Digital Sky Survey galaxies - III. How to distinguish AGN hosts}",
      journal = {\mnras},
         year = 2006,
        month = sep,
       volume = {371},
       number = {2},
        pages = {972-982},
          doi = {10.1111/j.1365-2966.2006.10732.x},
archivePrefix = {arXiv},
       eprint = {astro-ph/0606724},
 primaryClass = {astro-ph},
       adsurl = {https://ui.adsabs.harvard.edu/abs/2006MNRAS.371..972S}
}

@ARTICLE{Worthey1994,
       author = {{Worthey}, Guy},
        title = "{Comprehensive Stellar Population Models and the Disentanglement of Age and Metallicity Effects}",
      journal = {\apjs},
         year = 1994,
        month = nov,
       volume = {95},
        pages = {107},
          doi = {10.1086/192096},
       adsurl = {https://ui.adsabs.harvard.edu/abs/1994ApJS...95..107W}
}

@ARTICLE{Holt2007,
       author = {{Holt}, J. and {Tadhunter}, C.~N. and {Gonz{\'a}lez Delgado}, R.~M. and {Inskip}, K.~J. and {Rodriguez Zaurin}, J. and {Emonts}, B.~H.~C. and {Morganti}, R. and {Wills}, K.~A.},
        title = "{The properties of the young stellar populations in powerful radio galaxies at low and intermediate redshifts}",
      journal = {\mnras},
         year = 2007,
        month = oct,
       volume = {381},
       number = {2},
        pages = {611-639},
          doi = {10.1111/j.1365-2966.2007.12140.x},
archivePrefix = {arXiv},
       eprint = {0708.2605},
 primaryClass = {astro-ph},
       adsurl = {https://ui.adsabs.harvard.edu/abs/2007MNRAS.381..611H}
}

@ARTICLE{Gebhardt2000,
       author = {{Gebhardt}, Karl and {Bender}, Ralf and {Bower}, Gary and {Dressler}, Alan and {Faber}, S.~M. and {Filippenko}, Alexei V. and {Green}, Richard and {Grillmair}, Carl and {Ho}, Luis C. and {Kormendy}, John and {Lauer}, Tod R. and {Magorrian}, John and {Pinkney}, Jason and {Richstone}, Douglas and {Tremaine}, Scott},
        title = "{A Relationship between Nuclear Black Hole Mass and Galaxy Velocity Dispersion}",
      journal = {\apjl},
         year = 2000,
        month = aug,
       volume = {539},
       number = {1},
        pages = {L13-L16},
          doi = {10.1086/312840},
archivePrefix = {arXiv},
       eprint = {astro-ph/0006289},
 primaryClass = {astro-ph},
       adsurl = {https://ui.adsabs.harvard.edu/abs/2000ApJ...539L..13G}
}

@ARTICLE{Park2022,
       author = {{Park}, Daeseong and {Barth}, Aaron J. and {Ho}, Luis C. and {Laor}, Ari},
        title = "{A New Iron Emission Template for Active Galactic Nuclei. I. Optical Template for the H{\ensuremath{\beta}} Region}",
      journal = {\apjs},
         year = 2022,
        month = feb,
       volume = {258},
       number = {2},
          eid = {38},
        pages = {38},
          doi = {10.3847/1538-4365/ac3f3e},
archivePrefix = {arXiv},
       eprint = {2111.15118},
 primaryClass = {astro-ph.GA},
       adsurl = {https://ui.adsabs.harvard.edu/abs/2022ApJS..258...38P}
}

@ARTICLE{Park2012,
       author = {{Park}, Daeseong and {Woo}, Jong-Hak and {Treu}, Tommaso and {Barth}, Aaron J. and {Bentz}, Misty C. and {Bennert}, Vardha N. and {Canalizo}, Gabriela and {Filippenko}, Alexei V. and {Gates}, Elinor and {Greene}, Jenny E. and {Malkan}, Matthew A. and {Walsh}, Jonelle},
        title = "{The Lick AGN Monitoring Project: Recalibrating Single-epoch Virial Black Hole Mass Estimates}",
      journal = {\apj},
         year = 2012,
        month = mar,
       volume = {747},
       number = {1},
          eid = {30},
        pages = {30},
          doi = {10.1088/0004-637X/747/1/30},
archivePrefix = {arXiv},
       eprint = {1111.6604},
 primaryClass = {astro-ph.CO},
       adsurl = {https://ui.adsabs.harvard.edu/abs/2012ApJ...747...30P}
}

@ARTICLE{Barth2015,
       author = {{Barth}, Aaron J. and {Bennert}, Vardha N. and {Canalizo}, Gabriela and {Filippenko}, Alexei V. and {Gates}, Elinor L. and {Greene}, Jenny E. and {Li}, Weidong and {Malkan}, Matthew A. and {Pancoast}, Anna and {Sand}, David J. and {Stern}, Daniel and {Treu}, Tommaso and {Woo}, Jong-Hak and {Assef}, Roberto J. and {Bae}, Hyun-Jin and {Brewer}, Brendon J. and {Cenko}, S. Bradley and {Clubb}, Kelsey I. and {Cooper}, Michael C. and {Diamond-Stanic}, Aleksandar M. and {Hiner}, Kyle D. and {H{\"o}nig}, Sebastian F. and {Hsiao}, Eric and {Kandrashoff}, Michael T. and {Lazarova}, Mariana S. and {Nierenberg}, A.~M. and {Rex}, Jacob and {Silverman}, Jeffrey M. and {Tollerud}, Erik J. and {Walsh}, Jonelle L.},
        title = "{The Lick AGN Monitoring Project 2011: Spectroscopic Campaign and Emission-line Light Curves}",
      journal = {\apjs},
         year = 2015,
        month = apr,
       volume = {217},
       number = {2},
          eid = {26},
        pages = {26},
          doi = {10.1088/0067-0049/217/2/26},
archivePrefix = {arXiv},
       eprint = {1503.01146},
 primaryClass = {astro-ph.GA},
       adsurl = {https://ui.adsabs.harvard.edu/abs/2015ApJS..217...26B}
}

@ARTICLE{Runco2016,
       author = {{Runco}, Jordan N. and {Cosens}, Maren and {Bennert}, Vardha N. and {Scott}, Bryan and {Komossa}, S. and {Malkan}, Matthew A. and {Lazarova}, Mariana S. and {Auger}, Matthew W. and {Treu}, Tommaso and {Park}, Daeseong},
        title = "{Broad H{\ensuremath{\beta}} Emission-line Variability in a Sample of 102 Local Active Galaxies}",
      journal = {\apj},
         year = 2016,
        month = apr,
       volume = {821},
       number = {1},
          eid = {33},
        pages = {33},
          doi = {10.3847/0004-637X/821/1/33},
archivePrefix = {arXiv},
       eprint = {1603.00035},
 primaryClass = {astro-ph.GA},
       adsurl = {https://ui.adsabs.harvard.edu/abs/2016ApJ...821...33R}
}

@ARTICLE{Yu2023,
       author = {{Yu}, Zhibo and {Zou}, Fan and {Brandt}, William N.},
        title = "{Stellar Masses and Star Formation Rates of Galaxies and AGNs in the eFEDS GAMA09 Field}",
      journal = {Research Notes of the American Astronomical Society},
         year = 2023,
        month = nov,
       volume = {7},
       number = {11},
          eid = {248},
        pages = {248},
          doi = {10.3847/2515-5172/ad0ed7},
archivePrefix = {arXiv},
       eprint = {2311.16283},
 primaryClass = {astro-ph.GA},
       adsurl = {https://ui.adsabs.harvard.edu/abs/2023RNAAS...7..248Y}
}

@ARTICLE{Ricci_Trakhtenbrot2023,
       author = {{Ricci}, Claudio and {Trakhtenbrot}, Benny},
        title = "{Changing-look active galactic nuclei}",
      journal = {Nature Astronomy},
         year = 2023,
        month = nov,
       volume = {7},
        pages = {1282-1294},
          doi = {10.1038/s41550-023-02108-4},
archivePrefix = {arXiv},
       eprint = {2211.05132},
 primaryClass = {astro-ph.GA},
       adsurl = {https://ui.adsabs.harvard.edu/abs/2023NatAs...7.1282R}
}

@ARTICLE{Shen2013,
       author = {{Shen}, Yue},
        title = "{The mass of quasars}",
      journal = {Bulletin of the Astronomical Society of India},
         year = 2013,
        month = mar,
       volume = {41},
       number = {1},
        pages = {61-115},
          doi = {10.48550/arXiv.1302.2643},
archivePrefix = {arXiv},
       eprint = {1302.2643},
 primaryClass = {astro-ph.CO},
       adsurl = {https://ui.adsabs.harvard.edu/abs/2013BASI...41...61S}
}

\appendix
\onecolumn

\section{Fitting parameters}
\label{sec:app:emission_line}

\begin{table*}[t]
\caption{Parameters and references for the \texttt{pPXF} templates for the continuum.}
\centering
\begin{tabular}{ccc} 
\hline
Template & Parameters & References \\
\hline\hline
Stellar Population & \begin{tabular}[c]{@{}c@{}}Padova isochrone\\ Salpeter IMF\\ Ages: 0.0631, 0.0794, 0.1, 0.1259, 0.1585, 0.1995, \\ 0.2512, 0.3162, 0.3981, 0.5012, 0.631, 0.7943, 1.0, \\ 1.2589, 1.5849, 1.9953, 2.5119, 3.1623, 3.9811, 5.0119, \\ 6.3096, 7.9433, 10.0, 12.5893, and 15.8489 Gyr \\ Metallicities: -1.71, -1.31,
-0.71, -0.4, 0.0, 0.22 [M/H] \\ Velocity: -300 to 500 km s$^{-1}$ \\ Velocity dispersion: 10 to 500 km s$^{-1}$ \\ Skweness and Kurtosis: -0.3 to 0.3\end{tabular}  & \citet{Vazdekis2016} \\
\hline
Fe II pseudo-continuum & \begin{tabular}[c]{@{}c@{}}FWHM: 1000, 1200, 1400, 1600, 1800, 2000, \\ 2400, 2800, 3400, 4000, 4800, 5800, \\ 7000, 8400, 10000, and 11800 km s$^{-1}$\end{tabular} & \begin{tabular}[c]{@{}c@{}}\citet{BorosonGreen1992}, \\ \citet{VestergaardWilkes2001}\end{tabular}\\
\hline
AGN Featureless Continuum & \begin{tabular}[c]{@{}c@{}}Power-law normalization: 5100 \AA \\ Power-law slope: -3.0 to 0.0 \end{tabular} & \citet{Bernal2025} \\
\hline
Balmer Continuum & Optical depth $\tau$: 0.1 to 2.0 & \citet{Bernal2025} \\
\hline
Balmer High-order emission & FWHM: 100 to 11000 km s$^{-1}$ & \citet{Bernal2025} \\
\hline
\end{tabular}
\label{tab:ppxf_continuum}
\end{table*}

\begin{table*}[t]
\caption{Parameters and references for the \texttt{PyQSOFit} templates for the continuum.}
\centering
\begin{tabular}{ccc} 
\hline
Template & Parameters & References \\
\hline\hline
Fe II pseudo-continuum & \begin{tabular}[c]{@{}c@{}} Normalization: 0 to $10^{10}$ \\ FWHM: 1200 to 10000 km s$^{-1}$ \\ Velocity shift: -0.01 to 0.01 $\times$ c \end{tabular} & \begin{tabular}[c]{@{}c@{}}\citet{BorosonGreen1992}, \\ \citet{VestergaardWilkes2001}, \\ \citet{Tsuzuki2006}, \\ \citet{Salviander2007}\end{tabular}\\
\hline
Power-law & \begin{tabular}[c]{@{}c@{}}Normalization: 0 to $10^{10}$ \AA \\ Slope: -3.0 to 2.0  \end{tabular} & \\
\hline
\hline 
Third order polynomium & Unbounded & \\
\hline
\end{tabular}
\label{tab:pyqsofit_continuum}
\end{table*}

\setlength{\LTcapwidth}{\textwidth}
\begin{longtable}{cccccccccccccc}
\caption{Configurations for the emission line fit for \texttt{PyQSOFit}. Global fits on top and local fits on the bottom part of the table. For completeness, we added all lines, though in our redshift range we do not fit lines with wavelengths smaller than Mg II.
\label{tab:emission_lines_pyqsofit}} \\
     \hline\hline 
      $\lambda_0$ & Complex & $\lambda_{\rm{min}}$ & $\lambda_{\rm{max}}$ & Line & $N_{\rm{G}}$ & $\sigma_{\rm{i}}$ & $\sigma_{\rm{min}}$ & $\sigma_{\rm{max}}$ & $v_{\rm{off}}$ & $v_{\rm{idx}}$ & $w_{\rm{idx}}$ & $f_{\rm{idx}}$ & $f_{\rm{val}}$ \\
     \hline
     \endfirsthead
     
     \caption*{\textbf{Table \thetable:} continued.}\\
     \hline\hline
      $\lambda_0$ & Complex & $\lambda_{\rm{min}}$ & $\lambda_{\rm{max}}$ & Line & $N_{\rm{G}}$ & $\sigma_{\rm{i}}$ & $\sigma_{\rm{min}}$ & $\sigma_{\rm{max}}$ & $v_{\rm{off}}$ & $v_{\rm{idx}}$ & $w_{\rm{idx}}$ & $f_{\rm{idx}}$ & $f_{\rm{val}}$ \\
      \hline
     \endhead
     
     \hline 
     \endfoot
     
     \hline
     \endlastfoot

6732.67 & Ha & 6400 & 6800 & SII6732 & 1 & 0.001 & 0.00023 & 0.0017 & 0.005 & 1 & 1 & 0 & - \\
6718.29 & Ha & 6400 & 6800 & SII6718 & 1 & 0.001 & 0.00023 & 0.0017 & 0.005 & 1 & 1 & 0 & - \\
6585.28 & Ha & 6400 & 6800 & NII6585 & 1 & 0.001 & 0.00023 & 0.0017 & 0.005 & 1 & 1 & 1 & 0.003 \\
6564.61 & Ha & 6400 & 6800 & Halpha\_na & 1 & 0.001 & 0.00023 & 0.0014 & 0.005 & 1 & 1 & 0 & - \\
6564.61 & Ha & 6400 & 6800 & Halpha\_br & 3 & 0.005 & 0.0014 & 0.03 & 0.01 & 0 & 0 & 0 & - \\
6549.85 & Ha & 6400 & 6800 & NII6549 & 1 & 0.001 & 0.00023 & 0.0017 & 0.005 & 1 & 1 & 1 & 0.001 \\
5008.24 & Hb & 4630 & 5050 & OIII5007c & 1 & 0.001 & 0.00023 & 0.0017 & 0.006 & 1 & 1 & 1 & 0.003 \\
5008.24 & Hb & 4630 & 5050 & OIII5007w & 1 & 0.003 & 0.00023 & 0.004 & 0.006 & 2 & 2 & 2 & 0.003 \\
4960.30 & Hb & 4630 & 5050 & OIII4959c & 1 & 0.001 & 0.00023 & 0.0017 & 0.006 & 1 & 1 & 1 & 0.001 \\
4960.30 & Hb & 4630 & 5050 & OIII4959w & 1 & 0.003 & 0.00023 & 0.004 & 0.006 & 2 & 2 & 2 & 0.001 \\
4862.68 & Hb & 4630 & 5050 & Hbeta\_na & 1 & 0.001 & 0.00023 & 0.0014 & 0.005 & 1 & 1 & 0 & - \\
4862.68 & Hb & 4630 & 5050 & Hbeta\_br & 3 & 0.005 & 0.0014 & 0.03 & 0.01 & 0 & 0 & 0 & - \\
4685.68 & Hb & 4630 & 5050 & HeII\_na & 1 & 0.001 & 0.00023 & 0.0014 & 0.005 & 1 & 0 & 0 & - \\
4685.68 & Hb & 4630 & 5050 & HeII\_br & 1 & 0.005 & 0.0014 & 0.03 & 0.01 & 1 & 0 & 0 & - \\
2798.75 & MgII & 2700 & 2900 & MgII\_na & 1 & 0.001 & 0.0005 & 0.0014 & 0.01 & 0 & 0 & 0 & - \\
2798.75 & MgII & 2700 & 2900 & MgII\_br & 2 & 0.005 & 0.0014 & 0.05 & 0.01 & 0 & 0 & 0 & - \\
1908.73 & CIII & 1700 & 1970 & CIII\_br1 & 1 & 0.005 & 0.0014 & 0.05 & 0.015 & 3 & 0 & 0 & 0.01 \\
1908.73 & CIII & 1700 & 1970 & CIII\_br2 & 1 & 0.005 & 0.0014 & 0.05 & 0.015 & 3 & 0 & 0 & 0.01 \\
1892.03 & CIII & 1700 & 1970 & SiIII1892 & 1 & 0.002 & 0.001 & 0.015 & 0.003 & 1 & 1 & 0 & 0.005 \\
1857.40 & CIII & 1700 & 1970 & AlIII1857 & 1 & 0.002 & 0.001 & 0.015 & 0.003 & 1 & 1 & 0 & 0.005 \\
1816.98 & CIII & 1700 & 1970 & SiII1816 & 1 & 0.002 & 0.001 & 0.015 & 0.01 & 2 & 2 & 0 & 0.0002 \\
1750.26 & CIII & 1700 & 1970 & NIII1750 & 1 & 0.002 & 0.001 & 0.015 & 0.01 & 2 & 2 & 0 & 0.001 \\
1718.55 & CIII & 1700 & 1900 & NIV1718 & 1 & 0.002 & 0.001 & 0.015 & 0.01 & 2 & 2 & 0 & 0.001 \\
1549.06 & CIV & 1500 & 1700 & CIV\_br & 3 & 0.005 & 0.0014 & 0.05 & 0.015 & 0 & 0 & 0 & 0.05 \\
1640.42 & CIV & 1500 & 1700 & HeII1640 & 1 & 0.001 & 0.0005 & 0.0017 & 0.008 & 1 & 1 & 0 & 0.002 \\
1663.48 & CIV & 1500 & 1700 & OIII1663 & 1 & 0.001 & 0.0005 & 0.0017 & 0.008 & 1 & 1 & 0 & 0.002 \\
1640.42 & CIV & 1500 & 1700 & HeII1640\_br & 1 & 0.005 & 0.0017 & 0.02 & 0.008 & 2 & 2 & 0 & 0.002 \\
1663.48 & CIV & 1500 & 1700 & OIII1663\_br & 1 & 0.005 & 0.0017 & 0.02 & 0.008 & 2 & 2 & 0 & 0.002 \\
1402.06 & SiIV & 1290 & 1450 & SiIV\_OIV1 & 1 & 0.005 & 0.0014 & 0.05 & 0.015 & 1 & 1 & 0 & 0.05 \\
1396.76 & SiIV & 1290 & 1450 & SiIV\_OIV2 & 1 & 0.005 & 0.0014 & 0.05 & 0.015 & 1 & 1 & 0 & 0.05 \\
1335.30 & SiIV & 1290 & 1450 & CII1335 & 1 & 0.002 & 0.001 & 0.015 & 0.01 & 2 & 2 & 0 & 0.001 \\
1304.35 & SiIV & 1290 & 1450 & OI1304 & 1 & 0.002 & 0.001 & 0.015 & 0.01 & 2 & 2 & 0 & 0.001 \\
1215.67 & Lya & 1150 & 1290 & Lya\_br & 3 & 0.005 & 0.0014 & 0.05 & 0.02 & 0 & 0 & 0 & 0.05 \\
1240.14 & Lya & 1150 & 1290 & NV1240 & 1 & 0.002 & 0.001 & 0.01 & 0.005 & 0 & 0 & 0 & 0.002 \\
\hline \\
6376.27 & Fe\_OI & 5690 & 6400 & FeX6376 & 1 & 0.001 & 0.00023 & 0.0017 & 0.005 & 1 & 0 & 0 & - \\
6300.30 & Fe\_OI & 5690 & 6400 & OI6300 & 1 & 0.001 & 0.00023 & 0.0017 & 0.005 & 1 & 0 & 0 & - \\
6088.68 & Fe\_OI & 5690 & 6400 & FeVII6088 & 1 & 0.001 & 0.00023 & 0.0017 & 0.005 & 1 & 0 & 0 & - \\
5877.25 & Fe\_OI & 5690 & 6400 & HeI\_na & 1 & 0.001 & 0.00023 & 0.0014 & 0.005 & 1 & 0 & 0 & - \\
5877.25 & Fe\_OI & 5690 & 6400 & HeI\_br & 1 & 0.005 & 0.0014 & 0.02 & 0.01 & 1 & 0 & 0 & - \\
4364.44 & Hg\_Hd & 4000 & 4470 & OIII4363 & 1 & 0.001 & 0.00023 & 0.0017 & 0.005 & 1 & 0 & 0 & - \\
4341.68 & Hg\_Hd & 4000 & 4470 & Hgamma\_na & 1 & 0.001 & 0.00023 & 0.0014 & 0.005 & 1 & 1 & 0 & - \\
4341.68 & Hg\_Hd & 4000 & 4470 & Hgamma\_br & 1 & 0.005 & 0.0014 & 0.02 & 0.01 & 1 & 0 & 0 & - \\
4102.89 & Hg\_Hd & 4000 & 4470 & Hdelta\_na & 1 & 0.001 & 0.00023 & 0.0014 & 0.005 & 1 & 1 & 0 & - \\
4102.89 & Hg\_Hd & 4000 & 4470 & Hdelta\_br & 1 & 0.005 & 0.0014 & 0.02 & 0.01 & 1 & 0 & 0 & - \\
3968.59 & NeIII\_OII & 3650 & 4000 & NeIII3967 & 1 & 0.001 & 0.00023 & 0.0017 & 0.005 & 1 & 0 & 0 & - \\
3869.85 & NeIII\_OII & 3650 & 4000 & NeIII3869 & 1 & 0.001 & 0.00023 & 0.0017 & 0.005 & 1 & 0 & 0 & - \\
3759.99 & NeIII\_OII & 3650 & 4000 & FeVII3759 & 1 & 0.001 & 0.00023 & 0.0017 & 0.005 & 1 & 0 & 0 & - \\
3728.48 & NeIII\_OII & 3650 & 4000 & OII3728 & 1 & 0.001 & 0.00023 & 0.0017 & 0.005 & 1 & 0 & 0 & - \\
3426.86 & NeV & 3310 & 3480 & NeV3426\_na & 1 & 0.001 & 0.00023 & 0.0014 & 0.005 & 1 & 1 & 0 & - \\
3426.86 & NeV & 3310 & 3480 & NeV3426\_br & 1 & 0.005 & 0.0014 & 0.01 & 0.01 & 0 & 0 & 0 & - \\
3346.78 & NeV & 3310 & 3480 & NeV3346 & 1 & 0.001 & 0.00023 & 0.0017 & 0.005 & 1 & 1 & 0 & - \\
2423.59 & CII & 2300 & 2460 & NeIV2422 & 1 & 0.001 & 0.00023 & 0.0017 & 0.01 & 0 & 0 & 0 & 0.002 \\
2324.81 & CII & 2300 & 2460 & CII2326 & 1 & 0.001 & 0.00023 & 0.0017 & 0.01 & 0 & 0 & 0 & 0.002 \\
1034.76 & OIV & 1000 & 1060 & OIV1035 & 1 & 0.002 & 0.001 & 0.01 & 0.005 & 0 & 0 & 0 & 0.002 \\
\hline
\end{longtable}
\tablefoot{Column 1 ($\lambda_0$) specifies the central wavelength in vacuum of each specified line, in Angstroms. 
Column 2 (Complex) indicates the complex of lines, meaning that the lines belonging to the same complex will be fitted together due to their proximity in wavelength range.
Columns 3 ($\lambda_{\rm{min}}$) and 4 ($\lambda_{\rm{max}}$) indicate the minimum and maximum wavelengths in Angstroms for the fit of that line complex, respectively.
Column 5 (Line) shows the line component that is being fitted. The suffixes `\_br', `\_na', `c', and `w' indicate broad, narrow, core, and wing components, respectively.
Column 6 ($N_{\rm{G}}$) indicates the number of Gaussians considered for the fit of the line.
Columns 7 ($\sigma_{\rm{i}}$), 8 ($\sigma_{\rm{min}}$), and 9 ($\sigma_{\rm{max}}$) display the initial, minimum, and maximum values of sigma, the Gaussian width of the line, expressed as the natural logarithm of wavelength in Angstroms. 
In this scale, the FWHM is calculated by $2.3\times\sigma\times c$, where c is the speed of light in km s$^{-1}$. 
Column 10 ($v_{\rm{off}}$) shows the velocity offset as $v_{\rm{off}}=\rm{ln}(\lambda_0/\lambda)$.
Column 11 ($v_{\rm{idx}}$) is the tied velocity offset index, which sets fixed values of the velocity offset for all lines in the same line complex that have the same index.
Column 12 ($w_{\rm{idx}}$) is the tied width index.
Column 13 ($f_{\rm{idx}}$) is the tied flux index, and the ratio of the fluxes that have the same flux index within the same line complex is defined in Column 14 ($f_{\rm{val}}$) with the fixed flux value.}

\begin{table*}[h]
\caption{Width and offset parameters of the emission line fitting with \texttt{PyQSOFit} from Table \ref{tab:emission_lines_pyqsofit} in km s$^{-1}$.}
\centering
\begin{tabular}{cccc} 
\hline
Line & \begin{tabular}[c]{@{}c@{}}FWHM$_{\rm min}$\\ {[}km s$^{-1}${]}\end{tabular} & \begin{tabular}[c]{@{}c@{}}FWHM$_{\rm max}$\\ {[}km s$^{-1}${]}\end{tabular} & \begin{tabular}[c]{@{}c@{}}$v_{\rm{off}}$\\ {[}km s$^{-1}${]}\end{tabular} \\
\hline\hline
Most single narrow & 160 & 1\,200 & 1\,500 \\
Balmer, He, [Ne V] narrow & 160 & 1\,000 & 1\,500 \\\relax
H$\alpha$, H$\beta$, He II broad & 1\,000 & 21\,000 & 3\,000 \\\relax
H$\gamma$, H$\delta$, He I broad & 1\,000 & 14\,000 & 3\,000 \\\relax
[Ne V] broad & 1\,000 & 7\,000 & 3\,000 \\\relax
[O III] core & 160 & 1\,200 & 1\,800 \\\relax
[O III] wing & 160 & 2\,800 & 1\,800 \\
Mg II narrow & 350 & 1\,000 & 3\,000 \\
Mg II broad & 1\,000 & 35\,000 & 3\,000 \\
\hline
\end{tabular}
\label{tab:emission_lines_pyqsofit_kms}
\end{table*}

The continuum of \texttt{pPXF} fits was done with fewer details than the continuum fit of \texttt{PyQSOFit} due to some characteristics of the code, which performs a linear combination instead of the fitting of each parameter.
Therefore, the AGN continuum component added to \texttt{pPXF} fit has a more simplistic approach to account for a general trend of the AGN contribution, while the \texttt{PyQSOFit} one is more detailed to provide the measurements of different components contributing to the AGN continuum emission.
Table \ref{tab:ppxf_continuum} describes the parameters and references for the continuum fitting with \texttt{pPXF}, while Table \ref{tab:pyqsofit_continuum} does similarly for the continuum fitting using \texttt{PyQSOFit}.

The emission lines were fitted with \texttt{PyQSOFit} using the parameters according to Table \ref{tab:emission_lines_pyqsofit}.
The set of lines at the top set of Table \ref{tab:emission_lines_pyqsofit} was fitted together with the continuum and the fit for the whole wavelength range, while the set of lines at the bottom of the Table was fitted locally due to being fainter and often absent.
The same parameters are summarized in km s$^{-1}$ in Table \ref{tab:emission_lines_pyqsofit_kms}.
For the distinction between narrow and broad lines, we used the approximate value of FWHM$=1000$ km s$^{-1}$.
Some lines have a specific set of parameters (e.g., [O III] and [Ne V]) to avoid issues arising from possible blending.

\section{Deriving stellar mass}
\label{sec:app:stellar_mass}

The stellar mass is the direct product of the stellar luminosity and mass-to-light ratio (M/L). 
Since the luminosity depends linearly on flux, both measurements of the flux and of M/L carry similar weight in the calculation of stellar mass. 
The flux is integrated on the coverage of the \textit{r} DECam band in the fitted SSP spectral flux, while the M/L is an output from \texttt{pPXF} from the SSP analysis (also considering DECam \textit{r} photometric band coverage).

From the Monte Carlo runs, the average statistical error in the flux (and luminosity) is dependent on the AGN contribution, ranging from 2\% for the sample GAL to 12\% for the sample QSO. Regarding the M/L, the average statistical uncertainty is 3-10\% of the measurement, with lower values for the GAL sample than for the QSO sample. 
Therefore, both measurements have a similar stability in their statistical uncertainties after adding random noise. 

\begin{figure}[t]
     \includegraphics[width=0.5\columnwidth]{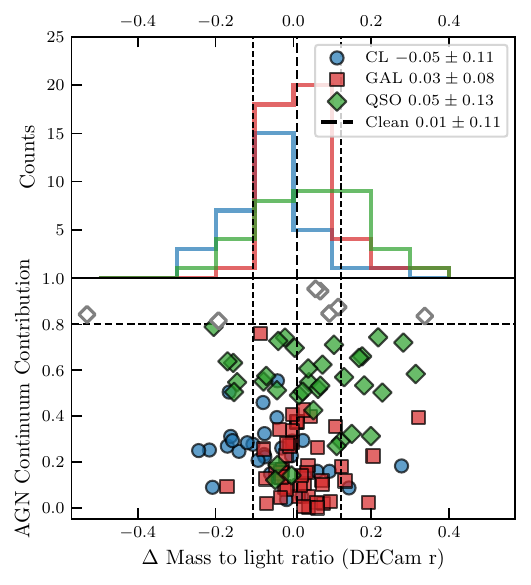}
    \includegraphics[width=0.5\columnwidth]{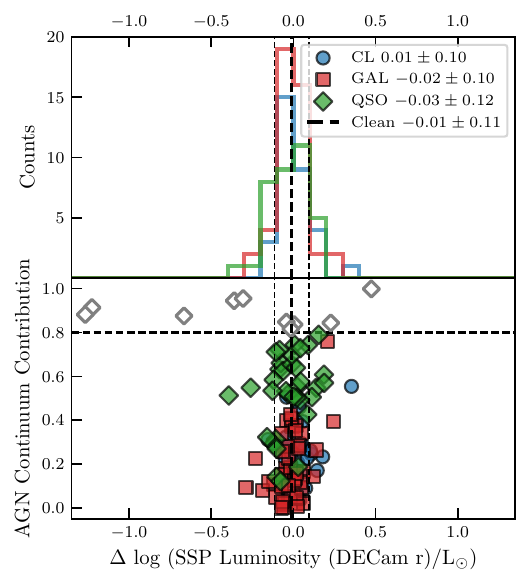}
     \caption{The difference of the mass-to-light ratio (left) and SSP luminosity (right), as in Fig. \ref{fig:results:hist_stellar_mass}.
     Both are based on spectral flux integrated over the photometric DECam \textit{r}-band coverage.}
\label{fig:app:hist_ML_SSPlum}
\end{figure}

To discuss and expand our estimates of the systematic uncertainties, we present the internal differences between the bright-dim and daily-allepoch observations for M/L and SSP luminosity in Fig. \ref{fig:app:hist_ML_SSPlum}. 
As expected, sample GAL has lower scatter in the internal comparison than sample QSO for both variables.
We also emphasize the importance of the cut for sources with $f_{\rm{AGN}}>0.8$, especially for the stellar luminosity, since this removes the main outliers from the QSO sample.
Both the M/L and the SSP luminosity have an overall scatter (after removing objects with $f_{\rm{AGN}}>0.8$) of 0.11 dex.
However, since the offsets of each sample have different signs when comparing M/L and SSP luminosity (positive for the M/L and negative for luminosity for samples GAL and QSO, and negative for M/L and positive for luminosity for sample CL), the scatter of the stellar mass comparisons is larger than that of each individual component of its calculation (0.14 dex, as in Fig. \ref{fig:results:hist_stellar_mass}).

Regarding the influence of the S/N on the results for the M/L and the SSP luminosity, we notice once again a stability of these methods, since both have a decrease of 0.01 dex when considering S/N$>5$, and a decrease of 0.03 dex for S/N$>10$.
Therefore, the offset and scatter for S/N$>10$ data is $-0.01
\pm0.08$ dex for M/L and $0.001 \pm0.07$ for the SSP luminosity.

Summarizing the uncertainties in stellar mass estimates, the Monte Carlo analysis yields an average statistical uncertainty of 3\% for sample GAL to 12\% for sample QSO, similar to the results obtained for M/L and SSP luminosity, indicating that this is a robust measurement across runs with random noise applied to the same sources. 
However, because this robustness is not as tight when comparing different observations of the same source, we argue that the systematic uncertainty of 0.14 dex obtained in the internal validation is more realistic than relying on the statistical uncertainty.
The choice to use 0.14 dex as the uncertainty is also supported by external validation tests (see Fig. \ref{fig:results:stellarmass_hsc_grahsp}), which yielded ~0.3-0.4 dex systematic differences across methods applied to the same sample.

\section{The Age-Metallicity Degeneracy}
\label{sec:app:degeneracies}

Reliably estimating the age and metallicity of stellar populations from optical data is complicated by the known degeneracy between these parameters, especially for old stellar populations \citep[e.g.,][]{Worthey1994, RodriguezMerino2020}.
Different combinations of age, metallicity, and dust can produce similar SEDs and spectral shapes; therefore, these properties will not be accurately measured unless the spectra have very high spectral resolution, S/N, and a large wavelength coverage \citep{Angthopo2024}.
This degeneracy occurs because older populations, metal-rich stars, and dust all contribute to reddening the observed emission.

\begin{figure}[t]
     \includegraphics[width=0.5\columnwidth]{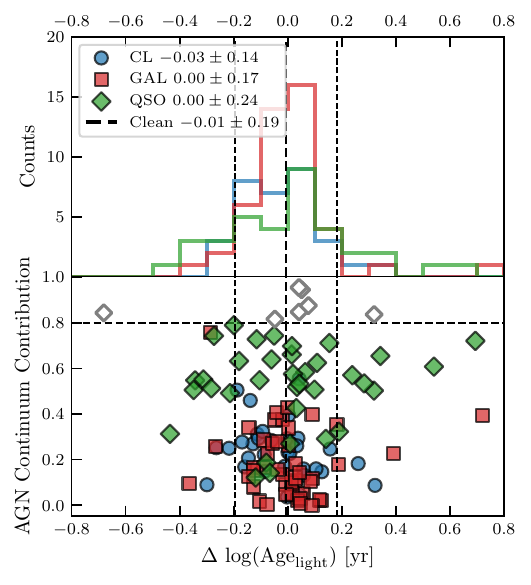}
    \includegraphics[width=0.5\columnwidth]{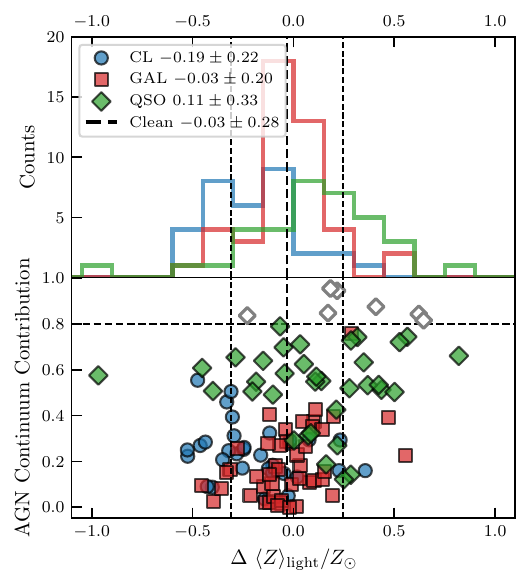}
     \caption{The difference of the average SSP light-weighted age (left) and metallicity (right), as in Fig. \ref{fig:results:hist_stellar_mass}.
     The bottom panel displays observations with $f_{\rm{AGN}}>0.8$, marked with empty markers; these are excluded from the Clean statistics estimate.}
\label{fig:app:hist_age_metal}
\end{figure}

Figure \ref{fig:app:hist_age_metal} shows the comparisons of the light-weighted average measurements of the age and metallicity for our samples.
However, the interpretation of such plots should not be based solely on the dispersion, since degeneracy could cause measurements from different observations to agree even if they are not accurate with respect to the object's stellar populations.
This behavior is evident in the age distributions of spectra with $f_{\rm{AGN}}>0.8$, with some cases showing dispersions of $<0.1$ dex, although they remain associated with unreliable measurements of the host galaxy's properties.
In cases dominated by AGN emission, the blue part of the continuum will be assigned mainly to the power-law featureless continuum, missing the contribution from young stellar populations \citep[e.g.,][]{Tadhunter1996}, and therefore assessing high ages and metallicities to the quasar-dominated spectra, which might indicate the AGN outshining the stellar population more than the accurate properties of the host galaxy.

Despite the above-mentioned degeneracies, improving the sample's S/N improves the results.
For the age offset and scatter, considering S/N$>5$ yields $-0.01\pm0.15$, while for S/N$>10$ we obtain $-0.03\pm0.13$.
For the metallicity, we have $-0.06\pm0.26$ for a sample of S/N$>5$ and $-0.06\pm0.21$ for S/N$>10$.

The fact that the quasar-dominated objects, which are in general bluer, will have their stellar populations estimated as redder, since the young population contribution will be attributed to the AGN power-law emission, enforces the recommendation of only trusting the host-galaxy properties for $f_{\rm{AGN}}<0.8$.
We also emphasize that the output of \texttt{pPXF} provides the weights for all the 150 E-MILES SSPs considered for the fit; therefore, there is more information available regarding the contribution of stellar populations with different ages and metallicities than considering the averages alone.

\end{document}